\documentstyle[12pt,epsf,fleqn,twoside]{report}

\newcommand{\EQ}{\begin{equation}}
\newcommand{\EN}{\end{equation}}
\newcommand{\bea}{\begin{eqnarray}}
\newcommand{\ena}{\end{eqnarray}}
\newcommand{\beas}{\begin{eqnarray*}}
\newcommand{\enas}{\end{eqnarray*}}
\renewcommand{\a}{\alpha}
\renewcommand{\b}{\beta}
\renewcommand{\l}{\lambda}

\newcommand{\Tr}{\mbox{Tr}\,}
\renewcommand{\Re}{\mbox{Re}\,}
\newcommand{\D}{{\cal D}}
\renewcommand{\L}{{\cal L}}
\newcommand{\M}{{\cal M}}
\renewcommand{\O}{{\cal O}}

\newcommand{\ra}{\rightarrow}
\newcommand{\pa}{\partial}
\newcommand{\z}{\zeta}
\newcommand{\ep}{\epsilon}
\begin{document}
\pagestyle{empty}
\pagenumbering{roman}                    
\setcounter{page}{1}

\vspace{-1.5cm}
\begin{center}
{\Large UNIVERSIT\`{A} DEGLI STUDI DI TRENTO} \\
\vspace{4mm}

\noindent{\Large Facolt\`{a} di Scienze Matematiche Fisiche e Naturali } \\
\vspace{1.5cm}

\noindent{\large ANNO ACCADEMICO 1996/1997} \\
\vspace{1cm}

\noindent{\Large X Ciclo di Dottorato in Fisica} \\
\vspace{3cm}

\noindent{\Large Tesi di Dottorato di Ricerca in Fisica} \\
\vspace{2cm}       

\noindent{\Large ASPECTS AND APPLICATIONS OF} \\
\vspace{7mm}                                    

\noindent{\Large QUANTUM FIELD THEORY ON SPACES} \\
\vspace{4mm}                                    

\noindent{\Large WITH CONICAL SINGULARITIES} \\
\vspace{2.5cm}

\noindent{\Large by Devis Iellici}
\end{center}

\newpage 
\

\newpage
\begin{tabbing}
\hspace{8cm}\={\large\em a Manuela }\\
\>{\large\em ai miei genitori}
\end{tabbing}
\vspace{8cm}

\noindent{\large\bf Ringraziamenti}\\

\noindent
Desidero ringraziare il mio supervisore prof. Sergio
Zerbini per avermi dato la possibilit\`a di studiare 
gli interessanti argomenti che sono oggetto di questa tesi,
per le numerose discussioni ed utili consigli.

Vorrei inoltre ringraziare il mio collega e coautore dr. 
Valter Moretti per la sua collaborazione fondamentale per 
ottenere molti dei risultati esposti in questa tesi. 

Sono grato al mio controrelatore prof. Emilio Elizalde
per i suggerimenti e le critiche costruttive.

Ci sono numerose altre persone che vorrei ringraziare, in particolare
il dr. Guido Cognola, il prof. Ruggero Ferrari, il dr. Dimitri V.
Fursaev, il dr. Edisom S. Moreira Jnr., il dr. Giuseppe Nardelli, il
prof. Marco Toller, il dr. Luciano Vanzo e il dr. Andrei Zelnikov.
Altri ringraziamenti per alcuni suggerimenti specifici si trovano in
fondo ai lavori pubblicati.

Infine, vorrei ringraziare i miei compagni del X ciclo di
dottorato Stefano Dusini,  Riccardo Giannitrapani
e Franco  Toninato per questi tre anni irripetibili.

\newpage
\

\newpage
\pagestyle{myheadings}
\noindent {\Large\bf Prefazione}
\bigskip

\noindent Questa tesi tratta di alcuni aspetti della teoria
quantistica dei campi su spazi che contengono singolarit\`a
di tipo conico, con particolare riferimento al caso della
stringa cosmica, della teoria a temperatura finita nello
spazio di Rindler e  vicino all'orizzonte
di un buco nero di Schwarzschild.

La tesi riassume i risultati ottenuti durante gli ultimi due anni
del corso di Dottorato di Ricerca, sia gi\`a pubblicati 
in alcuni articoli che del materiale inedito. I lavori pubblicati,
o comunque accettati per la pubblicazione al momento della
stesura della presente tesi,  corrispondono alla bibliografia
\cite{ielmo,sigrav,moiel,massive,fluctuations}.

Il lavoro \`e organizzato nel modo seguente. Dopo un'introduzione
generale, il Capitolo \ref{ONTHECONE} ha carattere introduttivo
alla singolarit\`a conica e alle tecniche di regolarizzazione basate
sul nucleo del calore e la funzione $\zeta$. Il Capitolo
\ref{FLUCTUATIONS}, basato su \cite{fluctuations}, ha interesse
generale per la teoria quantistica dei campi sugli spazi curvi,
introducendo un nuovo metodo per la regolarizzazione delle
fluttuazioni del vuoto usando la funzione $\zeta$. Nel Capitolo
\ref{massivechap}, basato su \cite{massive}, viene esteso il metodo
della funzione $\zeta$ sul cono al caso di un campo con massa. Il
risultato ottenuto viene poi usato per calcolare il tensore
energia-impulso, le fluttuazioni del campo e la contro-reazione delle
fluttuazioni quantistiche sulla metrica di una stringa cosmica. Il
Capitolo  \ref{BHENTROPY}, che contiene la maggior parte del
materiale inedito, tratta del calcolo delle correzioni quantistiche
all'entropia di Bekenstein-Hawking usando il metodo della
singolarit\`a conica. Nel Capitolo \ref{PHOTONS}, basato su
\cite{ielmo,sigrav}, viene esteso il metodo della singolarit\`a conica
al caso del campo del fotone e del gravitone, risolvendo alcuni
problemi non banali legati all'invarianza di gauge. Nel
Capitolo \ref{OPTICCHAP}, che \`e una rielaborazione di \cite{moiel} con
dei contributi inediti, la teoria del campi sullo spazio conico viene
studiata usando il metodo della trasformazione conforme alla metrica
ottica, sia per il campo scalare che per il campo elettromagnetico,
mettendo in evidenza alcuni problemi della teoria termica nello
spazio conico. Infine, nelle conclusioni vengono sottolineati i
principali problemi aperti della teoria sul cono e degli altri
argomenti trattati.

La tesi \`e scritta in inglese. A questo proposito vorrei ringraziare
il Ministero dell'Universit\`a e della Ricerca Scientifica e
Tecnologica per aver finalmente dato la possibilt\`a di scrivere le
tesi di dottorato in lingue diverse dall'italiano, favorendo in questo
modo lo scambio dell'informazione nella comunit\`a scientifica e
l'utilit\`a  stessa della tesi a livello internazionale.

\newpage
\
\tableofcontents

\newpage

\pagenumbering{arabic}
\setcounter{page}{1}

\chapter*{Introduction}

\markboth{Introduction}{Introduction}

The aim of this thesis is to investigate some aspects of the
behaviour of quantum fields propagating in curved or non-trivial
spacetimes. The gravitational field, namely the metric of the
background, is treated classically, while the matter quantum fields are
quantized. This approach is very useful when the interaction between
fields and gravity is important, but not the quantum nature of the
gravitational field itself. It can be reasonably assumed that
non-trivial gravitational effects occur  when one considers
quantum modes with  a wavelength comparable with the characteristic
curvature radius of the background spacetime. This approach resembles
the semiclassical calculations in the early days of quantum theory, in
which the electromagnetic field was considered as a classical
background field, interacting with the quantized matter.

Although one cannot expect that this approach gives final answers
about unification and quantum gravity, it is very important and
appealing because of the results obtained in the past twenty-five years
and, in particular, the open problems that has raised. Indeed, in some
of the topics faced in QFT in curved spacetimes one touches with the hand
the limits of the present theories of gravitation and quantum fields.
Furthermore, it is up to now the only way to say something about the
influence of the gravitational field on the quantum phenomena.

The Hawking radiation and the Unruh effect are among the most
important results of quantum field theory in curved spacetime and
provide a beautiful unification of aspects of quantum theory,
gravitation, and thermodynamics. Furthermore,  the lack of
an explanation of the huge entropy of a black hole in terms
of statistical mechanics, namely counting the number of possible
underlying quantum states,  is probably the clearest indication that
the present theories are close to their limits. We remind that, in
units $c=\hbar=G=k_B=1$, a ball of thermal radiation has mass $M\sim
T^4R^3$ and entropy $S\sim T^3R^3$, $T$ being the temperature and $R$
the radius of the ball. The radiation will form a black hole when
$R\sim M$, and so when $T\sim M^{-1/2}$ and $S\sim M^{3/2}$. In
contrast, the Bekenstein-Hawking entropy of the resulting black hole
is $S_{BH}\sim M^2$. Therefore, a black hole with a mass much larger
than the Planck mass, $M\gg 1$, has an entropy which is much larger
than the entropy of the radiation that formed it. It is worth
stressing that the thermal radiation has the highest entropy of
ordinary matter. Furthermore, when one tries to compute the
entropy of the quantum fluctuations living outside the horizon of a
black hole, which can be considered as the first quantum correction to
the thermodynamical  Bekenstein-Hawking entropy, one finds that it is
divergent and some cutoff is needed, such as a `fuzzy' horizon: in the
present theory it is difficult to find a satisfactory way to introduce
such a cutoff.

All these problems make the black-hole entropy and its quantum
corrections a very  interesting subject to study. In some sense,
there is an analogy with the ultraviolet catastrophe in the black-body
radiation, the study of which gave rise to the quantum theory.

Regarding the the problem of the explanation of the microscopical
origin of black hole entropy, it has be mentioned that much
progress has been done in the last two years in the context of
string theory. Indeed, a few years ago Susskind \cite{susskind93}
suggested that there should be a one to one correspondence
between strings and black holes, and that within this correspondence
the enormous black-hole entropy could be explained counting
the number of excited string states. Only very recently
some apparent contradictions in the correspondence have been
resolved, but now within string theory it is possible to reproduce
the correct dependence on the mass and charges for essentially
all black holes, although only for certain black holes the numerical
coefficient of the entropy can be computed and shown to agree
(for a recent review see \cite{horowitz97}\cite{peet97}). This is clearly
a subject of great interest to discuss and which
opens new prospects for the understanding of the black hole
entropy and related phenomena, but it is outside the
reach of this thesis and I am not discussing it here.

Many different methods have been developed for computing the
Bekenstein-Hawking entropy and its quantum corrections. Among
these, the method of the conical singularity, introduced by
Gibbons and Hawking \cite{GH77},  has been proved
useful and effective, especially for computing the quantum
corrections, but has also produced some contradictory results.
As we will see, in order to apply this method it is necessary
to employ one-loop physics for quantum fields
living on  manifolds with conical singularities: indeed, the aim of
this thesis is to discuss some aspects of quantum field theory on this
kind of manifolds.

The computation of the black-hole entropy is not the only case in
which one comes across  conical singularities. Another case that I
will consider is the spacetime around an idealized cosmic string,
whose study can have relevant cosmological applications.  Indeed,
the Rindler metrics,  which one usually considers when computing the
black-hole entropy, and the cosmic string metric are closely related,
and most of the results obtained for one spacetime can be translated
for the other by means of an appropriate identification. Other
examples of conical singularities, that will not be discussed in the
present thesis, are the orbifolds occurring in string compactification
\cite{GSW}, the configuration space of gravitation theory
\cite{cobra87} and the quantum gravity in $2+1$-dimensions
\cite{thooft88}.

Most of the original results exposed in this thesis have been obtained
during the years '96-'97 by myself or by  dr. Valter Moretti and
myself and published in five papers
\cite{ielmo,sigrav,moiel,massive,fluctuations}. The present thesis
contains also some new results which have not been published before.
In particular, these results concerns the computation of regularized
one-loop quantities for massive scalar fields on the cone from the
integral representation of the heat kernel, and the relation among
local and global approach for the regularization of  quantum field
theory on the cone. These new results are  reported in Chapters
\ref{ONTHECONE} and \ref{BHENTROPY}.

The thesis is organized as follows. In Chapter I will review
some of the techniques  that will be used in the following Chapters.
After a short introduction to the Rindler space and to the cosmic
string, I will review the one-loop regularization of the effective
action, with particular care for the $\z$-function regularization.
Then I will discuss the possibility of regularizing the conical
manifold by means of smooth manifolds which approximate it.
It will be shown that this procedure is reliable only in the limit
of vanishing deficit angles. Anyway, the smoothed singularity
has the advantage of allowing the use of standard Riemannian
geometry and is useful to gain some understanding of the
conical geometry. In the subsequent two sections I will review the
computation of the heat kernel and the $\z$ function of the
Laplace-Beltrami operator on the flat cone, which will be the main
tools used throughout the thesis for computing one-loop effects. In
the final subsection I will introduce a new procedure to compute the
integrated $\z$ function on the cone from the local one. This
procedure gives results in agreement with those the integrated heat
kernel on the cone and will be useful for clarifying the mysterious
relation between local and integrated quantities. This is not a
secondary point: it is my opinion that the relation between local and
integrated quantities on the cone is the main point which remains to
be clarified on quantum field theory on the cone. In Chapter
\ref{BHENTROPY} I will argue that the local approach is physically
more reasonable, but many authors still use the integrated approach
and the question has not been settled down yet.

Chapter \ref{FLUCTUATIONS} has a general character and it is
essentially based on \cite{fluctuations}. Here a new
prescription is introduced for computing the vacuum fluctuations
$\langle\phi^2\rangle$. This prescription, apart for being closer
to the general spirit of the $\z$ function regularization, has the
advantage of not requiring any subtraction of a reference state for
defining the fluctuations, a subtraction which is necessary with the
usual prescription. The new prescription is then applied to some
examples, in particular to the closed static Einstein universe for
general coupling and mass. An useful expression for the  trace of
stress tensor of a non-conformal invariant scalar field is also given.

In Chapter \ref{massivechap} I consider massive fields on the cone.
The massive case on the cone was an open problem when I started this
thesis: although there were representations for the heat-kernel and
the Green functions for massive fields, these were so complicate that
only few explicit results were available. Actually, only in
\cite{moreira95} some explicit results were given. In paper
\cite{massive}, on which this Chapter is based, the $\z$ function for
a massive scalar field was finally computed as a power series  of
$(mr)^2$, where $m$ is the mass of the field and $r$ is the proper
distance from the apex of the cone. The approximation is the right one
for computing quantum effects near a cosmic string or near the horizon
of a black hole. By means of the $\z$ function it was not only
possible to confirm the results of \cite{moreira95}, but also
compute important quantities such as the effective action and,
in particular, the energy-momentum tensor. The results
obtained have been recently confirmed by new computations,
partially reported in Chapter \ref{BHENTROPY}, and based
on the heat-kernel technique. Although in this Chapter I
mainly focus on the cosmic string case, the results are
easily translated for the Rindler case. In the final
section of this Chapter I present an interesting application:
the energy-momentum tensor of a massive field is employed
for computing the back reaction of the quantum fluctuations
on the background metric, showing how the conical geometry
is changed by quantum effects.

Chapter \ref{BHENTROPY} is devoted to the study of the quantum
corrections to the black hole entropy that, as I have said above, is
the main motivation for studying quantum field theory on manifold with
conical singularities. After a short review of the tree level black
hole entropy and the one-loop corrections, I discuss the conical
singularity method by comparing the local and integrated approaches,
and showing that they give different results. In particular, the
horizon divergences in the thermodynamical quantities, expected from
simple arguments based on the the behaviour of the local temperature,
seem to arise in a completely different way in the two approaches.
However, it is shown the the divergences in the integrated approach
should be more properly interpreted as usual ultraviolet divergences,
and in fact they can be renormalized away by redefining the Newton
constant in the standard way, while the horizon divergences are not
present in this approach. It is then argued that the local approach,
in which the horizon divergences arise in a natural way, is physically
more reasonable, although the question has not completely settled
down yet. Then, in the last two sections, I briefly discuss
the  conjecture, due to Susskind and
Uglum \cite{SU}, of a possible renormalization of the horizon
divergences in the local approach and the case of a non-minimally
coupled field.

In Chapter \ref{PHOTONS} I discuss the generalization of the local
$\z$ function on the cone to the case of the Maxwell  field  and the
graviton field. The final result is just  what one expects by counting
the number of degrees of freedom of the fields, so that, for instance,
one sees that the contribution of the photons or of the gravitons to
the one-loop quantum corrections to the entropy in the Rindler space
is just twice that of the massless scalar field. However, to derive
this result one has to deal with the presence in the theory of
gauge-dependent terms. These terms arise as surface terms and would
disappear on regular manifolds, but remain in the theory because of
the conical singularity. Actually, in section \ref{secthree} the
presence of such gauge-dependent terms is conjectured on more general
manifolds. In \cite{ielmo} it was argued that the
gauge-dependent surface terms should be discarded on the conical
manifold as would  happen on regular manifolds, since this is the only
possible procedure to obtain physically reasonable results. In
particular, this procedure gives an entropy which not only is
independent on the gauge chosen, but is positive for any value of the
temperature. If fact, the paper \cite{ielmo}, on which this Chapter is
based, originated from discussions about an interesting paper by Kabat
\cite{kabat}. In such paper it was shown that the one-loop quantum
corrections to the black-hole entropy due to the electromagnetic field
could be negative, and this negative entropy was identified with a
low-energy relic of string theory effects conjectured by Susskind and
Uglum \cite{SU}. Kabat's result was regarded as very interesting in
literature, but the negative result in \cite{ielmo} showed that, after
all,  photons do not remember much about the possible underlying
string theory.

In Chapter \ref{OPTICCHAP} I discuss the optical approach,
both on some aspects of general interest and on the
specific case of the Rindler space. In the Introduction
to this Chapter I review the relation between the canonical
definition of the free energy of a system  and the path
integral approach, showing that the canonical free energy
is equivalent to the path integral formulation in
the optical related manifold rather than in the physical
manifold. The difference between the free energies
computed in the physical manifold and in the optical manifold is the
logarithm of the  functional Jacobian of the conformal transformation
which relates the two manifolds. On regular manifolds this
Jacobian changes only the value of the zero-temperature
energy, which does not affect the thermodynamics.
On the conical manifold the question more subtle,
since the conical singularity could introduce a less
trivial dependence on the temperature: in four
dimension no explicit computation has been possible yet, and
this issue, discussed in Appendix B, is still under
investigation. In section \ref{secdue} the results
of the optical results and the results obtained in
Chapter \ref{BHENTROPY} in the physical static manifold
are compared in the case of a scalar field: although
the term proportional to $T^4$ agrees, the results
differ for the coefficient of the term proportional
to $T^2$, and the reason for this difference is still
obscure. Furthermore, it is shown that there are some
inconsistencies in the statistical-mechanical relations
if one employs the one-loop quantities computed
in the physical static manifold, while no such
inconsistencies appear for the optical results.
In any case, these problems  have no direct
influence on the discussion on the quantum
corrections to the entropy of a black hole
done in Chapter \ref{BHENTROPY}.
In sections \ref{sectre}  the optical
approach is extended to the case of the electromagnetic
field,  and in section \ref{secquattro} this approach
is applied to the particular case of the Rindler space.
Here an important result is that the gauge-dependent
terms in the effective action, analogous to those
encountered in Chapter \ref{PHOTONS}, are canceled out
by the ghost fields, and so the gauge-invariance
is restored in a more satisfactory way.
The free energies of the gas of photons computed
with the direct and the optical methods are compared,
and the conclusions are similar to those for the
scalar fields.

To finish with, I would like to comment about the
bibliography. I have tried to do my best to
acknowledge the importance and the priority of each
article. However, the number of papers is so large that
probably I have not succeeded completely. Therefore, I would like to
beg the pardon form all the authors that I have forgot or not
considered with the right weight.

In this thesis I will use units such that $c=\hbar=G=k_B=1$.
However,  in order to stress some physical fact, in some
case I will write explicitly the relevant constants.

\chapter{On the conical space}
\label{ONTHECONE}

\section*{Introduction}

In this first Chapter we review some well-known facts
and tools that we will use throughout the thesis.

In the next two sections we start introducing and discussing some
basic facts about two well-known spacetimes which show conical
singularities, namely the Rindler wedge and the space around a cosmic
string. Actually, as we will see,  while the cosmic string background
shows the conical singularity also in the Lorentzian section, in the
Rindler space the singularity appears only when one considers
the finite-temperature theory within the periodic Euclidean
time formalism.

In section \ref{shorthkz} we review the heat-kernel and
the $\z$-function regularizations of the one-loop
quantum field theory on a general curved manifold.
We also discuss the relation among different regularization
procedures, an important topic that will be used in
Chapter \ref{BHENTROPY}.

The last three sections are devoted to conical space.
In section \ref{riemangeom} we discuss the Riemannian
geometry of the cone by approximating it by means
non-singular manifolds. In this way it is possible
to understand some interesting facts about the
cone geometry, but the results must be used with
care  since, as we will see, they are correct only
for small deficit angles. In section \ref{conickernel}
we will review the computation of the heat kernel
for the Laplace-Beltrami operator on the cone,
a subject which has a very long story initiated
in the last century by the works of Sommerfeld.
Finally, in the last section we will review the computation
of the local $\z$ function on the cone. There we give also
some new results about the tricky relation among
local and integrated quantities on the cone.

\section{A short introduction to the Rindler space}
\markboth{The conical space}{The Rindler space}
\label{therindlerspace}

Finite temperature field theory in the Rindler wedge
will be the main subject of this thesis and, therefore,
we start by reminding some definitions and some
basic facts about the Rindler wedge. Since this space
has been discussed in details in thousands of beautiful
papers and reviews, we refer to them for a deeper
discussion (see, for instance,
\cite{unruh76,fulling77,birrel,takagi86,waldlibro}).

The Rindler wedge is a submanifold of the Minkowski
spacetime which can be considered as the part of
Minkowski spacetime causally accessible to an accelerated
observer. Let us consider the $D$-dimensional
Minkowski spacetime with Cartesian coordinates $(t,x,{\bf x}_\perp)$
and line element
\beas
ds^2&=&-dt^2+dx^2+d{\bf x}_\perp^2.
\label{theminkmetric}
\enas
Consider then the Rindler coordinates $(\tau,r)$ in place
of $(t,x)$:
\bea
x&=&r\,\mbox{cosh }\tau,\nonumber\\
t&=&r\,\mbox{sinh }\tau,
\label{therindlercoordinates}
\ena
with $0<r<+\infty$ and $-\infty<\tau<\infty$. This can be seen
as the Minkowskian version of the transformation from Cartesian
to cylindrical coordinates in Euclidean space. Then the line
element takes the form
\bea
ds^2&=&-r^2\,d\tau^2+dr^2+d{\bf x}_\perp^2.
\label{therindlermetric}
\ena
The above Rindler metric is static but not ultrastatic.
Clearly, the Rindler coordinates do not cover the whole
Minkowski spacetime, but only what is called the {\em right}
Rindler wedge $W_R$: $x>|t|$. One could also consider
the {\em left}  Rindler wedge $W_L$, $x<-|t|$, which
can be obtained by reflecting $W_R$ by means of the
transformation $(t,x,{\bf x}_\perp)\ra (-t,-x,{\bf x}_\perp)$.
For sake of brevity, by Rindler wedge we will always
mean the right wedge.

The components of the metric in the Rindler coordinates do
not depend on $\tau$, and so $\pa_\tau$ is a time-like Killing
vector: in Cartesian coordinates it is written as $x\pa_t+t\pa_x$,
showing that the symmetry related to $\pa_\tau$, $\tau\ra\tau+\tau_0$,
has the character of a boost rather than of an ordinary time
translation. Furthermore, the Rindler wedge is a globally
hyperbolic manifold \cite{fullingbook} in the sense that the
Cauchy problem is well posed assigning the initial data on a
surface of constant $\tau$, and so field theory is well founded
in $W_R$.

Consider now a world line such that
\beas
r(\eta)&=&\mbox{constant}=a^{-1},\\
{\bf x}_\perp(\eta)&=&\mbox{constant},
\enas
where $\eta$ is the proper time along the world line.
From the line element follows that $\tau(\eta)=a\eta$
and in the Minkowski coordinates this world line takes the form
\beas
t(\eta)&=&a^{-1}\mbox{ sinh }a\eta,\\
x(\eta)&=&a^{-1}\mbox{ cosh }a\eta,\\
{\bf x}_\perp(\eta)&=&\mbox{constant},
\enas
which is the world line of an observer who is accelerating
uniformly with respect to the proper time $\eta$ with
proper acceleration $a$. Notice that the hypersurfaces
with constant $r$ are hyperbolae asymptotic to the
hypersurfaces $t=\pm x$.

The two null hypersurfaces $t=\pm x$ form a bifurcate
Killing horizon for a Rindler observer: the hypersurface
$t=-x$ is a past horizon which divides the Rindler observers
from the region of Minkowski spacetime in which they cannot
send information, and the future horizon $t=x$ divides them
from the region from which they cannot receive information.

It is important to notice that the Rindler metric is an approximation
of the metric of the region near the horizon of a Schwarzschild
black hole, as we will see in Chapter  \ref{BHENTROPY}. Physically
this is related to the fact that an observer at rest in the static
gravitational field outside the horizon of the black hole must
undergo a constant acceleration to avoid being pulled
into the horizon: this situation is physically close to the
flat space as experienced by an accelerated observer
as the Rindler ones.

Passing to the Euclidean section, namely analytically
continuing the Rindler metric to imaginary values
of the Rindler time, $\tau\ra i\tau$, we obtain the line
element
\bea
ds^2&=&r^2\,d\tau^2+dr^2+d{\bf x}_\perp^2.
\label{therindlermetricE}
\ena
Particularly interesting is the case when we pass to the Euclidean
section for considering a finite temperature
field theory: in that case we must also make the imaginary time
periodic: $0\leq\tau\leq\b$ with $\tau=0$ and $\tau=\b$ identified,
where $\b=T^{-1}$ is the inverse temperature. In doing this
the Euclidean Rindler manifold becomes the manifold
$C_\b\times R^{D-2}$, where $C_\b$ is the simple
two-dimensional flat cone with deficit angle $2\pi-\b$: the
manifold has a conical singularity at $r=0$ unless $\b=2\pi$.

Let us now remind some important results of quantum field theory in
the Rindler wedge. The most important fact, known as Unruh effect
\cite{unruh76}, is that a Rindler observer experiences the usual Minkowski
vacuum as a thermal state with temperature $T=1/2\pi$, the
Unruh-Hawking temperature. Notice that this is the only temperature at
which the metric (\ref{therindlermetricE}) is non singular and that
the local temperature is $T_L(x)=1/2\pi r$. This effect can be seen
from the Bogolubov coefficients \cite{birrel} or by comparing the
two-point functions \cite{fullingreport}. The physical interpretation
of this phenomenon can be summarized as follows \cite{SU95}. In the
Minkowski space there are the usual fluctuations of the vacuum. This
fluctuations are closed loops in spacetime. Some of these loops will
encircle the origin $t=x=0$ of  Minkowski spacetime and therefore they
lay partially inside and partially outside the Rindler wedge. While a
Minkowski observer would not distinguish these fluctuations from the
others, as the Rindler observer is concerned they are particles which
intersect the past and future horizon, and so are particles present
for all time: a Rindler observer sees these fluctuations as a bath of
thermal particles which are ejected from the horizon infinitely far in
the past and which will eventually fall back onto the horizon in the
infinite future. To this bath of thermal particles is associated an
entropy which is proportional to the area of the horizon and
divergent, as we will see in the next Chapters.

The origin of these thermal states can also be seen in an
elegant way as entropy of entanglement
\cite{bomb86,srednicki93,kabstrass94,CW94,kabat}.
Let us suppose to divide the Cauchy hypersurface at $t=0$
of Minkowski space into two halves, one with $x<0$ and one with $x>0$.
Assume then that the Hilbert space ${\cal H}$ on the
hypersurface factorizes into a product space ${\cal H}_L\otimes
{\cal H}_R$, with orthonormal basis $|b\rangle_L$ and  $|a\rangle_R$
respectively. Then a general ket  $|\psi\rangle$ can be written as
\beas
|\psi\rangle&=&\sum_{b,a}\psi(b,a)\,|b\rangle_L\otimes|a\rangle_R.
\enas
Since no causal signal from the hypersurface $t=0$, $x<0$
can reach the right Rindler wedge, the complete set of states
on the hypersurface $t=0,\,x>0$ is the complete set of states to
describe the physics in the Rindler space for all time. Therefore,
when we restrict the quantum theory to $W_R$ we must trace
over the degrees of freedom in ${\cal H}_L$ obtaining a density
matrix
\beas
\rho(a,a')&=&\sum_b\psi(b,a)\psi^\ast(b,a').
\enas
This means that a pure state for an inertial observer,
such as the Minkowski vacuum, is seen as a thermal state by
an accelerated observer.

We will come back to these and other aspects of quantum
field theory in the Rindler space in the following Chapters.

\section{A short introduction to cosmic strings}
\markboth{The conical space}{Cosmic strings}
\label{Cosmicstrings}

In this section we summarize very briefly some basic facts
about cosmic strings. The main reference here is the report
by Vilenkin \cite{vilenk85}. Cosmic strings are essentially
topological defects of the vacuum of a field which can form
when a continuous symmetry is broken during a cooling process.
Cosmic strings are of great importance for cosmological
models where, for instance,  they could act
as seeds for galaxy formation. The simplest model that
gives rise to cosmic strings is that of a self-interacting
complex scalar field $\phi$, for instance the Higgs field,
with symmetry group $U(1)$, $\phi\ra e^{i\a}\phi$:
\beas
{\cal L}&=&\pa_\mu\phi^\dagger\pa^\mu\phi-V_T(\phi).
\enas
The self-interaction potential contains a
temperature-dependent term, for example
\beas
V_T(\phi)&=&AT^2\phi^\dagger\phi+\frac{1}{2}
\l (\phi^\dagger\phi-\eta^2)^2,
\enas
where $\l$ is the self-coupling constant and $A>0$ is
a dimensionless constant. At temperature $T$ the effective mass
of the theory is
\beas
m^2(T)&=&AT^2-\l\eta^2,
\enas
and so we have the critical temperature $T_c=\eta\sqrt{\l/A}$ at
which the effective mass of the theory is zero and we have a
second-order phase transition. For $T>T_c$ the effective mass
$m^2(T)$ is positive and the vacuum expectation value of the
field is $\langle\phi\rangle=0$: the symmetry $U(1)$ is restored for
$T>T_c$. At low temperature, instead, the field acquires a vacuum
expectation value $\langle\phi\rangle=\eta e^{i\theta}$, where the
phase $\theta$ is arbitrary: the vacuum is degenerate and the
symmetry $U(1)$ is broken at low temperature. The arbitrary
phase $\theta$ varies in the space and, since $\langle\phi\rangle$
is a single valued function, the total change of $\theta$ around
a closed loop must be $\Delta\theta=2\pi n$, where $n$ is an
integer. If we consider a closed loop with, e.g., $\Delta\theta=2\pi$
we can continuously shrink it to a point, for which  $\Delta\theta=0$,
only if we encounter one point for which the phase is undefined,
namely $\langle\phi\rangle=0$. The set of such points must form
a closed loop or an infinite line, which are called cosmic strings,
otherwise it would be possible to contract the loop without
crossing a singular point. This shows that at least a cosmic string,
namely a tube of false vacuum $\langle\phi\rangle=0$,
should exist for any loop with $n\neq 0$.

The radius of the string core is of the order of the Compton
wave length of the Higgs boson, $r_0\sim m_\phi^{-1}=
(\eta\sqrt{\l})^{-1}$,
and the mass per unit length is $\mu\sim \eta^2$.

Since for strings of cosmological interest the length is much
greater that their width, if we are not interested in the internal
structure of the string we can represent its energy-momentum
tensor as proportional to a $\delta$-function peaked on the string.
For instance, for a static straight string lying along the $z$-axis,
Lorentz invariance for boosts along $z$ and the conservation laws
fix the energy-momentum tensor to
\beas
T_{\mu\nu}&=&\mu\delta(x)\delta(y)\mbox{diag}(1,0,0,-1).
\enas
The metric of the space around the string is then a solution
of the Einstein equations
\beas
R_{\mu\nu}-\frac{1}{2}g_{\mu\nu} R&=&8\pi GT_{\mu\nu}.
\enas
The solution has been found by several authors \cite{Stringmetric}
and in cylindrical coordinates is
\bea
ds^2&=&-dt^2+dz^2+dr^2+(1-4G\mu)^2 r^2d\theta^2,
\label{thestringmetric}
\ena
which, in the spatial section $(r,\theta)$ describes a simple
flat cone with deficit angle $2\pi-\a=8\pi G\mu$. For typical
GUT theories the symmetry breaking scale is around $10^{16}GeV$,
so that $\mu\sim 10^{-6}$ and $2\pi-\a\sim 10^{-6}$. It is
clear that the Euclidean version of the above metric
is the same as the Euclidean Rindler metric (\ref{therindlermetricE})
after a suitable identification of the coordinates.

It is well known that, even though the space around a infinitely long,
static and straight cosmic string is locally flat, the non-trivial
topology gives rise to remarkable gravitational and quantum phenomena.
For example, the classical trajectories of particles are deviated by
the cosmic string by the same (absolute) angle independently of the
radius of closest approach (see, e.g., \cite{vilenk85,AuEc91}).
Moreover, the presence of the string allows effects such as
particle-antiparticle pair production by a single photon and
bremsstrahlung radiation from charged particles \cite{HaSk90,SkHaJa94}
which are not possible in empty Minkowski space, due to conservation
of linear momentum. Finally, the string polarizes the vacuum
around it, in a way similar to the Casimir effect between two
conducting planes forming a wedge \cite{DC79,HellKon86}: this
last effect will be the our main interest in cosmic strings in this
thesis.

\section{Short introduction to $\z$ function and heat kernel}
\label{shorthkz}
\markboth{The conical space}{$\z$ function and heat kernel}

The $\z$-function and heat-kernel techniques are tools extensively
used in this thesis, and therefore in this section we summarize
some basic facts about the methods \cite{dokcrit76,hawking77,
birrel,libro,elizlibro,report}.

Let us start considering a matter quantum field $\phi$ living
on a $D$-dimensional background manifold $\M$ with a metric
$g_{\mu\nu}$. Moreover, let us suppose that the metric is
static, namely $\pa_0 g_{\mu\nu}=0$ and $g_{0i}=0$,
so that it is always possible to perform the analogue
of the Wick rotation and make the metric Euclidean.
The general case is more subtle and will not be
considered in this thesis. The physical properties of the
field can then be described by means of the Euclidean path
integral \cite{birrel}
\beas
Z[g]&=&\int\D\phi\, e^{-S_E[\phi,g]},
\enas
where $S_E[\phi,g]$ is the classical action of the field
and the functional integral is taken over the field configurations
satisfying suitable boundary conditions. The functional
integration measure is the usual covariant one,
$\D\phi=\prod_xd\phi(x)g^{1/4}(x)$. Finally, an infinite
renormalization constant has been neglected.

The physical interpretation of the quantity $Z[g]$ is that,
if the spacetime is asymptotically flat and the functional
integral is taken over fields infinitesimally close to
the classical vacuum, then $Z[g]$  is the vacuum to vacuum
transition amplitude. Moreover, the quantity
$W[g]=-\ln Z[g]$ generates the effective field equations,
namely the classical ones plus quantum corrections,
and therefore it is called effective action.

When the condition of asymptotically flatness is
not satisfied, then the physical meaning of $Z[g]$ is
less clear, but the functional $W[g]$ is still supposed to
describe the effective action.

The dominant contribution to the path integral $Z[g]$
will come from fields that are near the solution $\phi_0$
of the classical field equations, namely which extremizes
the action and satisfy the boundary conditions. Therefore,
it is possible to expand the action in a Taylor series
about the classical solution:
\beas
S_E[\phi,g]&=&S_E^c[\phi_0,g]+S_E^{(2)}[{\tilde{\phi}},g]+
\mbox{higher order terms in }{\tilde{\phi}},
\enas
where ${\tilde{\phi}}=\phi-\phi_0$ are the fluctuations and
 $S_E^{(2)}[{\tilde{\phi}},g]$ is quadratic in ${\tilde{\phi}}$.
The above approximation is sufficient for considering one-loop
effects, namely at first order in $\hbar$: the one-loop
generating functional, known also as zero-temperature
partition function or simply partition function, reads
\beas
Z[\phi,g]&\simeq& e^{-S_E^c[\phi_0,g]}\int \D{\tilde{\phi}}
\,e^{-\frac{1}{2}S_E^{(2)}[{\tilde{\phi}},g]}.
\enas
 For a neutral scalar fields the quadratic terms have the form
\beas
S_E^{(2)}[{\tilde{\phi}},g]&=&-\frac{1}{2}\int_M
{\tilde{\phi}}\,A\,{\tilde{\phi}},
\enas
where $A$ is known as the small disturbance or small
fluctuations operator. When the background metric is Euclidean,
$A$ is a second order, elliptic, self-adjoint (or with a self-adjoint
extension), and non-negative operator. For a quasi-free scalar field,
namely that interacts only with the background metric,
the operator $A$ reads $A=-\Delta+m^2+\xi R$,
where $\Delta$ is the Laplace-Beltrami operator
on the manifold $\M$, $m$ is the mass of the field,
$R$ is the scalar curvature of the manifold and $\xi$ is
an arbitrary constant. If the field is self-interacting,
$A$ will depend on the classical solution $\phi_0$ and
a more sophisticated treatment is needed, especially
if $\phi_0$ is not a constant configuration. The charged
scalar field case is very similar, while in the Dirac's field
case the small disturbance operator is of first order,
but the result can be easily extended to include this case.
Finally, in the case of gauge fields and graviton field
one has to consider the spin indices.

Lets consider, for simplicity, the case of a neutral scalar field.
The operator $A$ will have a complete set of eigenvectors
$\phi_n$, with real, non-negative eigenvalues $\l_n$:
\beas
A\phi_n&=&\l_n\phi_n
\enas
and the eigenvectors can be normalized so that
\beas
\int_\M d^Dx\,\sqrt{g}\,\phi_n(x)\phi_m(x)&=&\delta_{nm}.
\enas
We have used a discrete index, and this will be the case
if the manifold $\M$ is compact. In the non-compact case
the eigenvectors will carry also continue indices, and
what follows can be formally easily extended to include
this case. However, in general the quantities will include
extra divergences proportional to the volume of the manifold,
and a local $\z$ function approach will be more convenient
(see below).

Due to the completeness of the set of eigenvectors
$\{\phi_n\}$, it is possible to expand the fluctuations
${\tilde{\phi}}$ in terms of the eigenfunctions,
\beas
{\tilde{\phi}}(x)&=&\sum_n a_n\,\phi_n(x),
\enas
and therefore in the path integral the measure $\D{\tilde{\phi}}$
can be recast in terms of the coefficients $a_n$:
\beas
\D{\tilde{\phi}}&=&\prod_n\mu da_n,
\enas
where $da_n$ is a standard integration measure. The
constant parameter $\mu$ is needed in order to match
the dimensions: it has the dimension of a mass or an
inverse length. On its meaning we will come back later.
Then it follows that \cite{hawking77}
\beas
\int \D{\tilde{\phi}}e^{-\frac{1}{2}S_E^{(2)}[{\tilde{\phi}},g]}&=&
\prod_n\int_{-\infty}^\infty \mu\,da_n \,e^{-\frac{1}{2}\l_n
a^2_n}=\prod_n\left(\frac{\l_n}{2\pi \mu^2}\right)^{-\frac{1}{2}}\\
&=&\left[\det(A\mu^{-2})\right]^{-\frac{1}{2}},
\enas
where we have rescaled $2\pi\mu^2\ra\mu^2$. The fundamental
relation just written relates the one-loop effective action of
the field with the determinant of the small fluctuations operator.

In the same way, it is possible to derive similar expressions
for charged scalar and Dirac's fields, and all these expressions
can be summarized as \cite{report}
\bea
Z^{(1)}[g]&=&\left[\det(A\mu^{-2})\right]^{-\nu},\nonumber\\
W^{(1)}[g]&=&\nu\ln\,\det (A\mu^{-2}),
\label{ZWdet}
\ena
where $\nu=-1,+1,+\frac{1}{2}$ for Dirac's, charged scalar and
neutral scalar fields respectively.

\subsection{$\z$-function regularization}
\label{zetaregul}

We have seen above that the one-loop effective action
can be expressed as the logarithm of the determinant of the
small disturbances operator. This is, of course, a divergent
quantity, since in the na\"{\i}ve definition as the product of the
eigenvalues, these grow without bound. It is therefore
necessary to regularize it in some way. The following
step would be to show that the physical results are
independent of the regularization prescription, but
we will not deal with this topic here (see, e.g.,
\cite{libro,elizlibro,report}).

A very convenient way of regularizing the determinant is the
$\z$-\-func\-tion me\-thod, studied long ago by mathematicians
\cite{minak49,raysing71} and introduced in the physical context by
Hawking \cite{hawking77} (see also \cite{dokcrit76}).

The basic idea behind the $\z$ function evaluation of the
determinant of the operator $A$ is the following. Let
us suppose that $A$ is invertible, namely it has no zero
modes, and consider the following quantity, known as
$\z$ function associated to the operator $A$:
\bea
\z(s|A)=\sum_n\frac{1}{\l_n^s},
\label{detnaif}
\ena
where $s$ is a complex number. From this definition, we can
formally write
\beas
\frac{d}{ds}\z(s|A)|_{s=0}&=&-\sum_n\ln\l_n\,
e^{-s\ln \l_n}|_{s=0}=-\sum_n\ln\l_n=-\ln\prod_n\l_n\\
&=&-\ln\,\det A.
\enas
This writing is just formal, since the sums diverge.
Nevertheless, we will see below that $\z(s|A)$
converges in a region of the complex plane $s$
and that it can be analytically continued as a meromorphic
function to the whole complex plane.  Moreover,
for many operators and spacetimes of physical
interest, $\z(s|A)$ is analytic in $s=0$ (for a contrary
example see \cite{bycoze97}) and so we can consider the above
formal identity as a regularized expression for $\det A$:
\beas
\det A&=&e^{-\frac{d}{ds}\z(s|A)|_{s=0}}.
\enas

For making our discussion more concrete, notice that Eq.
(\ref{detnaif}) is just $\Tr A^{-s}$, where the inverse complex power
of the operator $A$ can be defined as the Mellin transform of the
operator $e^{-tA}$:
\beas
A^{-s}&=&\frac{1}{\Gamma(s)}\int_0^\infty t^{s-1}\,
e^{-tA}\,dt,\nonumber\\\
\z(s|A)&=&\Tr\,A^{-s}=
\frac{1}{\Gamma(s)}\int_0^\infty t^{s-1}\Tr e^{-tA}\,dt
\enas
This allows us to relate the $\z$ function to another
very important quantity, namely the heat kernel of the
operator $A$, defined as
\beas
K_t(A)= e^{-tA}.
\enas
It is clearly a formal solution of the heat equation,
\beas
\pa_t K_t+A\,K_t=0,
\enas
with the boundary condition $\lim_{t\ra 0_+}K_t=1$.

The heat kernel of an elliptic operator and its relation with the
$\zeta$ function has been much studied by mathematicians and
physicists, a work still in progress producing remarkable results
(see, e.g., \cite{gilkey76,boelikir96,bogekireli96,elilyvass96,elilyvass97}
and references therein).  This fundamental
theorem is due to Minakshisundaram and Pleijel \cite{minpleij49}:

\smallskip

\noindent {\bf Theorem:} Let $A$ be a elliptic, self-adjoint,
positive differential operator of second order on a closed
manifold $\M$ of dimension $D$.
Then $\forall\, t>0$, $K_t(A)$ is an integral operator with trace,
namely it can be written as
$$
K_t(A)f(x)=\int_\M K_t(x',x|A) f(x')\sqrt{g} d^Dx,
$$
and the kernel $K_t(x',x|A)$ is a smooth function of $t$ and
$$
\Tr\, e^{-tA}=\int_\M K_t(x,x|A)\sqrt{g} d^Dx.
$$
Moreover, for $t\ra 0_+$ the following asymptotic expansion holds:
\beas
K_t(x,x|A)&\simeq&\frac{1}{(4\pi t)^{D/2}}\sum_{j=0}^\infty
k_j(x|A) t^{j/2},\\
\Tr\, e^{-tA}&\simeq&\frac{1}{(4\pi t)^{D/2}}\sum_{j=0}^\infty
K_j(A) t^{j/2},\hspace{7mm}K_j(A)=\int_\M k_j(x|A)\sqrt{g}d^Dx.
\enas
The coefficients $k_j(x|A)$ are known as Seeley-DeWitt coefficients.
They are local invariants built out of the curvature tensor
of the manifold and the extrinsic and intrinsic curvature
of the boundary. If the manifold is without boundary then only
the coefficients $k_j(x|A)$ with even index are present,
$k_{2j+1}(x|A)=0$, and so we can call $k_{2n}(x|A)=a_n(x)$
and write the expansion as
\bea
K_t(x,x|A)&\simeq& \frac{1}{(4\pi t)^{D/2}}\sum_{n=0}^\infty
a_n(x) t^{n}.
\label{heatexpa}
\ena
Only the first four coefficients $a_n(x)$ are known, and the
computation of the first three is due to DeWitt
\cite{dewitt65,dewitt75}: if $A=-\Delta+m^2+\xi R$ then
the first three coefficients read \cite{birrel}:
\bea
a_0(x)&=& 1\nonumber\\
a_1(x)&=&\frac{1}{6} (1-6\xi) R(x)-m^2\nonumber\\
a_2(x)&=&\frac{1}{2}[a_1(x)]^2+\frac{1}{6}\Delta a_1(x)+
\frac{1}{180}\left(\Delta R+R_{\a\b\mu\nu}R^{\a\b\mu\nu}-
R_{\a\b}R^{\a\b}\right)\nonumber\\
&=&\frac{1}{180}\left(R_{\a\b\mu\nu}R^{\a\b\mu\nu}-
R_{\a\b}R^{\a\b}\right)-\frac{1}{30}(1-5\xi)\Delta R\nonumber\\
&&+\frac{1}{72}(1-6\xi)^2R^2-\frac{m^2 R}{6}(1-6\xi)+\frac{m^4}{2}.
\label{heatcoeff}
\ena
The case with boundary is more complicate and the coefficients
are distributions on the boundary \cite{brangil90}.
For the computation of higher-order heat-kernel coefficients see
\cite{boelikir96} and references therein. Off the diagonal
an expansion similar to (\ref{heatexpa}) holds:
\beas
K_t(x,x'|A)&\simeq& \frac{e^{-\sigma(x,x')/2t}}{(4\pi t)^{D/2}}
\sum_{n=0}^\infty a_n(x,x') t^{n},
\enas
where $\sigma(x,x')$ is half the square of the proper distance from
$x$ to $x'$ and the first coefficients  $a_n(x,x')$ can be found in
\cite{birrel}.

Coming back to the $\z$ function, since the operator
$K_t(A)=e^{-tA}$ has a kernel, we see that also the operator
$A^{-s}$ has the same property:
\bea
A^{-s}f(x)&=&\int_\M A^{-s}(x',x) f(x') \sqrt{g} d^Dx,\nonumber\\
A^{-s}(x',x)&=&\frac{1}{\Gamma(s)}\int_0^\infty dt\,t^{s-1}
\left[K_t(x',x|A)-P_{0}(x',x|A)\right],
\label{zetakernel}
\ena
where $P_{0}$ is the projector onto the zero-modes, so that
we can drop the hypothesis that $A$ has no zero modes.
However, although $A^{-s}(x',x)$, $x'\neq x$ is an entire
function of $s$, the same does not hold for the trace:
in fact, if we consider the above relation among $A^{-s}(x,x)$
and $K_t(x,x|A)$ and use the Seeley-DeWitt expansion,
we see that the integral converges in $t=0$ only if
$\Re s>D/2$, since the main singularity of $K_t(x,x|A)$
 is $t^{-D/2}$. At infinity, because of the positiveness of
the eigenvalues, $K_t(x,x|A)$ is exponentially decreasing,
and the integral converges for any value of $s$.
Since $A^{-s}(x,x)$ exists for $\Re s>\frac{D}{2}$, where it is
an analytic function, it follows that
\beas
\Tr A^{-s}&=&\int_\M A^{-s}(x,x)\sqrt{g}d^Dx\\
&=&\sum_n\l_n^{-s}\equiv \z(s|A)
\enas
exists and is an analytic function in the semiplane
$\Re s>\frac{D}{2}$. This result was announced above.
We also call $A^{-s}(x,x)=\z(s|A)(x)$ the local $\z$ function
associated to $A$.

The next step is to analytically continue the $\z$ function.
The analytic continuation and the meromorphic structure
can be obtained in the following way (Seeley theorem):
we split the integration over $t$ in the Mellin transform
which relates the heat kernel to the $\z$ function in two
parts, from $0$ to $1$ and from $1$ to $\infty$. Since
$K_t(A)$ is a smooth function of $t$ as $t\ra\infty$ we can write
\bea
\z(s|A)(x)&=&\frac{1}{\Gamma(s)}\int_0^1 t^{s-1} K_t(x,x|A) dt+
\frac{1}{\Gamma(s)}\int_1^\infty t^{s-1} K_t(x,x|A) dt\nonumber\\
&=&\frac{1}{(4\pi)^{\frac{D}{2}}\Gamma(s)}\sum_{j=0}^\infty
\int_0^1 t^{s-1+\frac{j-D}{2}} k_j(x|A) dt+
\frac{h(s;x)}{\Gamma(s)}\nonumber\\
&=&\frac{1}{(4\pi)^{\frac{D}{2}}\Gamma(s)}\sum_{j=0}^\infty
\frac{k_j(x|A)}{s+\frac{j-D}{2}}+\frac{h(s;x)}{\Gamma(s)},
\label{seeleyth}
\ena
where $h(s;x)$ is a smooth function of $s$. It follows that
$\z(s|A)=\Tr A^{-s}$ is a meromorphic function in the complex
plane with at most simple poles on the real axis in
\beas
\left\{\begin{array}{l}s_j=\frac{D-j}{2}, \hspace{1cm}
j=0,1,2,\dots,\\
s_j\neq 0,-1,-2,\dots.\end{array}\right.
\enas
If the manifold is without boundary the poles are
\beas
\begin{array}{ll}s=\frac{D}{2},\frac{D}{2}-1,\dots,2,1\hspace{5mm}
&D\mbox{ even},\\
\  \\
s=\frac{D}{2}-j&D \mbox{ odd}.
\end{array}
\enas
In four dimensions the possible poles are in $s=2$ and $s=1$.
Furthermore, the residues of the poles of the $\z$ function
are related to the Seeley-DeWitt coefficients by \cite{voros87}
\beas
\mbox{Res}\left(\z(s|A)(x),s_j\right)=
\frac{k_j(x|A)}{(4\pi)^{\frac{D}{2}}\Gamma\left(\frac{D-j}{2}\right)}.
\enas
An important consequence is that $\z(s|A)$ is analytic in $s=0$ and
its value is
\bea
\z(s=0|A)=\frac{1}{(4\pi)^{\frac{D}{2}}}k_D(A),
\label{zetainzero}
\ena
and the corresponding local relation. If the manifold is without
boundary and $D$ is odd then $\z(s=0|A)=0$.

Now that we know the analytic properties of $\z(s|A)$, reminding
the na\"{\i}ve definition (\ref{detnaif}) we can define,
in the sense of the above analytic continuation, the determinant
of the elliptic operator $A$ as
\bea
\det A=\exp\left [-\frac{d}{ds}\z(s|A)|_{s=0}\right].
\label{zetadefinition}
\ena
From this definition it follows that
\beas
\det \left(A^r\right)=\left(\det A\right)^r,
\enas
since $\z(s|A^r)=\z(sr|A)$. Furthermore, since $\z(s|aA)=a^{-s}\z(s|A)$
we have that
\beas
\det(aA)=a^{\z(s=0|A)}\det A.
\enas
Considering the fact that for a $n\times n$ matrix we would have
$\det (aA)=a^n\det A$, we see that $\z(0|A)$ can be seen as the
dimension of the operator $A$. This latter property is particularly
important considering that, as we have seen above, for computing
the one-loop effective action it is necessary to introduce
an arbitrary parameter $\mu$ with the dimension of a mass, so that
(for a neutral scalar field)
\beas
\ln Z^{(1)}=-W^{(1)}=-\frac{1}{2}\ln\,\det (A\mu^{-2}).
\enas
By using the $\z$ function regularization we obtain
\bea
W^{(1)}&=&-\frac{1}{2}\frac{d}{ds}\z(s|A\mu^{-2})|_{s=0}\nonumber\\
&=&-\frac{1}{2}\frac{d}{ds}\z(s|A)|_{s=0}+\frac{1}{2}
\z(s=0|A)\ln\mu^2.
\label{basiczeta}
\ena
The ambiguity introduced by the parameter $\mu$ is related
with the scale dependence of the theory. For instance, if $A=-\Delta$
and we rescale the coordinates by a factor $\sqrt{k}$, then metric
changes as  ${\bar {g}}_{\mu\nu}=k g_{ab}$ and
${\bar{A}}=k^{-1}A$ and so
\beas
\ln\,\det {\bar{A}}&=&\ln\,\det A-k\z(s=0|A).
\enas
Hence, if $\z(s=0|A)\neq 0$ the determinant depends on
the scale (scale anomaly). From the physical point of view,
this means that, although the classical theory is scale invariant,
the one-loop quantum effects break the invariance. The arbitrary
scale $\mu$ can be fixed from the physical requests, such as
the energy scale at which the experiment is performed.

On a manifold without boundary, $\z(s|A)=0$ if the dimension is odd,
while in even dimension $\z(s=0|A)=(4\pi)^{-D/2}a_{D/2}(A)$.

There are a few more comments in order to finish this short
introduction to heat kernel and $\z$ function. First of all, we
see from Eq. (\ref{zetakernel}) that the local $\z$ function
can be written also by using the spectral representation of
the inverse complex power of the kernel of $A$:
\bea
\z(s|A)(x)={\sum_n}'\l_n^{-s}\phi_n(x)^\ast \phi_n(x),
\label{spectralzeta}
\ena
where the prime indicates that the zero modes have to be
omitted in the sum and the expression must be intended in the sense
of the above analytic continuation. This can be a convenient
starting point in some cases, at least when it is possible to
sum and analytically continue the series. On the other hand,
writing the $\z$ function as the Mellin transform of the heat
kernel has the advantage that it is then possible to use the
asymptotic expansion of the heat kernel to study the
divergences of the theory.

Another difference among the heat kernel and the $\z$ function
is that, although they seem to carry the same information,
the heat kernel $K_t(A)$ involves a dimensional parameter $t$,
with the dimension of a length squared, while in the $\z$ function
there is only the dimensionless variable $s$. This has important
consequences if one wants to use dimensional analysis to
predict the forms the two function may take \cite{AAA95}.

\subsection{Relation among different regularizations}
\label{regulrelations}

It is important to notice that if we employ the $\z$-function
method to compute the effective action of a theory, then no
explicit renormalization of the theory is necessary, since
the ultraviolet divergences are automatically canceled
by the analytic continuation procedure. Even though
this is clearly a computational advantage, sometimes it can
be misleading, since it is not a regularization independent feature.
If we used a different regularization procedure, such as
dimensional or Schwinger-DeWitt, we would obtain in the
effective action terms which are divergent as the regularization
parameter is removed, and these divergent quantities must be
renormalized by redefining the bare quantities in the action.

In this section we want to discuss the relation among different
possible regularization of $\ln\det A$ \cite{dokcrit76,report}. From the
definition of $\z$-function regularization we have
\beas
(\ln\det A)_\z&=&-\left.\frac{d}{ds}\right|_{\ep=0}\z(\ep|A)\\
&=&-\lim_{\ep\ra 0}\frac{d}{d\ep}\frac{1}{\Gamma(\ep)}
\int_0^\infty\frac{dt}{t}t^\ep\,\Tr e^{-tA}.
\enas
We see that the $\z$ function can be seen as a particular
regularization of the divergent expression
\beas
\int_0^\infty\frac{dt}{t}\Tr e^{-tA}
\enas
where by inserting the function
\beas
\rho(\ep,t)&=&\left.\frac{d}{d\ep}\right|_{\ep=0}
\frac{t^\ep}{\Gamma(\ep)}
\enas
we have regularized the singularity at $t=0$ coming from the asymptotic
expansion of $K_t(A)=\Tr\exp (-tA)$. In general, we can choose
another regularizing function $\rho(\ep,t)$ and define the
regularized determinant of the operator $A$ as (general
Schwinger regularization)
\beas
(\ln\det A)_{\rho_\ep}&=&-\lim_{\ep\ra 0}
\int_0^\infty\frac{dt}{t}\rho(\ep,t)\Tr e^{-tA}.
\enas
The regularizing function $\rho(\ep,t)$ has to satisfy some
requirements:
\begin{enumerate}
\item $\lim_{\ep\ra 0}\rho(\ep,t)=1$ for $t>0$ fixed.\\
\item For fixed and sufficiently large $\ep$, $\rho(\ep,t)$
has to regularize the divergence in $t=0$.\\
\item For small $\ep$ we must have
$$-\int_0^\infty\frac{dt}{t}\rho(\ep,t)\, e^{-ty}
\simeq\ln y+a+b(\ep)+\O(\ep),
$$
where $a$ and $\b(\ep)$ are independent of $y$ and
$b(\ep)$ is divergent as $\ep\ra 0$.
\end{enumerate}

Besides the $\z$-function regularization, other regularizations
which satisfy the above requirements and are often used are the
Schwinger proper-time, $\rho^{PT}(\ep,t)=\Theta(t-\ep)$, and the
Dowker-Critchley \cite{dokcrit76}, $\rho^{DC}(\ep,t)=t^\ep$.
Other examples can be found in \cite{report}.
The Dowker-Critchley regularization is related to the
dimensional regularization \cite{birrel}, in which we add
$\ep'$ flat dimensions to the spacetime, since this amounts
to multiply the heat kernel by $(4\pi t)^{-\ep'/2}$.

Let us now see the relation among the different regularizations.
We start considering the $\z$ function:
\beas
(\ln\det A)_{\rho^\z_\ep}&=&-\frac{d}{d\ep}
\frac{1}{\Gamma(\ep)}\sum_n\int_{0}^\infty dt\,
t^{\ep-1}e^{-t\l_n}\\
&=&-\frac{d}{d\ep}\sum_n \l_n^{-\ep}=
-\frac{d}{d\ep}\z(\ep|A)\\
&=&-\z'(\ep|A)+\O(\ep)
\enas
since $\z(\ep|A)$ is analytic in $\ep=0$. Now we consider the
Dowker-Critchley regularization:
\beas
(\ln\det A)_{\rho^{DC}_\ep}&=&-\sum_n\int_0^\infty dt\,
t^{\ep-1}e^{-t\l_n}\\
&=&-\Gamma(\ep)\z(\ep|A)=
-\z'(\ep|A)-\frac{1}{\ep}\z(0|A)+\O(\ep)\\
&=&-\z'(\ep|A)-\frac{K_D(A)}{(4\pi)^{D/2}\ep}+\O(\ep),
\enas
where we have used Eq. (\ref{zetainzero}). Finally, in the general
case
\beas
(\ln\det A)_{\rho_\ep}&=&(\ln\det A)_{\rho^{DC}_\ep}
-\int_0^\infty \frac{dt}{t}\left[\rho(\ep,t)-t^\ep\right]
\Tr e^{-tA}\\
&\simeq&(\ln\det A)_{\rho^{DC}_\ep}
+\sum_{j=0}^\infty\frac{K_j(A)}{(4\pi)^{D/2}}
\int_0^1 dt\left[\rho(\ep,t)-t^\ep\right]t^{\frac{j-D}{2}-1}\\
&=&-\z'(0|A)-K_D(A)B_D(\ep)+\sum_{j<D}K_j(A)B_j(\ep)+\O(\ep),
\enas
where
\beas
B_D(\ep)&=&\frac{1}{(4\pi)^{D/2}}\int_0^1 \frac{dt}{t}\rho(\ep,t),\\
B_j(\ep)&=&\frac{1}{(4\pi)^{D/2}}\int_0^1 dt\,t^{\frac{j-D}{2}-1}
\left[\rho(\ep,t)-t^\ep\right].
\enas
Notice that the constant terms can be reabsorbed in the
definition of $\mu$.

We see that in the $\z$-function regularization all the divergent
terms $B_j(\ep)$ vanish, while in the Dowker-Critchley only
$B_D(\ep)$ survives. Finally, for the proper time
regularization we get
\beas
(\ln\det A)_{\rho^{PT}_\ep}&=&
-\z'(0|A)+\frac{K_D(A)}{(4\pi)^{D/2}}\ln \ep+\\
&&+\frac{2}{(4\pi)^{D/2}}\sum_{j<D}
\frac{\ep^{(j-D)/2}}{j-D}K_j(A)
+\mbox{cts.}+\O(\ep).
\enas
In the case of a four dimensional manifold without boundary
\beas
(\ln\det A)_{\rho^{PT}_\ep}&=&
-\z'(0|A)+\frac{A_2(A)}{16\pi^2}\ln \ep-
\frac{A_1(A)}{16\pi^2\ep}-\frac{A_0}{32\pi^2\ep^2}+
\mbox{cts.}+\O(\ep).
\enas

It is clear from the above examples that the divergent terms
depend on the regularization functions: being proportional
to the Seeley-DeWitt coefficients they can be renormalized
by redefining the bare gravitational coupling constants and
the cosmological constant in the gravitational action
by means of the standard renormalization procedure
\cite{birrel}.

\section{Riemannian geometry on smoothed cones}
\label{riemangeom}
\markboth{The conical space}{Smoothed cone}

In this thesis we are more interested in functional analysis on
manifolds with conical defects rather than in the geometry of the manifolds
itself. It is however of great interest to have some understanding
of the geometrical properties of the manifolds and, in particular, of
quantities built out of the Riemannian curvature. In fact, these
quantities are essential for discussing the classical theory of
gravity on the cone, namely for writing down the Hilbert-Einstein
action. Moreover, as far as the quantum theory is concerned, it is
well known \cite{birrel} that the divergent terms in the one loop
effective action are proportional to the Seeley-DeWitt coefficients,
which are geometrical objects constructed from the curvature tensor.

The description of the Riemannian geometry in presence of conical
defects is quite complicate, because of the singular behavior of the
curvature. In particular, the standard formulas of Riemannian geometry
fail in describing the singularity, and alternative methods must be
used to get the correct results. When this thesis was almost completed,
a very interesting paper has been published by Dahia and Romero
\cite{dahrom98}, in which the problem of describing the conical
geometry is addressed employing the distribution theory, obtaining
a very clear and effective approach. It would be worth investigating the
potentiality of this method. Other interesting ideas can be found in
\cite{moreira97}.

A method developed to deal with the problem of describing the
Riemannian geometry in presence of conical defects problem is that
based on the ``smoothed singularity'', in which the singular manifold
$\M_\a$ is represented as the limit of a converging sequence of smooth
manifolds, on which the standard Riemannian formulae hold. This method
has been considered by various authors
(see, for instance
\cite{AlOt90,HayLou,SchWitt,SOL95,FUSOL95,FUSOL96,hotta}). The most
systematic approach has been given in the paper by Fursaev and
Solodukhin \cite{FUSOL95}. In this section we will summarize their
method and their results, since they shall be useful in the rest of
the thesis.

Let us first consider the two-dimensional case to illustrate the method.
Let $\M_\a$ be a two dimensional manifold whose metric is conformally
related to that of a flat cone $C_\a$: then its metric reads
\beas
ds_{\M_\a}^2=e^\sigma(dr^2+r^2d\theta^2)=e^\sigma ds^2_{C_\a},
\enas
where\footnote{Notice that our $\a$ corresponds to $2\pi\a$ in the
notation of \cite{FUSOL95}.} $0\leq\theta\leq 2\pi$, and the conformal
factor $\sigma$ is assumed to have the following expansion near $r=0$:
\beas
\sigma&=&\sigma_1(\theta)r^2+\sigma_2(\theta)r^4+\dots.
\enas
This asymptotic secures that the only singularity comes from the conical
metric.

Now we replace the the singular manifold $C_\a$ with metric $ds^2_{C_\a}$
by a regularized manifold ${\tilde C}_\a$ with metric
\beas
ds^2_{{\tilde C}_\a}=u(r;a)dr^2+r^2d\theta^2,
\enas
where the smooth function $u(r;a)$ depends on a regularization parameter
$a$ and obeys the following conditions:
\beas
\lim_{a\ra 0}u(r;a)&=&1,\nonumber\\
u(r;a)|_{r\gg a}&=&1,\nonumber\\
u(r;a)|_{r=0}&=&(\a/2\pi)^2.
\enas
The meaning of the first condition is obvious. The second one means that
far from the singularity the metric is unchanged by the regularization.
The last condition means that near the singularity the metric of ${\tilde C}$
is conformal to that of a plane and therefore regular.

The simplest example of the above regularization is that corresponding to the
change of $C_\a$ to a hyperbolic space:
\beas
ds^2_{{\tilde C}_\a}=\frac{r^2+a^2(\a/2\pi)^2}{r^2+a^2}dr^2+r^2d\theta^2.
\enas
As an alternative regularization one can change $g_{\theta\theta}$
instead of $g_{rr}$ \cite{hotta}.

In place of the singular manifold $\M_\a$ it is now possible to employ
the smooth manifold ${\tilde{\M}}_\a$ with topology ${\tilde C}_\a$,
taking the limit $a\ra 0$ at the end.
The first quantity considered in \cite{FUSOL95} is the integral curvature
on ${\tilde{\M}}_\a$: it turns out that this quantity is independent of the
regularizing function $u$ as $a\ra 0$, and it is possible to write
\beas
\lim_{{\tilde{\M}}_\a\ra\M_\a}\int_{{\tilde{\M}}_\a}R\,e^\sigma\sqrt{u}\,r
dr\,d\theta=2(2\pi-\a)+\int_{\M_\a/\Sigma} R.
\enas
On the right hand side, $R$ is the curvature computed in the standard way
on the smooth domain $\M_\a/\Sigma$ of $\M_\a$, and $\Sigma$ is the
set of singular points of the manifold; in the present case it is the point
$r=0$. The first term in the above equation is due to the singularity and
does not depend on the behavior of the manifold at regular points.
One can therefore represent the {\em local} curvature of the manifold
$\M_\a$ as \cite{sokstar77}
\beas
\sqrt{g}{}^{(\a)}R=2\left(\frac{2\pi}{\a}-1\right)\delta(r)+\sqrt{g}R.
\enas

By means of the method outlined above it is also possible
to consider higher order curvature polynomials or functionals. However,
in general these quantities do not have a strict mathematical meaning,
being divergent as $a\ra 0$ or depending on the particular regularizing
function $u(r;a)$. The fact that the integral curvature is independent
of $u(r;a)$ is related to the Gauss-Bonnet theorem: in two dimensions
$(4\pi)^{-1}\int_{\M_\a}R=\chi[\M_\a]$ is the Euler number, which is a
topological invariant of the manifold. The divergences are instead related
to the presence of ill-defined quantities such as $\delta^2(r)$.

An example of this behavior is given by the integral of $R^2$, for which
Fursaev and Solodukhin \cite{FUSOL95} obtain (for small $a$)
\beas
\int_{{\tilde{\M}}_\a}R^2=\int_{\M_\a/\Sigma}R^2+\frac{4\a}{3}R(0)
\left[\left(\frac{2\pi}{\a}\right)^3-1\right]+\chi(\a;a),
\enas
where $\chi(\a;a)$ depends on the function $u(r;a)$ and is
singular in the limit $a\ra 0$. An important fact about the function
$\chi(\a;a)$, because of it implications in the computation of the
black-hole entropy, is that at small deficit angles it vanishes as
$(2\pi-\a)^2$.

Let us now pass to the $D$-dimensional case. It is possible to consider the
general case of a two-dimensional cone $C_\a$ embedded in a $D$-dimensional
manifold $\M_\a$, so that near the singularity $(r=0)$ the metric is
represented as
\bea
ds_{\M_\a}^2=e^\sigma[dr^2+r^2d\theta^2+
\sum_{i,j=1}^{D-2}\left[\gamma_{ij}(x)+h_{ij}(x)r^2\right]dx^i dx^j+\dots],
\label{Dmetric}
\ena
where the ellipsis mean higher powers in $r^2$. The singular set is now
a $(D-2)$-dimensional surface $\Sigma$ with coordinates $\{ x^i\}$ and
metric $\gamma_{ij}(x)$. Near this surface the manifold $\M_\a$ looks
as the direct product $C_\a\times\Sigma$. We assume that the metric
do not depend on $\theta$, at least in a small region near $\Sigma$.

The metric (\ref{Dmetric}) can be regularized in the same way as the
two-dimensional case:
\beas
ds^2_{{\tilde{\M}}_\a}=e^\sigma[u(r;a)dr^2+r^2d\theta^2+
\sum_{i,j=1}^{D-2}\left[\gamma_{ij}(x)+h_{ij}(x)r^2\right]dx^i dx^j+\dots].
\label{RDmetric}
\enas
One can then proceed by computing the geometrical quantities for the smoothed
manifold ${\tilde{\M}}_\a$ and then taking the limit $a\ra 0$. As before,
only the two-dimensional conical part gives rise to the singular contributions.
Moreover, as before there are integral quantities which are independent
of the regularization in the limit $a\ra 0$, and others which are not.

Well defined quantities are the components of the Riemannian tensor,
that can be represented near $\Sigma$ in the following local form:
\bea
{}^{(\a)}{R^{\mu\nu}}_{\a\b}&=&{R^{\mu\nu}}_{\a\b}+
(2\pi-\a)\left[(n^\mu n_\a)(n^\nu n_\b)-(n^\mu n_\b)(n^\nu n_\a)\right]
\delta_\Sigma,\nonumber\\
{}^{(\a)}{R^\mu}_\nu&=& {R^\mu}_\nu+(2\pi-\a)(n^\mu n_\nu)\delta_\Sigma,
\nonumber\\
{}^{(\a)}R&=& R+2(2\pi-a)\delta_\Sigma,
\label{RimTens}
\ena
where $\delta_\Sigma$ is the $\delta$-function
$\int_{\M_\a}f\delta_\Sigma= \int_\Sigma f$, and $n^k=n^k_\mu dx^\mu$
are two orthonormal vectors orthogonal to $\Sigma$, $(n_\mu
n_\nu)=\sum_{k=1}^2n^k_\mu n^k_\nu$. The quantities
${R^{\mu\nu}}_{\a\b}$, ${R^\mu}_\nu$, and $R$ are computed at the
regular points $\M_\a/\Sigma$ by means of the standard methods.

An important consequence of the above equations is the following formula for
the integral curvature on $\M_\a$:
\beas
\int_{\M_\a}{}^{(\a)}R=2(2\pi-\alpha){\cal A}_\Sigma+
\int_{\M_\a/\Sigma} R,
\enas
where ${\cal A}_\Sigma$ is the area of $\Sigma$.

As an example of quantities which are mathematically ill-defined on the cone,
it is worth considering integrals of terms quadratic in the curvature. These
contain a part which is well defined as the regularization is removed, and
terms which are divergent or regularization dependent. However, it is
an important result \cite{FUSOL95} that these ill-defined terms vanish
as $(2\pi-\a)^2$ for small deficit angles, so that we can write ($D=4$)
\bea
&&\int_{\M_\a} R^2=4(2\pi-\a)\int_\Sigma R+\frac{\a}{2\pi}
\int_{\M_\a/\Sigma}R^2+\O\left((2\pi-\a)^2\right)\nonumber\\
&&\int_{\M_\a} R^{\mu\nu}R_{\mu\nu}=2(2\pi-\a)\int_\Sigma R_{\mu\nu}
n^\mu_i n^\nu_i+\frac{\a}{2\pi}\int_{\M_\a/\Sigma}R^{\mu\nu}R_{\mu\nu}
\nonumber\\&&\hspace{3.5cm}+\O\left((2\pi-\a)^2\right)\nonumber\\
&&\int_{\M_\a} R^{\mu\nu\a\b}R_{\mu\nu\a\b}=4(2\pi-\a)
\int_\Sigma R_{\mu\nu\a\b}n^\mu_i n^\a_i n^\nu_j n^b_j\nonumber\\
&&\hspace{3.5cm}+\frac{\a}{2\pi}
\int_{\M_\a/\Sigma}R^{\mu\nu\a\b}R_{\mu\nu\a\b}+
\O\left((2\pi-\a)^2\right).
\label{QRiem}
\ena

Another important result  in four-dimensions regards
the Euler number, which for the regularized manifold ${\tilde{\M}}_\a$
reads  \cite{FUSOL95}
\beas
\chi[{\tilde{\M}}_\a]&=&\frac{1}{32\pi^2}\int_{{\tilde{\M}}_\a}
\left[R^2-4R^{\mu\nu}R_{\mu\nu}+R^{\mu\nu\a\b}R_{\mu\nu\a\b}\right].
\enas
Although each term contains parts which are divergent or depend on the
regularization, it turns out that these terms cancel each other, leaving
a well-defined result as $a\ra 0$:
\beas
\chi[\M_\a]&=&\frac{1}{2\pi}(2\pi-\a)\chi[\Sigma]\\
&&+\frac{1}{32\pi^2}\int_{\M_\a/\Sigma}
\left[R^2-4R^{\mu\nu}R_{\mu\nu}+R^{\mu\nu\a\b}R_{\mu\nu\a\b}\right],
\enas
where $\chi[\Sigma]=(4\pi)^{-1}\int_\Sigma R_\Sigma$ is the Euler number
of the surface $\Sigma$. It is clear that the first term on the right hand
side is the contribution of the singular points. This topological result
can be generalized to higher even dimensions $D=2d$.

Finally, we can compute the heat kernel coefficients for the operator
$A=-\Delta+m^2+\xi R$ on the manifold $\M_\a$. The general expressions
of the first three coefficients was given in Eq. (\ref{heatcoeff}),
and using the formulae given above we get ($D=4$)
\beas
a_0(A)&=&\int_{\M_\a},\\
a_1(A)&=&\frac{1}{3}(1-6\xi)(2\pi-\alpha)\int_\Sigma+\frac{1}{6}
(1-6\xi)\int_{\M_\a}R-m^2\int_{\M_\a},\\
a_2(A)&=&\frac{1}{180}\left\{\frac{\alpha}{2\pi}\int_{\M_\a}
\left[R^{\mu\nu\a\b}R_{\mu\nu\a\b}-R^{\mu\nu}R_{\mu\nu}+
\frac{5}{2}(1-6\xi)^2R^2\right.\right.\\
&&\hspace{-1cm}\left.-30m^2(1-6\xi)R+\frac{180\pi}{\a}m^4\right]\\
&&\hspace{-1cm}+2(2\pi-\a)\int_\Sigma
\left[2R_{\mu\nu\a\b}n^\mu_i n^\a_i
n^\nu_j n^\b_j-R_{\mu\nu}n^\mu_i n^\nu_i+2R-30m^2(1-6\xi)\right]\\
&&\hspace{-1cm}\left.+\O\left((2\pi-\a)^2\right)\right\}.
\enas
We have used the fact that on the smooth manifold $\int \Delta R$ is
a surface term that vanishes if $R\ra 0$ at infinity.

In order to check these results we consider the important case of a
simple flat cone, $\M_\a=C_\a\times R^2$, so that the regular part of
the curvature vanishes. In the case $\xi=0$ and $m=0$ we get
\beas
a_0(A)&=&{\cal A}_\Sigma V(C_\a),\\
a_1(A)&=&\frac{{\cal A}_\Sigma}{3}(2\pi-\a),\\
a_2(A)&=&\O\left((2\pi-\a)^2\right).
\enas
The short proper-time asymptotic expansion of the heat kernel
then reads
\beas
K_t(A)&\simeq&\frac{{\cal A}_\Sigma V(C_\a)}{16\pi^2 t^2}+
\frac{{\cal A}_\Sigma}{48\pi^2 t}(2\pi-\a)+\O\left((2\pi-\a)^2\right).
\enas
On the simple flat cone $C_\a\times R^2$ we know the exact integrated
heat kernel (see Eq. (\ref{HKtrace}), which reads
\beas
K_t(A)&=&\frac{{\cal A}_\Sigma V(C_\a)}{16\pi^2 t^2}+
\frac{{\cal A}_\Sigma}{48\pi t}
\left(\frac{2\pi}{\a}-\frac{\a}{2\pi}\right)\\
&\simeq&\frac{{\cal A}_\Sigma V(C_\a)}{16\pi^2 t^2}+
\frac{{\cal A}_\Sigma}{48\pi^2 t}(2\pi-\a)+\O\left((2\pi-\a)^2\right),
\enas
where in the last row we expanded for $2\pi-\a$ small. Thus, we
see that for small deficit angles the two results are equivalent.
However, for arbitrary deficit angles the smoothed singularity
method fails in predicting the correct heat kernel coefficients.
Indeed, if we consider just the $a_1$ coefficient, it is well
defined within the smoothed singularity method, but,
nevertheless it is equivalent to the exact one only
up to terms $\O\left((2\pi-\a)^2\right)$: the presence of a
term proportional to $\a^{-1}$ in this heat kernel
coefficient implies that it is not a locally computable
geometric invariant \cite{cheeger83}.

In conclusion, we can say that although the smoothed
singularity method is a valid tool to understand some aspects
of the conical geometry, it must be used with great care,
since it is reliable only for small deficit angles.

\section{Heat kernel of the scalar Laplacian on the cone}
\label{conickernel}
\markboth{The conical space}{Heat kernel on the cone}

 In this section we study the heat kernel of minus the Laplace-Beltrami
operator on a simple two-dimensional flat cone, $C_\a$, with metric
\bea
ds^2=dr^2+r^2d\tau^2.
\label{twocone}
\ena

\noindent This problem has been considered by many authors
(see, e.g., \cite{sommerfeld,carslaw19,DO77,cheeger83,brunsee87,
DO87a,DesJack88,GiRuVa90,fursaev94,CKV94}) and here we will
summarize some of the results.

In order to construct the heat kernel of the operator $A=-\Delta_\a$
we need a complete set of eigenfunctions of the self-adjoint extension
of the operator of the manifold $C_\a$. It was shown by Kay and Studer
\cite{KS91} that this extension is not unique, but rather there is a family
of self-adjoint extensions labeled by a parameter $R$, $R\in [0,\infty)$.
If $R=0$ the extension of $A$ is positive, while when $R\neq 0$ there is a
bound state in the spectrum. Essentially, these extensions correspond to
different boundary conditions of the eigenfunctions at the apex of the cone,
$r=0$: regular, and therefore vanishing, in the case $R=0$, or divergent as
$\ln r/R$ for $R\neq 0$. The $R=0$ case corresponds to the Friedrichs
extension of the operator (see, for instance, \cite{reedsimonII}).

Notwithstanding this, in the same paper it was shown that only the
Friedrichs extension should be physically relevant, unless
we want to consider some non trivial interaction between the field
and some singular $\delta$-like potential at the conical singularity
(see \cite{KS91} and \cite{AlKaOt96} for a complete discussion of this
problem). Therefore we limit ourselves to the case of the Friedrichs
extension, which
corresponds to the problem of finding the normalized eigenfunctions
of $A=-\Delta_\a=-(\pa_r^2+\frac{1}{r}\pa_r+\frac{1}{r^2}\pa_\tau^2)$
which vanish at $r=0$. These are easily found and read
\beas
\phi_{n\l}(x)&=&\frac{1}{\sqrt{\a}}e^{i\frac{2\pi n}{\a}\tau}J_{\nu_n}(\l r),
\hspace{1cm}n=0,\pm 1,\pm 2,\dots,\hspace{5mm}\l\in R^+,\nonumber\\
\Delta_\a\phi_{n\l}(x)&=&-\l^2\phi_{n\l}(x),
\enas
where $J_\nu(z)$ are the Bessel functions. We have also introduced
the useful notation  $\nu_n=2\pi |n|/\a$, which will be used
throughout this thesis. Often we will also use $\nu=\nu_1=2\pi/\a$. The
normalization can be checked using the relation
\bea
\int_0^\infty dr\, r J_\nu(\l' r) J_\nu(\l r)=\frac{1}{\l}\delta(\l-\l').
\label{besselnormal}
\ena

\begin{figure}
\centerline{\epsffile{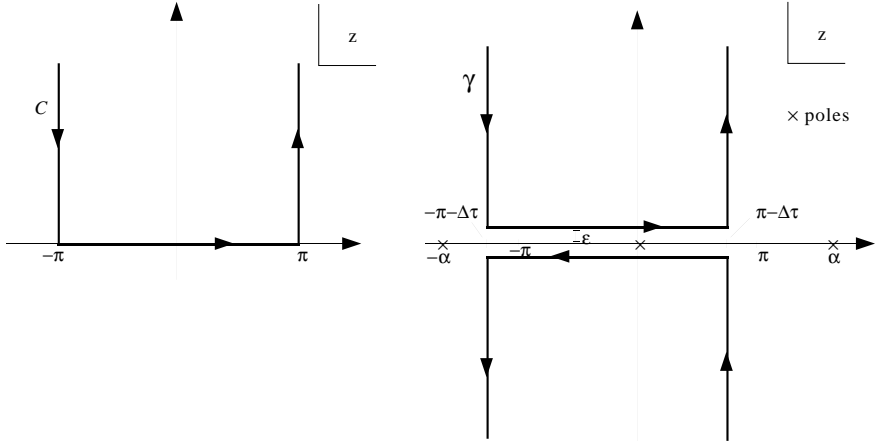}}
\caption{{\small Contours $C$ and $\gamma$}.}
\label{contour1}
\end{figure}

Using the above eigenfunctions we can write the heat kernel of the
operator $A$ as
\bea
K_t^\a(x,x'|A)&=& \langle x|e^{-t A}|x'\rangle\nonumber\\
&=&\frac{1}{\a}\sum_{n=-\infty}^{\infty}\int_0^\infty d\l\,\l\,
e^{-t\l^2}e^{i\nu n\Delta\tau}J_{\nu_n}(\l r')J_{\nu_n}(\l r),
\label{spectralkernel}
\ena
where $\Delta\tau=\tau-\tau'$. The integration measure is a consequence
of the normalization chosen. The integration over $\l$ can be easily
performed (see \cite{GR}, page 718) to give
\beas
K_t^\a(x,x'|A)&=&\frac{1}{2\a t}e^{-\frac{r^2+{r'}^2}{4t}}
\sum_{n=-\infty}^{\infty} I_{\nu_n}\left(\frac{rr'}{2t}\right)
e^{i\nu n\Delta},
\enas
where $I_\nu(z)$ are the Bessel functions of imaginary argument. The
summation over $n$ can now be performed with the help of the Schalafi
representation of $I_\nu$ (see \cite{GR}, page 952 and 954):
\beas
I_\nu(x)=\frac{1}{2\pi}\int_C dz\, e^{x\cos z+i\nu z},
\enas
where the integration contour is given in figure \ref{contour1}.

\noindent Using this representation we have
\beas
K_t^\a(x,x'|A)&=&\frac{1}{4\pi\a t}e^{-\frac{r^2+{r'}^2}{4t}}
\int_Cdz\, e^{\frac{rr'}{2t}\cos z}
\sum_{n=-\infty}^{\infty} e^{i\nu_n z+i\nu_n\Delta\tau},
\enas
The series is divergent and so it must be regularized: we can use an
$\epsilon$-prescription multiplying by $e^{-\epsilon |n|}$:
\beas
\sum_n e^{i\nu |n| z+i\nu n\Delta\tau}&\Rightarrow&
\sum_n e^{i\nu |n| z+i\nu n\Delta\tau-\ep |n|}\\
&&\hspace{-15mm}=1+\sum_{n=1}^\infty
 \left[e^{i(\nu z+\nu\Delta\tau+i\ep}\right]^n
+\sum_{n=1}^\infty \left[e^{i(\nu z-\nu\Delta\tau+i\ep}\right]^n\\
&&\hspace{-15mm}=\frac{i}{2}{\mbox{ctg}\,}
\left[\frac{1}{2}(\nu(z+\Delta\tau)+i\ep)\right]+
\frac{i}{2}{\mbox{ctg}\,}\left[\frac{1}{2}(\nu(z-\Delta\tau)+i\ep)\right].
\enas
Then we have
\beas
K_t^\a(x,x'|A)&=&\frac{i}{8\pi\a t}e^{-\frac{r^2+{r'}^2}{4t}}
\int_C dz\, e^{\frac{rr'}{2t}\cos z}\times\nonumber\\
&&\hspace{-10mm}\times\left\{{\mbox{ctg}\,}
\left[\frac{1}{2}(\nu(z+\Delta\tau)+i\ep)\right]+
{\mbox{ctg}\,}\left[\frac{1}{2}(\nu(z-\Delta\tau)+i\ep)\right]\right\}.
\enas
Now we can split the integral in two parts: in the first we change variable
to $w=-(z+\Delta\tau+i\ep/\nu)$, while in the second
$w=z-\Delta\tau+i\ep/\nu$. The final result is \cite{DO77,DesJack88}
($\ep/\nu\ra\ep$)
\bea
K_t^\a(x,x'|A)&=&\frac{i}{8\pi\a t}\int_\gamma dw\,
e^{-\frac{r^2+{r'}^2-2r r' \cos(w+\Delta\tau)}{4t}}
{\mbox{ctg}\,}\frac{\pi w}{\alpha}.
\label{HKreprA}
\ena
Since the flat-space ($\nu=1$) heat kernel is just
\beas
K_t^\a(x,x'|A)=\frac{1}{4\pi t} e^{\frac{(x-x')^2}{4t}},
\hspace{1cm} x=(r\cos\tau, r\sin\tau),
\enas
we can also write the heat kernel on the cone as
\bea
K_t^\a(x,x'|A)&=&\frac{i}{2\a }\int_\gamma dw\,
{\mbox{ctg}\,}\left(\frac{\pi w}{\alpha}\right)\,K_t^{2\pi}(x(w),x'|A),
\label{HKreprB}
\ena
where $x(w)=\left(r\cos(\tau+w), r\sin(\tau+w)\right)$.

It is also possible to deform the integration contour $\gamma$ in
various ways. For example \cite{DesJack88}, we can separate the
contour into the sum of vertical lines plus the closed Cauchy contour
around the poles of ${\mbox{ctg}\,}(\pi w/\a)$:
\bea
K_t^\a(x,x'|A)&=&\frac{1}{4\pi t}{\sum_n}'
e^{-\frac{r^2+{r'}^2-2r r' \cos(n\a+\Delta\tau)}{4t}}\nonumber\\
&&+\frac{1}{8\pi\a t}\int_{-\infty}^{\infty} dy\,
e^{-\frac{r^2+{r'}^2+2r r' {\mbox{\tiny ch}\,} y}{4t}}\times\nonumber\\
&&\times\left\{{\mbox{ctg}\,}\left[\frac{\pi}{\a}(iy-\pi-\Delta\tau)\right]+
{\mbox{ctg}\,}\left[\frac{\pi}{\a}(iy+
\pi-\Delta\tau)\right]\right\}.\nonumber\\
&&
\label{HKreprC}
\ena
The primed sum is over the $n$ such that $-\pi<n\a+\Delta\tau<\pi$.
In the coincidence limit $x'\ra x$ the above expression can be written
as
\bea
K_t^\a(x,x'|A)&=&\frac{1}{4\pi t}{\sum_n}'
e^{-\frac{r^2}{t}\sin^2\frac{n\a}{2}}\nonumber\\
&&\hspace{-20mm}+\frac{1}{8\pi\a t}\int_{-\infty}^{\infty} dy\,
e^{-\frac{r^2}{t}{\mbox{ch}\,}^2 \frac{y}{2}}
\left\{{\mbox{ctg}\,}\left[\frac{\pi}{\a}(iy-\pi)\right]+
{\mbox{ctg}\,}\left[\frac{\pi}{\a}(iy+\pi)\right]\right\}.
\nonumber\\
&& \label{HKreprD}
\ena
Some comments on the behavior of this kernel can be found in
\cite{DesJack88}.

Another useful representation of the heat kernel (\ref{HKreprB}) can
be obtained for $\Delta\tau<\pi$ by modifying the contour $\gamma$
into a contour $\Gamma$ which consists of two branches, one going from
$-\pi-\Delta\tau+i\infty$ to $-\pi-\Delta\tau-i\infty$ and
intersecting the real axis very close to the origin, and the other one
from $\pi-\Delta\tau+i\infty$ to $\pi-\Delta\tau-i\infty$ in the same
way. Then we have \cite{carslaw19,CKV94,fursaev94}
\bea
K_t^\a(x,x'|A)&=&\frac{1}{4\pi t}e^{-\frac{(x-x')^2}{4t}}
+\frac{i}{8\pi\a t}\int_\Gamma dw\,
e^{-\frac{r^2+{r'}^2-2r r' \cos(w+\Delta\tau)}{4t}}
{\mbox{ctg}\,}\frac{\pi w}{\a}\nonumber\\
&=&K_t^{2\pi}(x,x'|A)+\frac{i}{2\a}\int_\Gamma dw\,
{\mbox{ctg}\,}\frac{\pi w}{\a}\, K_t^\a(x(w),x'|A)
\label{HKreprE}
\ena

\begin{figure}
\centerline{\epsffile{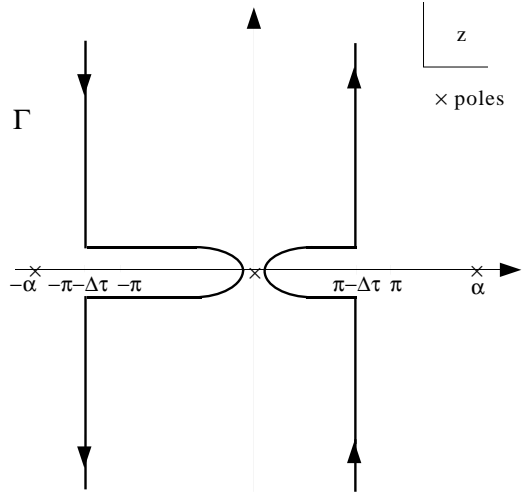}}
\caption{{\small Contour $\Gamma$}.}
\label{contourGamma}
\end{figure}
The advantage of this representation is that we have isolated the
flat-space heat kernel from the conical singularity contribution.

Up to we have considered the local heat kernel. The integrated trace of
the kernel can be computed integrating (see Eq. (\ref{ciennealpha}))
the above representation (\ref{HKreprE})  on the cone
\cite{kac66,cheeger83,DO87a,CKV94}:
\bea
\Tr e^{t\Delta_\a}=K_t^\a(A)=\frac{V(C_\a)}{4\pi t}+
\frac{1}{12}\left(\frac{2\pi}{\a}-\frac{\a}{2\pi}\right),
\label{HKtrace}
\ena
where $V(C_\a)=\a R^2/2$ is the (infinite) volume of the cone
of radius $R$. The constant term is the conical singularity contribution,
first obtained by Kac \cite{kac66}: the presence of the term $\a^{-1}$
implies that it is not a locally computable geometric invariant
\cite{cheeger83}.

\subsection{Singular heat-kernel expansion}
\label{singularsection}

In this section we consider the asymptotic expansion of the heat
kernel on the cone considered in the previous section. As a consequence
of the singular nature of the curvature at the apex of the cone,
we expect that the terms of such an expansion are distributions
concentrated at the tip of the cone. Indeed, the expansion has been
computed by Cognola {\em et al.} \cite{CKV94} and Fursaev
\cite{fursaev94} (see also \cite{fursaev94b} and \cite{DO94b}).
Because of the distributional nature of the asymptotic expansion
we consider it as a functional acting on test functions $f(r,\tau)$
periodic in $\tau$ that we suppose integrable on the cone and such
that $rf(r,\tau)$ in infinitely differentiable at $r=0$. Then the
asymptotic expansion of the trace reads ($t\ra 0_+$)
\bea
K_t^\a(A)&=&\Tr\left(e^{-tA}f\right) =
\int d^2x\sqrt{g}\, K^\a_t(x,x|A) f(x)\nonumber\\
&\simeq&\frac{1}{4\pi t}\sum_{n=0}^\infty a_{\a,n}(f)\, t^n.
\label{singexp}
\ena
The first coefficient is just the flat space one, and so
it reads
\beas
a_{\a,0}(f)&=&\int d^2x\sqrt{g}\, f(x).
\enas
The other contributions act as Dirac's delta functions and derivatives
at $r=0$ \cite{fursaev94}:
\bea
a_{\a,n}(f)&=&\frac{\Gamma(n)}{(2n-1)!}C_{2n}(\a)
\int_0^\a d\tau\,\frac{d^{2n-1}}{dr^{2n-1}}rf(r,\tau)|_{r=0},
\hspace{5mm}n\geq 1,
\label{singcoeff}
\ena
where the coefficients $C_{2n}(\a)$ are given by the integrals
\bea
C_{2n}(\a)=\frac{i}{4\a} \int_\Gamma dw \,
\mbox{ctg}\left(\frac{\pi w}{\a}\right)
\left(\sin^2\frac{w}{2}\right)^{-n},
\label{ciennealpha}
\ena
and the contour $\Gamma$ is that in figure (\ref{contourGamma}).
These integrals have been computed by Dowker \cite{DO87a}, and the
first two are
\bea
C_2 (\a)&=&\frac{1}{6}\left[\left(\frac{2\pi}{\a}\right)^2-1\right]
=\frac{\pi}{3\a}\left(\frac{2\pi}{\a}-
\frac{\a}{2\pi}\right)=\frac{2\pi}{\a}I_\a(0)\nonumber\\
C_4 (\a)&=&\frac{1}{15}C_2(\a)
\left[\left(\frac{2\pi}{\a}\right)^2+11\right]=
\frac{2\pi}{\a}I_\a(-1),
\label{C2C4}
\ena
where we have also reported the relation with the
function $I_\a(z)$ that will be defined in the next section.
It is also possible to write a local form for the above
expansion \cite{CKV94}:
\beas
K_t^\a(x,x|A)&=&\frac{1}{4\pi t}\left[1+\a\sum_{k=0}^\infty
\frac{C_{2k+2}(\a)}{2^k k!}{\Delta_\a}^k\hat{\delta} (r)\,
 t^{k+1}\right],
\enas
where $\hat{\delta}(r) =\delta(r)/r$.

From the above equations we see that the heat-kernel coefficients
act like delta functions and derivatives of delta functions at $r=0$,
and so they do not depend on the value of $f$ at regular points of the
cone: they would never appear if the integration over $r$ is stopped
short before $r=0$, no matter how close. This problem, however,
affects only the asymptotic expansion, not the integral
representations of the local heat kernel we have seen in the
previous section.

Another aspect to be remarked of the above expansion is
that the half-integer powers of $t$ are absent
\cite{CKV94,fursaev94b}: this means that,
as far as the asymptotic expansion of the heat kernel is concerned,
the cone $C_\a$ does not behave as a two-dimensional manifold with a
boundary at $r=0$.

Let us now consider a function which is $1$ in a domain of radius $R$, and
zero outside, $f(r,\tau)=\Theta(R-r)$, then the asymptotic series is
truncated and one gets the expression exact up to terms
that vanish exponentially as $t\ra 0_+$ \cite{fursaev94}
\beas
K_t^\a(A)_{R}&=&\frac{1}{4\pi t}\left(\frac{\a R^2}{2}+
\a C_2(\a)\,t\right)+\mbox{E.S.}\nonumber\\
&=&\frac{V(C_\a)}{4\pi t}+\frac{1}{12}\left(\frac{2\pi}{\a}-
\frac{\a}{2\pi}\right)+\mbox{E.S.}
\enas
which is just the expression (\ref{HKtrace}) obtained integrating the
integral representation of the heat kernel \cite{kac66,cheeger83}, and
so we conclude that the exponentially small terms are actually zero.

\section{$\z$ function of the scalar Laplacian on the cone}
\label{Coniczeta}
\markboth{The conical space}{$\z$ function on the cone}

In the previous section we have studied the heat kernel of the scalar
Laplacian on the simple cone $C_\a$, and we have seen that there are
various integral representation of the local heat kernel. In this
section we turn our attention to the local $\z$ function that,
as we have seen in section (\ref{shorthkz}), is related to the local heat 
kernel
by a Mellin transform. Nevertheless, we will follow a completely
different way to compute the $\z$ function, and the final result,
instead of an integral representation, will be a meromorphic function.
This part is based on the paper \cite{ZCV} by Zerbini {\em et al.}.
In the final part of this section we will compare the global
heat kernel and $\z$ function, introducing a new procedure
to define the global $\z$ function which solves the apparent
discrepancy of the two methods.

In the case of the $\z$ function on the cone the bibliographic
references are much more limited than for the heat kernel.
The general theory was developed by Cheeger \cite{cheeger83}
for manifolds and operators more general than those of interest
for us. Other relevant papers in the mathematical literature
are those of Br\"{u}ning and Seeley \cite{brunsee85,brunsee87}.
Finally, the method of Cheeger was introduced and developed in
the physical context by Zerbini {\em et al.}\cite{ZCV}.

We start by reviewing the Cheeger-Zerbini method for computing
the $\z$ function on a simple cone $C_\a\times \mbox{R}^{D-2}$,
with metric $ds^2=r^2 d\tau^2+dr^2+\sum_{i=1}^{D-2}dx_i^2$,
${\bf x}$ being the transverse flat coordinates. This
method will be extensively used in the following Chapters,
where, in particular, it will be also extended to the case of a
massive scalar field, the Maxwell field and the graviton field.

As for the heat kernel, the starting point is a complete set
of normalized eigenfunctions of the Friedrichs extension of
minus the Laplace-Beltrami operator $A=-\Delta-\Delta_\a
-\Delta_{D-2}=-(\pa_r^2+\frac{1}{r}\pa_r+\pa_\tau^2+
\sum_{i=1}^{D-2}\pa_{x_i}^2)$. These eigenfunctions read
\bea
\phi_{n\l{\bf k}}(x)&=&\frac{1}{(2\pi)^{\frac{D-2}{2}}\sqrt{\a}}
e^{i{\bf k}\cdot{\bf x}-i\frac{2\pi n}{\a}\tau}
J_{\nu_n}(\l r),\nonumber\\
A\phi_{n\l{\bf k}}(x)&=&
(\l^2+{\bf k}^2)\phi_{n\l{\bf k}}(x)\nonumber\\
(\phi_{n'\l'{\bf k}'}(x),\phi_{n\l{\bf k}}(x))&=&\int d^Dx\sqrt{g}\,
\phi_{n'\l'{\bf k}'}^\ast(x)\phi_{n\l{\bf k}}(x)\nonumber\\
&=&\delta_{nn'}\delta({\bf k}-{\bf k'})\frac{\delta(\l-\l')}{\l}.
\label{eigenfuncionzeta}
\ena
By means of these eigenfunctions we can write the spectral
representation of the local $\z$ function:
\beas
\z_\a(s;x|A)&=&
\frac{2}{(4\pi)^{\frac{D-2}{2}}\a\Gamma\left(\frac{D-2}{2}\right)}
\int_0^\infty dk\,k^{D-3}\sum_{n=-\infty}^\infty \int_0^\infty
\frac{d\l\,\l}{(\l^2+k^2)^{s}}J_{\nu_n}^2(\l r)\nonumber\\
&=&\frac{\Gamma\left(s-\frac{D-2}{2}\right)}
{(4\pi)^{\frac{D-2}{2}}\a\Gamma(s)}
\sum_{n=-\infty}^\infty \int_0^\infty
d\l\,\l^{D-1-2s}J_{\nu_n}^2(\l r),
\enas
where we have used $\int d\Omega_{N-1}=2\pi^{N/2}/\Gamma(N/2)$.
Now we would have to perform the integration over $\l$ and the
sum over $n$, but just here we find the main obstacle in the definition
of the $\z$ function. In fact, for
$\frac{D}{2}-\frac{1}{2}<\Re s<\frac{D}{2}+\nu_n$ the integration
converges (see \cite{GR}, formula 6.574.2) and we
formally get
\beas
\z_\a(s;x|A)&=&
\frac{r^{2s-D}\Gamma\left(s-\frac{D-1}{2}\right)}
{2\sqrt{\pi}(4\pi)^{\frac{D-2}{2}}\a\Gamma(s)}
\sum_{n=-\infty}^\infty
\frac{\Gamma\left(\nu_n-s+\frac{D-2}{2}+1\right)}
{\Gamma\left(\nu_n+s-\frac{D-2}{2}\right)}.
\enas
However, as it is shown in the appendix of \cite{ZCV},
the series converges only for $\Re s>\frac{D}{2}$, and so
the region of convergence does not overlap with that of
the integral over $\l$.

The solution of this convergence problem has been suggested
by Cheeger \cite{cheeger83}, and it simply consists in a separate
treatment of the lower and higher eigenvalues, namely
$\nu_0=0$ (or $\nu_n$, $n< n_0$) and $\nu_n$, $n\geq 1$
(or $\nu_n$, $n\geq n_0$). So, one treats separately the terms
corresponding to $\nu_0$ and $\nu_n$, $n\geq 1$, and
only after the analytic continuation is performed the two
contributions are summed to give the local $\z$ function.
Of course, such definition of $\z$ function has all the requested
properties and coincides with the usual definition when the
manifold is smooth.

So, following the procedure outlined above, we isolate the term
with $n=0$ and define, for $\frac{D}{2}-\frac{1}{2}<\Re s<\frac{D}{2}$
\beas
\z_<(s;x|A)&=&\frac{r^{2s-D}\Gamma\left(s-\frac{D-1}{2}\right)}
{\sqrt{\pi}(4\pi)^{\frac{D-2}{2}}\a\Gamma(s)}
\frac{\Gamma\left(-s+\frac{D-2}{2}+1\right)}
{2\Gamma\left(s-\frac{D-2}{2}\right)}\\
&=&-\frac{r^{2s-D}\Gamma\left(s-\frac{D-1}{2}\right)}
{\sqrt{\pi}(4\pi)^{\frac{D-2}{2}}\a\Gamma(s)}
G_{2\pi}\left(s-{\scriptsize \frac{D-2}{2}}\right).
\enas
Then we consider all the other terms, performing the integration
over $\l$ and the summation over $n$. In order to have a common
strip of convergence we have to restrict to
$\frac{D}{2}<\Re s<\frac{D}{2}+\nu_1$. The result reads
\beas
\z_>(s;x|A)
&=&\frac{r^{2s-D}\Gamma\left(s-\frac{D-1}{2}\right)}
{\sqrt{\pi}(4\pi)^{\frac{D-2}{2}}\a\Gamma(s)}
G_\a\left(s-{\scriptsize \frac{D-2}{2}}\right).
\enas
where we have set
\beas
G_\a(s)&=&\sum_{n=1}^\infty\frac{\Gamma(\nu_n-s+1)}
{\Gamma(\nu_n+s)},\\
G_{2\pi}&=&-\frac{\Gamma(1-s)}{2\Gamma(s)}.
\enas
The series $G_\a(s)$ is studied in the appendix of \cite{ZCV},
where it is shown that it is convergent for $\Re s>1$ and
that it has an analytic continuation in the whole complex
plane showing only a simple pole in $s=1$.

Since both $\z_<$ and $\z_>$ can be analytically continued
in the whole complex $s$ plane, we can define the local $\z$
function as
\bea
\z_\a(s;x|A)&=&\z_<(s;x|A)+\z_>(s;x|A)\nonumber\\
&=&\frac{r^{2s-D}}{(4\pi)^{\frac{D-2}{2}}\a\Gamma(s)}
I_\a\left(s-{\scriptsize \frac{D-2}{2}}\right),
\label{coniczeta}
\ena
where
\bea
I_\a(s)&\equiv&\frac{\Gamma(s-1/2)}{\sqrt{\pi}}
\left[G_\a(s)-G_{2\pi}(s)\right].
\label{Ifunction}
\ena
The properties of the function $I_\a(s)$ are studied
in the appendix of \cite{ZCV} (see also \cite{massive}):
it has only a simple pole in $s=1$ with residue  ($\nu=2\pi/\a$)
\beas
\mbox{Res }I_\a(s)|_{s=1}&=&
\frac{1}{2}\left(\nu^{-1}-1\right),
\enas
and near the pole \cite{massive}
\beas
I_\a(s)&=&\frac{1}{2(s-1)}(\nu^{-1}-1)+\frac{1}{2}(\nu^{-1}-1)
(\gamma-2\ln 2)-\frac{\ln\nu}{\nu}+\O(s-1).
\enas
Finally, important values of $I_\a(s)$ are
\beas
I_\a(0)&=&\frac{1}{6\nu}(\nu^2-1)\\
I_\a(-1)&=&\frac{1}{90\nu}(\nu^2-1)(\nu^2+11),
\enas
and, by definition, $I_{2\pi}(s)=0$.

\subsection{Global $\z$ function}
\label{globalzeta}

Let us now consider the global $\z$ function. From the
mathematical point of view, only the local $\z$ function
has a precise meaning, because of the non-integrable
singularity in $r=0$. As a consequence, for defining global
quantities one has to introduce some kind of regularization
by means of a smearing function $f(r)$ with compact support
not containing the origin. Of course, it is also necessary to
control the volume divergences in the transverse dimensions,
which integrated give an infinite volume ${\cal A}_\Sigma$.
Thus, we define the smeared trace as \cite{ZCV}
\bea
\z_\a(s|A)_f&=&\a {\cal A}_\Sigma\int_0^\infty
dr\,r\,\z_\a(s;x|A) f(r)\nonumber\\
&=&\frac{{\cal A}_\Sigma}{(4\pi)^{\frac{D-2}{2}}\Gamma(s)}
I_\a\left(s-{\scriptsize\frac{D-2}{2}}\right) {\hat{f}}(2s-D+2),
\label{primadef}
\ena
where
\beas
{\hat{f}}(s)&=&\int_0^\infty dr\, r^{s-1}\,f(r)
\enas
is the Mellin transform of $f(r)$. The function ${\hat{f}}(s)$
is analytic since the integral in (\ref{primadef})
exists for all $s$ by definition. So, we see that the only
possible singularity of $\z_\a(s|A)_f$ is that coming from
$I_\a(s)$ in $s=1$. The simplest choice of smearing function
is $f(r)=\Theta(R-r)\Theta(r-\ep)$, which is convergent
to $1$ as $R\ra\infty$ and $\ep\ra 0$. Thus, we have
${\hat{f}}(s)=(R^s-\ep^s)/s$ and \cite{ZCV}
\beas
\z_\a(s|A)_f &=&
\frac{{\cal A}_\Sigma}{(4\pi)^{\frac{D-2}{2}}\Gamma(s)}
I_\a\left(s-{\scriptsize\frac{D-2}{2}}\right)
\frac{R^{2s-D+2}-\ep^{2s-D+2}}{2s-D+2}.
\enas

There is also an alternative procedure to define the integrated
$\zeta$ function on the cone, and it is the following. Let us consider
$\z_<(s;x|A)$ and $\z_>(s;x|A)$ {\em before} the analytic
continuation procedure. They are defined for
$\frac{D}{2}-\frac{1}{2}<\Re s<\frac{D}{2}$ and
$\frac{D}{2}<\Re s<\frac{D}{2}+\nu_1$ respectively.
In both these strips of the complex $s$ plane, $r^{2s-D}$
does not give rise to non-integrable singularities
in $r=0$, and so we can integrate them over $r$ separately,
provided we regulate the harmless volume divergence
at $r\ra\infty$, for instance by means of $\Theta(R-r)$.
After the integration we can analytically continue each piece
and sum. The final result is
\bea
\z_\a(s|A)_R &=&
\frac{{\cal A}_\Sigma}{(4\pi)^{\frac{D-2}{2}}\Gamma(s)}
I_\a\left(s-{\scriptsize\frac{D-2}{2}}\right)
\frac{R^{2s-D+2}}{2s-D+2}.
\label{secondadef}
\ena
An important property of this  $\z$ function is
that we have an extra pole in $s=\frac{D-2}{2}$, besides
the one in $s=\frac{D}{2}$ coming from $I_\a(s)$.

Let us now study the relation of these two definitions of the
integrated $\zeta$ function with the asymptotic heat kernel expansion.
 Notice that we will not try to compare the
local $\z$ function with the local asymptotic expansion of the heat
kernel, since the distributional nature of the local Seeley-DeWitt
coefficients cannot be seen from the local $\z$ function.

For the sake of simplicity we consider the
case $D=4$. The asymptotic expansion reads
\beas
K_t^\a(A)\simeq\frac{1}{(4\pi t)^2}\sum_{n=0}^\infty
A_n \,t^n,
\enas
where on $C_\a\times\mbox{R}^2$ (see Eq. (\ref{HKtrace}))
\beas
A_0&=&{\cal {A}}_\Sigma V(C_\a),\\
A_1&=&\frac{\pi{\cal {A}}_\Sigma}{3}
\left(\frac{2\pi}{\a}-\frac{\a}{2\pi}\right)=
2\pi {\cal {A}}_\Sigma I_\a(0),
\enas
and $A_n=0$, $n>1$. From Eq. (\ref{seeleyth}) we know that
the meromorphic structure of the $\z$ function is related to
the Seeley-DeWitt coefficients by
\beas
\z(s|A)=\frac{1}{16\pi^2\Gamma(s)}\frac{A_0}{s-2}+
\frac{1}{16\pi^2\Gamma(s)}\frac{A_1}{s-1}+h(s),
\enas
where $h(s)$ is an analytic function. Therefore, we expect
for $\z(s|A)$ two simple poles in $s=2$ and $s=1$, which is
not the case in $\z(s|A)_f$, that has only a simple pole in $s=2$.
Instead, the two poles correctly appear in the proposed
regularization $\z(s|A)_R$. In fact, the pole structure
of $\z(s|A)_R$ is
\bea
\z(s|A)_R&=&\frac{{\cal {A}}_\Sigma \a R^2}{32\pi^2(s-2)}
\left(1-\frac{2\pi}{\a}\right)+\frac{{\cal {A}}_\Sigma I_\a(0)}
{8\pi(s-1)}+g(s)\nonumber\\
&=&\frac{A_0}{16\pi^2(s-2)}\left(1-\frac{2\pi}{\a}\right)+
\frac{A_1}{16\pi^2(s-1)}+g(s),
\label{secondadefploes}
\ena
where $g(s)$ is an analytic function. The discrepancy in the residue
of the pole in $s=2$ is easily explained. It is shown in \cite{ZCV}
that the Mellin transform of the local heat kernel does
not correspond to $\z_\a(s;x|A)$, but rather to
$\z_\a(s;x|A)-\frac{2\pi}{\a}\z_{2\pi}(s;x|A)$, where the
Cheeger-Zerbini analytic procedure gives
$\z_{2\pi}(s;x|A)=0$. Hence the term
$$
-\frac{A_0}{16\pi^2(s-2)}\frac{2\pi}{\a}=
-\frac{{\cal {A}}_\Sigma \pi R^2}{16\pi^2(s-2)}
$$
is just the flat space contribution in the asymptotic
expansion.Therefore, we see that by means of this alternative
procedure for taking the trace of the local $\z$ function we get a
complete agreement with the integrated heat kernel.

One could wonder which one is the correct procedure, but this question
has not a simple answer. The problem can be summarized as follows. If
one is simply interested in the computation of the integrated heat
kernel from the local one, then there is no doubt that the correct
result is given in Eq. (\ref{HKtrace}). However, if one wants to
compute one-loop quantities from the local heat kernel, then the
question is more subtle and involves a conflict between the
regularization of the divergent integral over the manifold in $r=0$
and the regularization of  the ultraviolet divergences, namely the
analytic continuation procedure in the $\zeta$ function or some other
procedure employing the heat kernel, such as the dimensional one.
We will see that, if the ultraviolet divergences are regularized in
the local quantities, then the ultraviolet regularized local
quantities diverge in $r=0$ and some cut-off in the
integration over the manifold is necessary. This procedure corresponds
to the one which leads to Eq. (\ref{primadef}). On the other hand, if
we integrate the local quantities before the ultraviolet
regularization, then no cut-off in $r=0$ is needed.   This procedure
clearly corresponds to the second one, which leads to Eq.
(\ref{secondadef}).

It is important to stress that the problem is to choose the order in
which the two regularization have to be performed, and probably this
problem cannot be solved by mathematical reasoning, for which both
procedures are equally correct.  We postpone this topic to Chapter
\ref{BHENTROPY}, where we will use physical requirements to argue that
the correct procedure is the first one, Eq. (\ref{primadef}).

\cleardoublepage
\newpage

\chapter{$\zeta$-function regularization
and one-loop renormalization of
field fluctuations in curved space-time}

\label{FLUCTUATIONS}
\section*{Introduction}
\markboth{Field fluctuations}{Introduction}

This  Chapter, based on the paper \cite{fluctuations},
is devoted to improve the $\zeta$-function approach to
regularize and renormalize the averaged square field
$\langle\phi\sp{2}(x)\rangle$ in a general curved background.

The quantity $\langle\phi\sp{2}(x)\rangle$ has been studied by several
authors \cite{birrel} because its knowledge is an  important step to
proceed within the point-splitting renormalization of the stress
tensor and also  due to its importance in cosmological theories. The
knowledge of the  value of $\langle\phi\sp{2}(x)\rangle$ is also
necessary to get the renormalized Hamiltonian from the renormalized
stress tensor in non minimally coupled theories,  and this is
important dealing with thermodynamical considerations
\cite{frofur97,Zmoretti}. Furthermore, the quantity
$\langle\phi\sp{2}(x)\rangle$ gives a measure of the vacuum distortion due,
for instance, to a boundary, and so it is of great interest in the study of
Casimir's effects (see, e.g., \cite{blau88,AAA95}).

Let us consider a generic Euclidean field theory in a curved background
${\cal M}$ corresponding to the  Euclidean action
\begin{eqnarray}
S\sb{A}[\phi] = -\frac{1}{2}\int\sb{\cal M} d\sp{4}x \sqrt{g}\:
\phi A \phi,
\end{eqnarray}
where Euclidean motion operator $A$ is supposed self-adjoint
and positive-definite. The  local $\zeta$ function related to the
operator $A$, $\zeta(s,x|A)$, can be defined as the
analytic continuation of the series \cite{hawking77,libro,elizlibro}
\begin{eqnarray}
\zeta(s,x|A) = {\sum\sb{n}}'
 \lambda\sb{n}\sp{-s} \phi\sb{n}\sp{*}(x)\phi\sb{n}(x)
\label{zeta}.
\end{eqnarray}
where $\{\phi\sb{n}(x)\}$ is a complete set of normalized eigenvectors
of $A$ with eigenvalues $\lambda\sb{n}$, and $'$ means that, in the
sum, the null eigenvectors are omitted. In    (\ref{zeta}), the right
hand side is supposed to be analytically continued in the whole
$s$-complex plane. Indeed, it is well-known that the $\zeta$ function
is a meromorphic function with, at most, simple poles situated at
$s=1$ and $s=2$ in case of a four dimensional compact (Euclidean)
spacetime without boundaries.\footnote{When the (Euclidean) spacetime
is not compact some  suitable limit can be employed.} Since we assume
a compact spacetime, we work with a discrete spectrum, but all
considerations can be trivially extended to operators with continuous
spectrum. The importance of the $\zeta$ function is that its
derivative evaluated at $s=0$ defines a regularization of the one-loop
effective action:
\begin{eqnarray}
S\sb{\mbox{\scriptsize eff}}[\phi,g]=\left.
-\frac{1}{2}\frac{d}{ds}\right|\sb{s=0}
\zeta(s,x|A/\mu\sp{2}), \label{fct}
\end{eqnarray}
where $\mu$ is an arbitrary parameter with the dimensions of a  mass
necessary from dimensional considerations. Recently, the method has
been extended in order to define a similar $\zeta$-function
regularization directly for the renormalized stress tensor
\cite{Zmoretti}.

The usual approach to compute the field fluctuations by means of the
$\zeta$-function technique leads to the following  na\"{\i}ve definition
for the one-loop averaged square field
\begin{eqnarray}
\langle\phi\sp{2}(x)\rangle := \zeta(s,x|A)|\sb{s=1}
\label{naive},
\end{eqnarray}
This definition follows directly from (\ref{zeta}) taking account of
the spectral decomposition of the two-point function. Anyhow, barring
exceptional situations (e.g. see \cite{moretticonic}), this definition
is not available because the presence of a pole at $s=1$ in the
analytically continued $\zeta$ function, and so a further infinite
subtraction procedure seems to be necessary. Conversely, in the cases
of the effective action and  stress tensor regularization, the
$\zeta$-function approach leads naturally to the  complete
cancellation of divergences maintaining the finite $\mu$-parameterized
counterterms physically necessary \cite{Zmoretti}. We shall see
shortly that, also in the case of $\langle\phi\sp{2}(x)\rangle$, it is
possible to improve the $\zeta$-function approach to get the same
features: cancellation of all divergences and  maintenance of the
finite $\mu$-parameterized counterterms. This fact will allow us to
get the one-loop renormalized field fluctuations of a scalar field in
Einstein's closed universe for the massive/non-conformally-coupled
case. As we shall say, this result can be achieved in several other
ultrastatic spacetime with  constant spatial curvature. Finally, our
definition will allow us to give a formula for the trace of the stress
tensor of a generally non-conformally invariant field in which all
quantities are regularized  by means of the $\zeta$-function approach.

\section{The general approach}
\markboth{Field fluctuations}{The general approach}

Let us define the $\zeta$ function of $\langle\phi\sp{2}(x)\rangle$.
To get a {\em definition} of the $\zeta$ function regularization of
the field fluctuations we shall follow
an heuristic way already considered in other papers for different
aims \cite{CVZ90,Zmoretti,bo}. Let us consider the  one-loop effective
action in the presence of an external source $J(x)$
\begin{eqnarray}
S\sb{\scriptsize \mbox{eff}}[g\sb{ab},J] =
\ln\int{\cal D}\phi\:e\sp{S\sb{A}[J]}=
\ln \int {\cal D}\phi\: e\sp{S\sb{A} -
\frac{1}{2} \int d\sp{4}x \sqrt{g}\: J
\phi\sp{2}}
\label{effective'}.
\end{eqnarray}
We have
\begin{eqnarray}
\langle\phi\sp{2}(x)\rangle = -\left.\frac{2}{\sqrt{g(x)}}\:
\frac{\delta S\sb{\scriptsize \mbox{eff}}}{\delta
J(x)}\right|\sb{J(x) \equiv 0}. \label{formal}
\end{eqnarray}
{From} the $\zeta$-function regularization, and supposing to be possible to
interchange the order of the $J$ functional derivative and the
$s$ derivative  we can {\em formally} write, in a purely heuristic
view
\begin{eqnarray}
\langle\phi\sp{2}(x)\rangle =
\left.\frac{d\:\:}{ds}\right|\sb{s=0}
\frac{1}{\sqrt{g(x)}}
\frac{\delta \zeta(s|A/\mu\sp{2})}{\delta
J(x)}|\sb{J(x) \equiv 0}. \label{one}
\end{eqnarray}
Still  heuristically, we can also  write
\begin{eqnarray}
\frac{\delta \zeta(s|A/\mu^2)}{\delta J(x)}
={\sum_{n}}' \frac{\delta (\lambda\sb{n}/\mu^2)^{-s}}{\delta J(x)}
 = -s {\sum_{n}}' \frac{\delta (\lambda\sb{n}/\mu^2)}{\delta J(x)}
\left( \frac{\lambda_n}{\mu^2}\right)^{-(s+1)},
\label{mezzo}
\end{eqnarray}
where one can use the relation  (obtained as in Appendix of
\cite{Zmoretti})
\begin{eqnarray}
\left.\frac{\delta \lambda\sb{n}}{\delta J(x)}\right|_{J\equiv 0} =
-\phi\sb{n}\sp{*}(x)\phi\sb{n}(x)\sqrt{g(x)}.
\label{dl}
\end{eqnarray}
Supposing to analytically continue the right hand side of
(\ref{mezzo}) as far as is possible in the complex $s$ plane, through
(\ref{one}), (\ref{mezzo}), (\ref{dl}) and (\ref{zeta}) we are led
very naturally to the following {\em definition} of the
$\zeta$-function regularization of the field fluctuations, which we
shall assume without further justifications
\begin{eqnarray}
\langle\phi\sp{2}(x)\rangle := \left.\frac{d\:\:}{ds}\right|\sb{s=0}\:
\frac{s}{\mu\sp{2}}\:
\zeta(s+1,x|A/\mu\sp{2}) \label{phisquare}.
\end{eqnarray}
The identity above can be written down also as
\begin{eqnarray}
\langle\phi\sp{2}(x)\rangle := \left.\frac{d\:\:}{ds}\right|\sb{s=0}
Z(s,x|A/\mu\sp{2})\:  \nonumber,
\end{eqnarray}
where we have  {\em defined}
\begin{eqnarray}
Z(s,x|A/\mu\sp{2})
 := \frac{s}{\mu\sp{2}}\:
\zeta(s+1,x|A/\mu\sp{2}) \nonumber
\end{eqnarray}
the {\em local} $\zeta$ {\em function of the field fluctuations}. This
function has essentially the same meromorphic structure as the
function $\zeta(s+1,x|A/\mu^2)$ except for the fact that the simple
pole at $s=0$ in $\zeta(s+1,x|A/\mu\sp{2})$, whenever it exists, is
now canceled out by the factor $s$. Moreover, when
$\zeta(s+1,x|A/\mu\sp{2})$ is regular at $s=0$ the  definition in
(\ref{phisquare}) coincides with the naive one,  Eq. (\ref{naive}).
The scale $\mu$  represents the usual ambiguity due to a remaining
finite renormalization (already found concerning the effective action
and the stress tensor\cite{birrel,wald79,Zmoretti}) and it remains into the
final result whenever another {\em fixed} scale is already present in
the theory. Two final remarks are in order. First we notice that Eq.
(\ref{phisquare}) can be also written as
\begin{eqnarray}
\langle\phi\sp{2}(x)\rangle =\left. \frac{d\:\:}{ds}\right|\sb{s=0}\: s\:
\zeta(s+1,x|A) + \
 s \zeta(s+1,x|A)|\sb{s=0} \ln \mu\sp{2}.
\label{phisquare2}
\end{eqnarray}
Secondly, from the well-known (see, e.g., \cite{birrel,blau88,boelikir96})
asymptotic expansion of the local heat kernel on manifolds without
boundary
\begin{eqnarray}
K\sb{t}(x|A)\simeq \frac{1}{(4\pi)\sp{D/2}}\sum\sb{j=0}
\sp{\infty}a\sb{j}(x|A) t\sp{j-D/2},
\nonumber
\end{eqnarray}
it is easy to show that the $\zeta$ function has a pole in $s=1$ if
and only if the dimension $D$ is even and the coefficient
$a\sb{\frac{D}{2}-1}(x|A)$ is non-vanishing. Therefore, in odd
dimensions we can use the definition (\ref{naive}) of the
fluctuations, while in even dimensions we can also rewrite
(\ref{phisquare2}) as
\begin{eqnarray}
\langle\phi\sp{2}(x)\rangle :=\lim\sb{s\rightarrow 1}
\left[\zeta(s,x|A)-\frac{a\sb{\frac{D}{2}-1}
(x|A)}
{(4\pi)\sp{D/2}\Gamma(s)(s-1)}\right]+
\frac{\gamma +\ln \mu^2}{(4\pi)\sp{D/2}}a\sb{\frac{D}{2}-1}(x|A),
\nonumber
\end{eqnarray}
where $\gamma$ is Euler's constant. From this expression, it is clear
that the disappearance of $\mu$ from the final result is equivalent to
the possibility of using  the naive definition (\ref{naive}).  In four
dimensions we have that $a\sb{1}=\frac{1}{6}(1-6\xi)R(x)-m\sp{2}$, and
so (\ref{naive}) is available either for the  massless case with
vanishing scalar curvature or for the massless conformal coupling
case, $\xi=1/6$.

We finish this section giving a general formula for the trace of the
stress tensor of a non-conformal scalar field. It is well known
\cite{birrel} that in the case of a non-conformally invariant field
such trace is the sum of two contributions, an anomalous one and
a non-anomalous one. Classically, the trace can be computed from
the variation of the classical action under a conformal transformation
$g_{ab}\rightarrow\lambda(x)g_{ab}$,
$\phi(x)\rightarrow\lambda^{(2-D)/4}(x)\phi(x)$:
\begin{eqnarray*}
g^{ab}T_{ab}(x)&=&-\left.\frac{2\lambda(x)}{\sqrt{g}}
\frac{\delta S[m,\xi]}{\delta\lambda(x)}\right |_{\lambda=1}.
\end{eqnarray*}
Now we write the action as $S[m,\xi]=S[m=0,\xi=m]-\frac{1}{2}
\int \phi[m^2+\xi R]\phi\,\sqrt{g}d^Dx$: by making the
variation on both sides, using also the conformal invariance
of $S[m=0,\xi=\xi_D]$, we arrive to the following expression:
\begin{eqnarray*}
-\frac{2}{\sqrt{g}}g^{ab}\frac{\delta S[m,\xi]}{\delta g^{ab}}
&=&-\frac{D-2}{2}\phi^\ast A\phi+\left[\frac{1}{2\xi_D}
(\xi_D-\xi)\Delta-m^2\right]\phi^\ast\phi.
\end{eqnarray*}
Then, following the method introduced in \cite{Zmoretti},  we evaluate
this expression on the modes $\phi_n$ of the operator $A$, multiply by
$\lambda_n^{-s-1}$, sum over the non-null modes and analytically
continue the obtained expressions. The result reads
\begin{eqnarray}
g^{ab}\zeta_{ab}(s+1,x|A)&=&-\frac{D-2}{2}\zeta(s,x|A)\nonumber\\
&&+\left[\frac{1}{2\xi_D}(\xi_D-\xi)\Delta-m^2\right]\zeta(s+1,x|A),
\label{moloc}
\end{eqnarray}
where the tensorial $\zeta$ function, $\zeta_{ab}(s,x|A)$, is that
defined in \cite{Zmoretti}. Finally, we multiply Eq. (\ref{moloc})
by $s\mu^{-2}$ and take the derivative at $s=0$: remembering the
definitions of \cite{Zmoretti} and the definition (\ref{phisquare}),
we get the final expression for the trace of the stress tensor for a
non-conformally invariant scalar field:
\begin{eqnarray}
g^{ab}\langle T_{ab}(x)\rangle&=&\zeta(0,x|A)
+\left[\frac{1}{2\xi_D}(\xi_D-\xi)\Delta-m^2\right]\langle \phi^2(x)\rangle
\nonumber\\
&=&\frac{a_{D/2}(x)}{(4\pi)^{D/2}}
+\left[\frac{1}{2\xi_D}(\xi_D-\xi)\Delta-m^2\right]
\langle \phi^2(x)\rangle.
\label{traccia}
\end{eqnarray}
We clearly identify the contribution coming from the conformal anomaly
and the non-anomalous contribution which depends on the quantum state
chosen. The above expression is formally equivalent to a corresponding
formula containing also derivative terms given in \cite{birrel} which has
to be further regularized. Such a formal equivalence can be proven by
employing the formal motion equation for the field operator $\hat
\phi$ in the formula in \cite{birrel}. Conversely, we remark that in our
expression all quantities are already consistently regularized by
means of the same $\zeta$-function procedure taking also account of
the scale $\mu$ which appears in both sides of Eq. (\ref{traccia}).

\section{Some applications and comments}
\markboth{Field fluctuations}{Applications and comments}

Let us  consider now some simple applications in order to check on the
correctness of our procedure. First we  consider two cases in which
the simplest formula (\ref{naive}) may be used giving  correct
results. Let us consider  a scalar massless field in Minkowski
spacetime contained in a large box at the temperature $\beta$. The
local $\zeta$ function is easily obtained (see \cite{hawking77}) and
reads
\begin{eqnarray}
\zeta(s,x|A/\mu\sp{2} ) = \frac{\sqrt{\pi}\mu\sp4}{(2\pi)\sp3}
\left( \frac{2\pi}{\beta \mu}\right)\sp{4-2s}
\frac{\Gamma(s-3/2)}{\Gamma(s)}\zeta\sb{R}(2s-3)\nonumber,
\end{eqnarray}
where $\zeta\sb{R}(s)$ is the usual Riemann zeta function. Notice that
no pole appears at $s=1$, hence we could also use the naive definition
(\ref{naive}) instead of (\ref{phisquare}). In both cases the result
is
\begin{eqnarray}
\langle\phi\sp{2}(x)\rangle\sb{\beta} = \frac{1}{12 \beta\sp{2}}.
\nonumber
\end{eqnarray}
This result is the same which follows from other approaches, for
instance subtracting the Minkowski massless zero-temperature two-point
function from the thermal one and performing the limit of coincidence
of the arguments.

Consider now the Casimir effect due to two infinite parallel planes
on which the field is constrained to vanish, namely we consider a
massless scalar field in the Euclidean manifold $[0,L]\times R\sp3$.
In this case the local $\zeta$ function can be computed by taking the
Mellin transform of the corresponding heat kernel, which is given,
e.g., in \cite{AAA95}. A straightforward computation yields
($0<x<L/2$)
\begin{eqnarray*}
\zeta(s,x|A/\mu\sp{2})&=&\frac{L\sp{2s-4}\Gamma(2-s)}{16\pi\sp{2}
\Gamma(s)}\times\\
&&\times\left[2\zeta\sb{R}(4-2s)+\left(\frac{x}{L}\right)
\sp{2s-4}-2\zeta\sb{R}(4-2s,x/L)\right].
\end{eqnarray*}
We see that the $\zeta$ function is regular at $s=1$ and so the
fluctuations can be computed in the naive way:
\begin{eqnarray}
\langle\phi\sp{2}(x)\rangle\sb{\beta}
&=&\frac{1}{48L\sp{2}}-\frac{1}{8\pi\sp{2}L\sp{2}}
\left[\zeta\sb{R}(2,x/L)-\frac{L\sp{2}}{2x\sp{2}}\right]\nonumber\\
&=&\frac{1}{48L\sp{2}}\left[1-3\mbox{csc}\sp{2}
\frac{\pi x}{L}\right]\nonumber\\
&\sim&-\frac{1}{16\pi\sp{2}x\sp{2}}+{\cal O}(x\sp{0}).
\nonumber
\end{eqnarray}
This result is in agreement with the known one (see, e.g.,
\cite{fullingbook}), but $\zeta$-function approach saves much labour.

Now, let us consider other  cases where one is forced to use
(\ref{phisquare}). The simplest case is represented by a  scalar field
in Minkowski space time at zero temperature. As it is well known
\cite{AAA95}, for a massless field the corresponding $\zeta$ function
can be considered as vanishing, as follows also from the limit
$\beta\rightarrow\infty$ in first example considered,  and so we pass
directly to the massive case. The local $\zeta$ function in four
dimensions reads
\begin{eqnarray*}
\zeta(s,x|A/\mu\sp{2})=\frac{m\sp4(\mu/m)\sp{2s}}
{16\pi\sp{2}(s-1)(s-2)},
\end{eqnarray*}
which has a simple pole at $s=1$. Using (\ref{phisquare}) we immediately
obtain the correct Coleman-Weinberg-like \cite{colwein73} result
\begin{eqnarray}
\langle \phi\sp{2}(x)\rangle=\frac{m\sp{2}}
{16\pi\sp{2}}\left[2\ln\frac{m}{\mu}-1\right].
\label{mink}
\end{eqnarray}

As a simple application on a curved background we may start
considering the field fluctuations of a massless conformally coupled
scalar field on the closed static Einstein universe with spatial
section metric $a\sp{2} [dX\sp{2} + \sin\sp{2} X (d\theta\sp{2} +
\sin\sp{2}\theta d\varphi\sp{2})] $.
The spatial section of this manifold is
the  $3$-sphere $S\sp{3}$ with  curvature
$R = 6/a\sp{2}$, $a$ being the radius of the sphere.
This case has been considered in
several papers (see \cite{birrel}) and, more recently, in \cite{Zmoretti}
employing the $\zeta$-function approach for a (generally thermal)
{\em massless conformally coupled} scalar field. In this simplest case
\begin{eqnarray}
\zeta(s,x|A) = \frac{1}{4\pi\sp{1/2} V}\frac{\Gamma(s-3/2)}{\Gamma(s)}
(2s-3) a\sp{2s-1} \zeta\sb{R}(2s-3) \label{ECU},
\end{eqnarray}
where $V= V(a)= 2\pi\sp{2} a\sp{3}$ is the spatial volume of the
universe. This function has no pole at $s=1$ but takes a pole
at $s=2$. Through (\ref{naive}),  we have the well-known result
which can be achieved by other approaches with a similar amount of
work \cite{birrel}
\begin{eqnarray}
\langle \phi^2(x) \rangle = \zeta(1,x|A/\mu\sp{2}) = - \frac{1}{48\pi\sp{2}
a\sp{2}}.
\end{eqnarray}
Unlike other approaches and in a very direct and elegant
way, employing our improved procedure (\ref{phisquare}) and some
technical results found out in \cite{camporesi} we are able to
compute the vacuum fluctuations also in the case of a non-conformal
coupling and/or in the presence of a massive field, when  the $\zeta$
function does take a pole at $s=1$. The  Euclidean operator reads
\begin{eqnarray}
A = -\Delta + \xi R + m\sp{2}   =    A\sb{0}/ a\sp{2},
\end{eqnarray}
As any finite
value of the mass can be interpreted in terms of a redefinition of
$\xi$ ($R$ is constant) $\xi \rightarrow \xi' = \xi +  m\sp{2}/R$, we
shall explicitly consider the case $m=0$ only. From $\zeta(s,x|A) =
a\sp{2(s-2)} \zeta(s,x|A_{0})$ we have
\begin{eqnarray}
a\sp{2} \langle \phi^2(x) \rangle = \left.\frac{d\:\:}{ds}\right|\sb{s=0}\:
s\zeta(s+1,x|A\sb{0}) +
s\zeta(s+1,x|A_{0})|\sb{s=0} \ln(\mu\sp{2}a\sp{2} )
 \label{phisquare'},
\end{eqnarray}
where $\langle \phi^2(x) \rangle$ is  referred to $A$ and not to
$A\sb{0}$. Notice that the argument of the logarithm is adimensional:
the natural scale already  present in the theory is $a$. Using
particular cases of  the equations (11.78), (11.85) and (12.19) in
\cite{camporesi} and (\ref{ECU}) for the  the case $a=1$, we find
\begin{eqnarray}
s\zeta(s+1,x|A\sb{0}) &=&
\frac{1}{2V\sb{0}\pi\sp{1/2}}
 \frac{\Gamma(s+3/2)}{\Gamma(s)}
\left[ (1-6\xi) \zeta\sb{R}(2s+1) -\delta\sb{\xi 0} \right]\nonumber\\
&&\hspace{-2.5cm}+\frac{1}{2V\sb{0}\pi\sp{1/2}} \sum_{n=0,n\neq 1}^{+\infty}
\frac{\Gamma(s+n+1/2)}{n! \Gamma(s)}
\left[ (1-6\xi)\sp{n} \zeta\sb{R}(2s+2n-1) -\delta\sb{\xi 0} \right].
\nonumber\\
&&\label{lunga}
\end{eqnarray}
This equation holds for  $0 \leq 6\xi < 2$ (i.e., replacing  $\xi$
which appears in the equation above with $\xi'$ defined previously, $0
\leq 6\xi + a\sp{2} m\sp{2}< 2$), so it includes both the minimal
coupling $\xi=0$ and the conformal coupling $\xi=1/6$. The series
converges everywhere in $s$ complex plane away from the poles of the
function $\zeta$ and $\Gamma$ in the various numerators (see
\cite{camporesi} for a general discussion). Notice that the pole at
$s=1$ of  $\zeta(s,x| A\sb{0})$ in the case of a non-conformal
coupling arises from a corresponding pole in the term $n=1$ of the
series above which has been made explicit in the first term in the
right hand side of (\ref{lunga}). This pole at $s=1$ is nothing but
the pole at $s=2$ which is present in the conformal coupled $\zeta$
function. Finally, taking also account that (\ref{naive}) can be used
in the (massless) conformal case,  Eq. (\ref{phisquare'}) leads us to
the relationship between  $\langle \phi\sp{2}(x) \rangle\sb{\xi}$ and
$\langle \phi\sp{2}(x) \rangle\sb{\xi=1/6}$
\begin{eqnarray}
\langle \phi\sp{2}(x) \rangle\sb{\xi} &=&
\langle \phi\sp{2}(x) \rangle\sb{\xi=1/6}+
\frac{1-6\xi}{4V}(1 + \gamma -\ln 2)  \nonumber\\
&+& \frac{1}{2V \pi\sp{1/2}} \sum_{n=2}^{+\infty}
\frac{\Gamma(n+1/2)}{n!} \left[(1-6\xi)\sp{n} \zeta\sb{R}(2n-1)
-\delta\sb{\xi 0} \right] \nonumber \\
&-& \frac{3\delta\sb{\xi 0}}{4 V}+
\ln(a\sp{2}\mu\sp{2}) \frac{1-6\xi}{8V} \label{fineECU}.
\end{eqnarray}
The series in (\ref{fineECU}) converges for $0 \leq 6\xi< 2 $ at
least, as one can check employing Stirling's estimate of the $\Gamma$
function and $\zeta(s) = 1 + 2\sp{-s} + {\cal O} (3\sp{-s}) $ for
large positive values of  $s$.

To generalize the  relation (\ref{fineECU}),
let us consider a massless field (the mass can be restored by a trivial
redefinition of $\xi$, as $R$ is a constant) propagating in
an ultrastatic  manifold with the topology
of  $R \times {\cal M} $, where the second factor is a
$3$-dimensional compact rank one symmetric space (the procedure can be
generalized considering higher dimensions). The metric reads $ds\sp{2}
= -dt\sp{2} + a\sp{2} ds\sp{2}\sb{{\cal M}}$. Employing results found out in
\cite{camporesi}, we get
\begin{eqnarray}
s\zeta_\xi(s+1,x | A\sb{0})
&=& \frac{(\rho\sp{2} -\xi R\sb{0})
\Gamma(s+2)}{\Gamma(s)} \zeta\sb{\xi=1/6}(s+2,x | A\sb{0})\nonumber\\
&&\hspace{-1.5cm}
- \frac{\delta\sb{\xi 0}\Gamma(s+3/2)}{2V\sb{0}\pi\sp{1/2}\rho\sp{2s+1}}
\nonumber\\
&&\hspace{-1.5cm}+ \sum\sb{n=0,n\neq 1}\sp{+\infty}
\left[
 \frac{(\rho\sp{2} -\xi R\sb{0})\sp{n}
\Gamma(s+n+1)}{n!\Gamma(s)} \zeta\sb{\xi=1/6}(s+n+1,x | A\sb{0})\right.
\nonumber\\
&&\hspace{-1.5cm}\left.- \frac{\delta\sb{\xi 0}\Gamma(s+n+1/2)}{2V\sb{0} n!
\pi\sp{1/2}\rho\sp{2s+1}}
\right].
\end{eqnarray}
We have set $\rho\sp{2} := R\sb{0}/6 $ ($\rho>0$) and $R\sb{0},
A\sb{0}, V\sb{0}$ are referred to the metric with $a=1$. The series
converges for $0\leq \xi R\sb{0} < 2\rho\sp{2}$ away from the explicit
poles (see \cite{camporesi}). In spite of the notations, the various
local $\zeta$ functions which appear above do not depend on $x$
because of the symmetry of the space and so do the corresponding
fluctuations of the field. Also in this case, the pole at $s=1$ of the
general $\zeta$ function is related with that at $s=2$ in the
conformally coupled $\zeta$ function. Once again,  (\ref{phisquare})
and  (\ref{naive}) for the conformal case, lead us to
\begin{eqnarray}
\langle \phi\sp{2}(x) \rangle\sb{\xi} &=&
\langle \phi\sp{2}(x) \rangle\sb{\xi=1/6}- \frac{3}{4V}\frac{\delta
\sb{\xi 0}}{\rho} + \ln(\mu\sp{2}a\sp{2})
\left(\rho\sp{2}/a\sp{2}
- \xi R \right) P\nonumber\\
&&\hspace{-2cm}+ \sum\sb{n=2}\sp{+\infty}
\left[ \frac{(\rho\sp{2}/a\sp{2} -\xi R)\sp{n}
\Gamma(n+1)}{ n!} \zeta\sb{\xi=1/6}(n+1,x | A)
- \frac{\delta\sb{\xi 0}\Gamma(n+1/2)}{2V n! \pi\sp{1/2}\rho}  \right]
\nonumber\\
&&\hspace{-2cm}+  \left(\rho\sp{2}/a\sp{2} -\xi R\right) P
+\left(\rho\sp{2}/a\sp{2} -\xi R\right)
\frac{d\:\:}{ds}|_{s=0}
s\zeta\sb{\xi=1/6}(s+2,x|A), \label{fff}
\end{eqnarray}
where we have set $P = \lim\sb{s\rightarrow 0} s\zeta\sb{\xi=1/6}(s+2,x|A) =
a_{0}(x|A)/(4\pi)\sp{2}$ and, for $n>2$, $\zeta_{\xi=1/6}(n,x|A)$ is
the arguments coincidence limit of the Green function of
$A^n_{\xi=1/6}$. Up to the knowledge of the authors, this is the first
time that equations as (\ref{fineECU}) and (\ref{fff}) appear in the
literature and there is no analogous relation obtained employing
approaches different from the $\zeta$ function one.

\cleardoublepage
\newpage

\chapter{Massive scalar field}

\label{massivechap}
\section*{Introduction}
\markboth{The massive case}{Introduction}

In this Chapter, based on the paper \cite{massive}, we discuss
the extension of the local $\z$ function on the cone to the case
of massive fields. As we  have partially seen in Chapter
\ref{ONTHECONE}, many techniques have been developed to compute
relevant quantities in presence of conical singularities (see, among
the others,
\cite{DO77,cheeger83,DO87a,DesJack88,AlOt90,GiRuVa90,KS91,CKV94,fursaev94,FUSOL95,ZCV}).
With few exceptions
\cite{CKV94,fursaev94,linet87,ShiraHire92,GuimLin93,GuimLin94,moreira95},
most authors considered massless fields only. Indeed, the
introduction of the mass complicates  the problem considerably, so that
it becomes very difficult to obtain manageable forms for the physical
quantities. In this regard, it has to be noted that using the integrated
heat-kernel approach \cite{CKV94,fursaev94} it seems that all the
problems vanish: in fact, for a scalar field of mass $m$ the
integrated heat-kernel  is simply related to the massless one as
\begin{eqnarray}
 K_t^{m}=e^{-tm^2}K_t^{m=0},
\label{heat}
\end{eqnarray}
and, at least for a simple cone $R^{D-2}\times C_\alpha$,
the integrated massless heat-kernel is well known
\cite{cheeger83,CKV94,fursaev94} (see Eq. (\ref{HKtrace}):
\begin{eqnarray*}
K_t^{m=0}=\frac{\Sigma_{D-2}}{(4\pi t)^2}\left[\frac{V(C_\alpha)}
{4\pi t}+\frac{1}{12}\left(\frac{2\pi}{\alpha}-\frac{\alpha}{2\pi}
\right)\right],
\end{eqnarray*}
where $2\pi-\alpha$ is the deficit angle of the cone, $V(C_\alpha)$ is
the volume of the cone and $\Sigma_{D-2}$ is the volume of the
transverse dimensions. In contrast to the massless case, the
Mellin transform of Eq. (\ref{heat}) exists in the massive case, and
so it is easy to compute the massive global $\zeta$ function and from
it quantities like the effective action, which have a simple closed
form. However, it is well known \cite{ZCV} that this procedure does
not yield the correct dependence on the deficit angle of the cone in
the massless limit and, moreover, these global quantities are finite,
apart for the volume divergences, while the local quantities show a
non-integrable singularity near the tip of the cone.\footnote{In the
cosmic string case, this problem could be overcome renormalizing the
action of the string as in \cite{fursaev94}.} Therefore, it is at
least dubious that this simple result for the massive case is correct.
Besides, in order to compute important quantities such as the vacuum
fluctuations of the field  and the stress tensor, local rather than
global quantities are needed. The problems related to the
local and integrated approaches on the cone will be discussed
in Chapter \ref{BHENTROPY}.

The aim of this Chapter is to give a more manageable tool for
computing local quantities for a massive scalar field around a cosmic
string or at finite temperature in the Rindler wedge. To avoid any
confusion, we will focus on the cosmic string case, but all the
results we will obtain are easily translated to the case of fields at
finite temperature in the Rindler space or near the horizon of a
Schwarzschild black hole. We will accomplish this result  by computing
the local $\zeta$ function, which will be given as an expansion in
powers of $mr$, where $r$ is the distance form the string core.
Therefore, the obtained expressions are useful near the string, where
$mr\ll 1$.\footnote{Note that $r=\hbar/mc$ is the Compton wavelength
of the particle. When the result are translated for the Rindler case,
the approximation is valid close to the event horizon.} However, since
the local quantities diverge at the string core, this is just the
interesting region if, for instance, one wants to consider the
back-reaction of the energy-momentum tensor on the background metric
\cite{Hiscock87,MEX97}. The expression we obtain is simple enough to
allow us to compute the renormalized vacuum expectation value $\langle
\phi^2(x)\rangle$ and the the energy-momentum tensor, potentially up
to arbitrary order in $(mr)^2$. Notice that the most common tool used
to compute local quantities is the Green's function, which in the
massive case is given in a complicated integral  form
\cite{linet87,ShiraHire92,GuimLin94,CKV94} or as a sum of generalized
hypergeometric functions \cite{moreira95}.

As we said above, we are mainly concerned in the comic-string
background. We remind that the space-time around an infinitely long,
static and straight cosmic string \cite{vilenk85} has the topology
${\cal M}^4=R^2\times C_\alpha$, where $C_\alpha$ is the simple
two-dimensional cone with deficit angle $2\pi-\alpha$, and the metric
is
\begin{eqnarray}
ds^2&=&-dt^2+dz^2+dr^2+r^2 d\theta^2,\nonumber\\
&&\hspace{2cm} r\in(0,+\infty),\,
\,\theta\in[0,\alpha],\, \,t,z\in(-\infty,+\infty).
\label{metric}
\end{eqnarray}
The polar angle deficit $2\pi-\alpha$ is related to the mass per unit
length of the string $\mu$ by $2\pi-\alpha=8\pi G\mu$. Throughout this
chapter we shall assume positive deficit angle, so that
$\nu\equiv2\pi/\alpha>1$. For GUT strings $\nu-1\sim 10^{-6}$
\cite{vilenk85}. In the rest of the chapter we  adopt an Euclidean
approach and so we perform a Wick rotation $\tau=it$: the form of the
metric is the same as above with the replacement
$-dt^2\rightarrow+d\tau^2$. Moreover, for technical reasons we will
also consider the obvious generalization to the $D$-dimensional case,
with topology $R^{D-2}\times C_\alpha$.

We have assumed zero thickness for the string: this is clearly an
idealization, since an actual cosmic string should have a small  but
finite radius, of the order of the Compton wavelength of the Higgs
boson involved in the phase transition which gives rise to the cosmic
string \cite{vilenk85}. The internal structure of the string has
non-negligible effects even at large distances, as it has been shown
by Allen, Kay, and Ottewill \cite{AlKaOt96}. However, in the same work
it has been shown that for a minimally coupled scalar this dependence
on the internal structure is absent, and one can safely use the
idealized string. In order to avoid the complications discussed in
\cite{AlKaOt96}, which are related to the non-uniqueness of the
self-adjoint extension of the Laplace-Beltrami operator in the
idealized conical space-time, we shall consider the minimally coupled
case only. Actually, in Sec. IV we will compute the stress tensor
and, for sake of completeness, we will consider arbitrary coupling.
Therefore,  we have to remember that for $\xi\neq 0$ others effects
could be present in a realistic cosmic string.

The rest of this chapter is organized as follows. In section
\ref{sex2} we compute the $\zeta$ function of a massive scalar field
in the cosmic-string background. Then we use this result to compute,
in the region $mr\ll 1$, the vacuum fluctuations of the field in
section \ref{sex3} and the expectation value of the stress tensor in
section \ref{sex4}. In section \ref{sex5} we employ the stress tensor
to compute the backreaction on the metric of the idealized cosmic
string, in the context of of the semiclassical approach to quantum
gravity. Section \ref{sex6} contains the conclusions.

\section{Computation of the $\zeta$ function}
\markboth{The massive case}{Massive $\z$ function}
\label{sex2}

We consider a quasi-free real scalar field in the background given by
the (Wick-rotated) metric (\ref{metric}). The action of the theory is
the given by
\begin{eqnarray*}
S[\phi]&=&\frac{1}{2}\int d^Dx\sqrt{g}\,\phi A\phi,
\end{eqnarray*}
where $A$ is known as the small fluctuations operator and in our
case reads
\begin{eqnarray}
A&=&-\Delta_D+m^2+\xi R.
\end{eqnarray}
Here $\xi$ is a parameter which fixes the coupling of the field to the
gravity by means of the scalar curvature $R$. We shall consider the
minimally coupled case, $\xi=0$. Then, by means standard passages
one sees that the generating functional of the theory can be expressed
in terms of the functional determinant of the small fluctuations
operator, which can be conveniently defined by means of the $\zeta$
function regularization \cite{hawking77} (see Chapter \ref{ONTHECONE}):
\beas
\ln Z_\alpha&=&-\frac{1}{2}\ln\det\mu^{-2}A=\frac{1}{2}
\zeta'(0|A\mu^{-2})
\enas
where $\zeta(s|A)$ is the global $\zeta$ function related to the
operator $A$, the prime indicates the derivative with respect to $s$
and $\mu$ is an arbitrary parameter with the dimensions of a mass
needed for dimensional reasons and not to be confused with the mass
per unit length of the string. The global $\zeta$ function can be
formally written as the integral over the manifold of a local $\zeta$
function $\zeta(s|A)(x)$:
\begin{eqnarray*}
\zeta(s|A)&=&\int_{{\cal M}^D_\alpha}\zeta(s|A)(x)\sqrt{g} d^Dx.
\end{eqnarray*}
Actually, when the manifold is non-compact only the local $\zeta$
function has a precise mathematical meaning, since the integration
requires the introduction of cutoffs or smearing functions to avoid
divergences.

In the massless case, the $\zeta$ function on the cone has been
explicitly computed in \cite{ZCV} (see section \ref{Coniczeta}). In
order to compute it in the massive case, the starting point is the
well known relation between the local $\zeta$ function and the heat
kernel given by the Mellin
transform:
\begin{eqnarray*}
\zeta^{m}(s;x)=
\frac{1}{\Gamma(s)}\int_0^\infty dt\,t^{s-1}K_t(x).
\end{eqnarray*}
The heat kernel of the massive field is related to the massless
one by means of the following obvious decomposition:
\begin{eqnarray*}
K_t^{m}(x)=e^{-tm^2}K_t^{m=0}(x).
\end{eqnarray*}
Now we take into account the following property of the Mellin
transform of the product of two functions \cite{report}
\begin{eqnarray*}
\int_0^\infty t^{s-1}f(t)g(t)dt=
\frac{1}{2\pi i}\int_{\sigma-i\infty}^{\sigma-i\infty}
F(z)G(s-z)dz,
\end{eqnarray*}
where $F$ and $G$ are the Mellin transforms of $f$ and $g$
respectively and $\sigma$ is a real number in the common strip of
convergence of the two  Mellin transforms. In this way we can write
the massive $\zeta$ function of  in terms of the massless one:
\begin{eqnarray}
\zeta^{m}(s;x)=\frac{1}{2\pi i \Gamma(s)}
\int_{\sigma-i\infty}^{\sigma-i\infty}
\Gamma(z)\zeta^{m=0}(z;x)m^{2z-2s}\Gamma(s-z) dz,
\label{general}
\end{eqnarray}
where we have used the Mellin transform
\begin{eqnarray*}
\int_0^\infty t^{s-1}e^{-tm^2}dt=m^{-2s}\Gamma(s),
\end{eqnarray*}
which converges for $\mbox{Re} s>0$.

Since we are interested in the $\zeta$ function of a massive scalar
field in the (Euclidean) space $ R^{D-2}\times C_\alpha$, we consider
the corresponding massless $\zeta$ function. It has been shown in
\cite{cheeger83} and \cite{ZCV} that it is the sum of two parts which
converge in separate strips of the complex $s$-plane and which are
summed only after their analytic continuation:
\begin{eqnarray*}
\zeta^{m=0}(z;x)=\zeta_<(s;x)+\zeta_>(s;x),
\end{eqnarray*}
where
\begin{eqnarray*}
\zeta_<(s;x)&=&\frac{r^{2s-D}}{\alpha(4\pi)^{\frac{D-2}{2}}
\Gamma(s)} \frac{\Gamma(s-\frac{D-1}{2})
\Gamma(\frac{D}{2}-s)}{2\sqrt{\pi}\Gamma(s-\frac{D-2}{2})}, \\
&& \hspace{4cm}\mbox{for }\frac{D-1}{2}<\mbox{Re } s<\frac{D}{2},\\
\zeta_>(s;x)&=&\frac{r^{2s-D}}{\alpha(4\pi)^{\frac{D-2}{2}}
\Gamma(s)} \frac{\Gamma(s-\frac{D-1}{2})}{\sqrt{\pi}}
G_\alpha(s-{\scriptstyle\frac{D-2}{2}}),\\
&& \hspace{4cm}\mbox{for }
\frac{D}{2}<\mbox{Re } s<\frac{D}{2}+\nu,
\end{eqnarray*}
and the function $G_\alpha(s)$ is defined as
\begin{eqnarray*}
G_\alpha(s)=\sum_{n=1}^{\infty}
\frac{\Gamma(\nu_n-s+1)}{\Gamma(\nu_n+s)},
\hspace{1cm}\nu_n=\frac{2\pi}{\alpha}|n|,\hspace{0.5cm}
\nu\equiv \nu_1
\end{eqnarray*}
and can be analytically continued in the whole complex plane showing
simple poles in $s=1$, with residue $\alpha/4\pi$, and $s=\nu_n+k+1$,
$k=0,1,2,\dots,$ (if $\alpha\neq2\pi$) with obvious residue. Since
$\zeta_<$ and $\zeta_>$ do not have a common strip of convergence,
we must split Eq. (\ref{general}) in two parts: setting $\mbox{Re}
s>D/2$ we have
\begin{eqnarray*}
\zeta_>^{m}(s;x)&=&\frac{1}{2\pi i\Gamma(s)}
\frac{r^{-D}m^{-2s}}{(4\pi)^{\frac{D-2}{2}}\sqrt{\pi}\alpha}\times\\
&&\times\int_{\scriptsize \frac{D}{2}<{\mbox{\scriptsize Re}\,} z<
{\mbox{\scriptsize Re}\,} s}
dz\,\Gamma(z-{\scriptstyle\frac{D-1}{2}})
G_\alpha(z-{\scriptstyle\frac{D-2}{2}}) \Gamma(s-z)(mr)^{2z}
\nonumber\\
&&\hspace{-2.5cm}=\frac{r^{2s-D}}{(4\pi)^{\frac{D-2}{2}}
\sqrt{\pi}\alpha\Gamma(s)}
\left\{\sum_{n=0}^\infty\frac{(-1)^n}{n!}(mr)^{2n}
\Gamma(s+n-{\scriptstyle\frac{D-1}{2}})
G_\alpha(s+n-{\scriptstyle\frac{D-2}{2}})\right.
\nonumber\\
&&\hspace{-2.5cm}+\left.(mr)^{D-2s}
\sum_{n=1}^\infty\sum_{k=0}^\infty\frac{(-1)^k}{k!}
\frac{\Gamma(\nu_n+k+1/2)}{\Gamma(2\nu_n+k+1)}
\Gamma(s-\nu_n-k-{\scriptstyle\frac{D}{2}})
(mr)^{2\nu_n+2k}\right\},
\end{eqnarray*}
where we have performed the integral shifting the integration path to
the right and picking up the residues of the poles of the integrand.
The same procedure can be applied to $\zeta_<^{m}(s;x)$, but in this
case we have to consider the poles in $z=s+n$ and $z=2+n$:
\begin{eqnarray*}
\zeta_<^{m}(s;x)&=&\frac{1}{2\pi
i\Gamma(s)}\frac{r^{-D}m^{-2s}}{(4\pi)^{\frac{D-2}{2}}\sqrt{\pi}\alpha}\times\\
&&\times\int_{\scriptsize\frac{D-1}{2}<{\mbox{\scriptsize Re}\,} z<
{\mbox{\scriptsize Re}\,} \frac{D}{2}}
\hspace{-5mm}dz\,\frac{\Gamma(z-\frac{D-1}{2})
\Gamma(\frac{D}{2}-z)}
{2\Gamma(z+1-D/2)}\Gamma(s-z)(mr)^{2z}\nonumber\\
&&\hspace{-1.5cm}=
\frac{r^{2s-D}}{(4\pi)^{\frac{D-2}{2}}\sqrt{\pi}\alpha\Gamma(s)}
\left\{\sum_{n=0}^\infty\frac{(-1)^n}{n!}(mr)^{2n}
\frac{\Gamma(s+n-\frac{D-1}{2})\Gamma(\frac{D}{2}-s-n)}
{2\Gamma(s+n+1-D/2)}\right.
\nonumber\\
&&\hspace{-1.5cm}\left.+(mr)^{D-2s}
\sum_{n=0}^\infty\frac{(-1)^n}{n!}(mr)^{2n}
\frac{\Gamma(n+1/2)\Gamma(s-n-D/2)}{2\Gamma(n+1)}\right\}.
\end{eqnarray*}
Now we analytically continue each term and sum to get the final
expression of the local $\zeta$ function of a massive scalar field:
\begin{eqnarray}
\zeta^{m}(s;x)&=&\frac{r^{2s-D}\mu^{2s}}{(4\pi)^{\frac{D-2}{2}}
\alpha\Gamma(s)}\left\{\sum_{n=0}^\infty\frac{(-1)^n}{n!}(mr)^{2n}
I_\alpha(s+n-{\scriptstyle\frac{D-2}{2}})
\right.\nonumber\\
&&\hspace{-25mm}+\frac{(mr)^{D-2s}}{\sqrt{\pi}}
\sum_{n=0}^\infty\frac{(-1)^n}{n!}(mr)^{2n}
\frac{\Gamma(n+1/2)\Gamma(s-n-D/2)}{2\Gamma(n+1)}\nonumber\\
&&\hspace{-25mm}\left.+\frac{(mr)^{D-2s}}{\sqrt{\pi}}
\sum_{n=1}^\infty\sum_{k=0}^\infty\frac{(-1)^k}{k!}
\frac{\Gamma(\nu_n+k+1/2)}{\Gamma(2\nu_n+k+1)}
\Gamma(s-\nu_n-k-{\scriptstyle\frac{D}{2}})
(mr)^{2\nu_n+2k}\right\},\nonumber\\
&&
\label{zetafun}
\end{eqnarray}
where the function $I_\alpha(s)$ is that defined in Eq. (\ref{Ifunction}).
In expression (\ref{zetafun})
we have also reintroduced the arbitrary mass $\mu$ using the formal
relation $\zeta(s|A\mu^{-2})=\mu^{2s} \zeta(s|A)$, and which has been
omitted in the derivation for simplicity. It must be noted that the
above $\zeta$ function can be obtained in a more direct way, albeit
much longer, performing the integrations in the Mellin transform of
the spectral representation of the massive heat kernel, and then
rearranging the sums and the Euler gamma functions in the generalized
hypergeometric functions obtained by the integrations. We prefer the
used method since it can be more easily applied to the computation of
others quantities, as we will see in the next sections. Moreover, this
method based of the Mellin transform could be useful in other cases to
treat the massive case when the massless $\zeta$ function is known.

Although expression (\ref{zetafun}) looks awful, it is very simple
when $mr\ll 1$: considering the physically interesting case
$D=4$ and terms up to\footnote{Note that we are interested in the
values of the $\zeta$ function at $s=0,1$, and we consider positive
deficit angles only, so that $\nu>1$.} $(mr)^4$
\begin{eqnarray*}
\zeta^{m}(s;x)&=&
\frac{r^{2s-4}\mu^{2s}}
{4\pi\alpha\Gamma(s)}\left[I_\alpha(s-1)-
(mr)^2 I_\alpha(s)\right.\\
&&\left.+\frac{1}{2}(mr)^4 I_\alpha(s+1)+
{\cal O}\left((mr)^6\right)\right]\nonumber\\
&&+\frac{m^4(m/\mu)^{-2s}}{8\pi\alpha\Gamma(s)}
\left[\Gamma(s-2)-\frac{(mr)^2}{2}\Gamma(s-3)\right.\\
&&\left.+2(mr)^{2\nu}
\frac{\Gamma(\nu+1/2)\Gamma(s-\nu-2)}
{\sqrt{\pi}\Gamma(2\nu+1)}+{\cal O}\left((mr)^4\right)\right].
\end{eqnarray*}
In the first term of the first row we recognize the massless $\zeta$
function on $R^2\times C_\alpha$ \cite{ZCV}, while the first term in
the second row becomes the $\zeta$ function of a massive scalar field
in the Minkowski space-time when $\alpha=2\pi$ \cite{AAA95}. The
others terms are clearly corrections due to the presence of the
conical singularity. When the singularity is absent, namely when
$\alpha=2\pi$, the first row vanishes, since $I_{2\pi}(s)=0$, while in
the second and third rows only the first term survives, since the
others cancel two by two.

At the same order in $mr$, the effective lagrangian density is
given by
\begin{eqnarray*}
{\cal L}(x)&=&-\frac{1}{2}\frac{d}{ds}\zeta^m(s;x)|_{s=0}\nonumber\\
&=&-\frac{1}{8\pi\alpha r^4}\left\{I_\alpha(-1)
-(mr)^2 I_\alpha(0)\right.\nonumber\\
&&-\frac{(mr)^4}{2}
\left[\frac{\nu-1}{\nu}\left(\gamma+\ln\frac{{r\mu}}{2}\right)+
\frac{\ln\nu}{\nu}+\ln\frac{m}{\mu}-\frac{3}{4}\right]\\
&&-\frac{(mr)^6}{24}\left[2G_\alpha(2)+2\gamma+
2\ln\frac{mr}{2}-\frac{11}{6}\right]\nonumber\\
&&\left.+2(mr)^{2\nu+4}
\frac{\Gamma(\nu+\frac{1}{2})\Gamma(-\nu-2)}
{\sqrt{\pi}\Gamma(2\nu+1)}\right\}.
\end{eqnarray*}
where  $G_\alpha(n)$, $n\geq 2$, are readily computed being in the
region of convergence of the series defining $G_\alpha$. For example
\begin{eqnarray*}
G_\alpha(2)&=&-\frac{1}{2\nu}\left[2\gamma+
\psi\left(\nu^{-1}\right)+
\psi\left(-\nu^{-1}\right)\right],
\nonumber\\
G_\alpha(3)&=&-\frac{1}{6\nu}\left[3\gamma-
\psi\left(\nu^{-1}\right)-
3\psi\left(-\nu^{-1}\right)
+\psi\left(-2\nu^{-1}\right)\right],
\end{eqnarray*}
where $\psi(x)=\Gamma'(x)/\Gamma(x)$. In the limit
$\alpha\rightarrow 2\pi$ the above effective lagrangian reduces to the
usual Coleman-Weinberg potential. It is important to note that the
computation of higher orders in the above expansion in powers of
$(mr)^2$ does not involve particular complications.

\section{Vacuum fluctuations}
\markboth{The massive case}{Vacuum fluctuations}
\label{sex3}

Now we use the above $\zeta$ function to compute the (renormalized) value
of the vacuum fluctuations of the field. We start considering the na\"{\i}ve
definition (\ref{naive}) \cite{AAA95}:
\beas
\langle\phi^2(x)\rangle&=&\lim_{s\rightarrow 1}\zeta(s;x).
\enas
Since even in the Minkowski space the vacuum expectation value
diverges, we renormalize the vacuum expectation value on the cone
subtracting the Minkowski value: in this way we have a remarkable
cancellation of the poles, yielding a finite result:
\begin{eqnarray}
\langle\phi^2(x)\rangle_\alpha-\langle\phi^2(x)\rangle_{2\pi}&=&
\nonumber\\
&&\hspace{-2cm}
\frac{1}{4\pi\alpha r^2} \left\{I_\alpha(0)+\frac{(mr)^2}{\nu}
\left[(\nu-1)\left(\ln\frac{mr}{2}+\gamma-\frac{1}{2}\right)+
\ln\nu\right]\right.\nonumber\\
&&\hspace{-2cm}+\frac{(mr)^4}{8}\left[2G_\alpha(2)+2\gamma-1+
2\ln\frac{mr}{2}\right]\nonumber\\
&&\hspace{-2cm}\left.+(mr)^{2\nu+2}
\frac{\Gamma(\nu+1/2)\Gamma(-\nu-1)}{\sqrt{\pi}\Gamma(2\nu+1)}+
{\cal O}\left((mr)^6\right)\right\}.
\label{vacuum}
\end{eqnarray}
The computation of higher corrections is straightforward. One can
verify that Eq. (\ref{vacuum}) correctly vanishes in the limit
$\alpha\rightarrow 2\pi$, and that up to order $(mr)^2$ it is
identical to the result obtained by Moreira Jnr. \cite{moreira95}. In
particular, we notice the additional logarithmic divergences of the
vacuum fluctuations at the conical singularity due to the massive
corrections.

Let us now compute the same vacuum fluctuations employing
the improved definition (\ref{phisquare}): in this case no explicit
subtraction is needed and, considering only terms up to order
$(mr)^2$,  the result is
\begin{eqnarray}
\langle\phi\sp{2}(x)\rangle_\a&=&\frac{1}{4\pi\a r\sp{2}}
\left\{I_\a(0)+\frac{(mr)\sp{2}}{\nu}\left[(\nu-1)
\left(\gamma+\ln\frac{r\mu}{2}\right)\right.\right.\nonumber\\
&&\left.\left.+\ln\nu+
\frac{\nu}{2}\left(2\ln\frac{m}{\mu}-1\right)\right]\right\}.
\label{cosmic}
\end{eqnarray}
The above expression is different from the result ob\-tai\-ned
by sub\-trac\-ting the Min\-kow\-ski val\-ue and in fact reduces to
the Minkowski value, Eq. (\ref{mink}), rather than vanishing when the
conical singularity is removed, namely $\nu\rightarrow 1$. We note
that in the massless case both procedures give the same result, since
the massless $\zeta$ function is regular at $s=1$. On the other hand,
the arbitrariness in the choice of the reference state to be subtracted
appears  in (\ref{cosmic})
in the form of a dependence on the arbitrary mass parameter
$\mu$: changing $\mu$ is equivalent to a change in the
subtraction point. Indeed, the difference between
two results for the vacuum fluctuations can be written as
\begin{eqnarray*}
\frac{(mr)^2}{8\pi^2 r^2}\left[\ln\frac{m}{\mu}-\frac{1}{2}\right].
\end{eqnarray*}
So, e.g., if one chooses $\mu=m/\sqrt{e}$ then Eq. (\ref{cosmic})
reproduces exactly Eq. (\ref{vacuum}).

\section{Energy-momentum tensor}
\markboth{The massive case}{Energy-momentum tensor}
\label{enermomtens}
\label{sex4}

Another important vacuum average is that of the energy-momentum
tensor. While in the massless case it is quite easy to be computed,
since its form is fixed from symmetry arguments, in the massive case
the computation is much more difficult and, to our knowledge, the
massive corrections near the string have never  been shown explicitly.
Only in  \cite{ShiraHire92} and \cite{GuimLin94} the explicit form of
the energy-momentum tensor has been given for $mr\gg 1$, where they
found the expected exponential damping factor $\exp(-2mr)$.

We use a point-splitting approach, in which the vacuum expectation
value of the energy momentum tensor is given by the coincidence
limit of a non-local differential operator applied to the regularized and
renormalized propagator of the field \cite{birrel}:
\begin{eqnarray*}
\langle T_{\mu\nu}(x)\rangle=i\lim_{x'\rightarrow x}D_{\mu\nu}(x,x')
G_{\cal F}(x,x'),
\end{eqnarray*}
where
\begin{eqnarray}
D_{\mu\nu}(x,x')&\equiv&(1-2\xi)\nabla_\mu\nabla_{\nu'}+
-2\xi \nabla_\mu\nabla_\nu
+(2\xi-\frac{1}{2})g_{\mu\nu}\left[\nabla_\alpha \nabla^{\alpha'}
-m^2\right],\nonumber\\
&&\label{dimunu}
\end{eqnarray}
and the prime indicates that the derivative has to be taken with
respect to $x'$.   Since the propagator can be obtained from the
off-diagonal $\zeta$ function \cite{birrel} as
\begin{eqnarray*}
G_{\cal F}(x,x')=i\lim_{s\rightarrow 1}\zeta(s;x,x'),
\end{eqnarray*}
we can compute the energy momentum tensor from the
$\zeta$ function as (see also \cite{hawking77,CVZ90})
\begin{eqnarray*}
\langle T_{\mu\nu}(x)\rangle=-\lim_{s\rightarrow 1}
\lim_{x'\rightarrow x}
D_{\mu\nu}(x,x')\zeta(s;x,x').
\end{eqnarray*}

The partial derivatives of the off-diagonal $\zeta$ function which
appear in the above expression can be computed from the spectral
representation in $D$ dimensions (${\bf x}\equiv(\tau,{\vec{z}})$,
$k=|{\bf k}|$)
\begin{eqnarray*}
\zeta_D(s;x,x')&=&\frac{2 (4\pi)^{-\frac{D-2}{2}}}
{\alpha\Gamma(\frac{D-2}{2})}
\int_0^\infty dk\, k^{D-3}\sum_{n=-\infty}^\infty\int_0^\infty
d\lambda\,\lambda [\lambda^2+k^2+m^2]^{-s}\nonumber\\
&&\times J_{\nu_n}(\lambda r')
J_{\nu_n}(\lambda r)e^{i{\bf k}\cdot({\bf x}-{\bf x}')+
i\frac{2\pi}{\alpha}n(\theta-\theta')},
\end{eqnarray*}
and then taking the coincidence limit. In this way one can easily show
that
\begin{eqnarray*}
\partial_\theta\partial_{\theta'}\zeta_D(s;x,x')|_{x=x'}&=&
-\partial_\theta^2\zeta_D(s;x,x')|_{x=x'}
\nonumber\\
\partial_{z_i}\partial_{{z_i}'}\zeta_D(s;x,x')|_{x=x'}&=&
-\partial_{z_i}^2\zeta_D(s;x,x')|_{x=x'}=
2\pi\zeta_{D+2}(s;x)\nonumber\\
\partial_\tau\partial_{\tau'}\zeta_D(s;x,x')|_{x=x'}&=&
-\partial_\tau^2\zeta_D(s;x,x')|_{x=x'}=2\pi\zeta_{D+2}(s;x)
\nonumber\\
\partial_{r'}\zeta_D(s;x,x')|_{x=x'}&=&\frac{1}{2}\partial_r
\zeta_D(s;x).
\end{eqnarray*}
As far as $\partial_\theta^2\zeta_D(s;x,x')|_{x=x'}$ is concerned,
in the $m=0$ case it is easy to see that we have
\begin{eqnarray*}
\partial_\theta^2\zeta_D^{m=0}(s;x,x')|_{x=x'}=
-\frac{r^{2s-D}\Gamma(s-\frac{D-1}{2})}
{(4\pi)^{\frac{D-2}{2}}\sqrt{\pi}\alpha\Gamma(s)}
H_\alpha(s-{\scriptstyle\frac{D-2}{2}}),
\end{eqnarray*}
where the function $H_\alpha(s)$ is defined and studied in the
Appendix {\bf A}. The massive case can then be treated using the
off-diagonal version of Eq. (\ref{general}) with the partial
coincidence limit $r=r'$, $z=z'$ and $t=t'$:
\begin{eqnarray*}
\partial_\theta^2\zeta_4(s;\theta,\theta')|_{\theta=\theta'}&=&
\frac{1}{2\pi i\Gamma(s)}\int \Gamma(z)
\partial_\theta^2\zeta_4^{m=0}(z;\theta,\theta')|_{\theta=\theta'}
\Gamma(s-z) m^{2z-2s}dz\nonumber\\
&&\hspace{-3cm}=
\frac{-1}{2\pi i\Gamma(s)} \int_{\mbox{\scriptsize Re } z>3}
\frac{r^{2z-4}}{4\pi\sqrt{\pi}\alpha}
\Gamma(z-3/2)H_\alpha(z-1)\Gamma(s-z) m^{2z-2s}dz.\\
&&\hspace{-3cm}=
-\frac{r^{2s-4}\mu^{2s}}{4\pi\sqrt\pi\alpha\Gamma(s)}\left\{
\sum_{n=0}^\infty\frac{(-1)^n}{n!}(mr)^{2n}\Gamma(s+n-3/2)
H_\alpha(s+n-1) \right.\nonumber\\
&&\hspace{-2cm}\left.\hspace{-2cm}+(mr)^{4-2s}
\sum_{n=1}^\infty\sum_{k=0}^\infty\frac{(-1)^k}{k!}
(mr)^{2\nu_n+2k}\nu_n^2\frac{\Gamma(s-\nu_n-k-2)
\Gamma(\nu_n+k+1/2)}
{\Gamma(2\nu_n+k+1)}\right\}.
\end{eqnarray*}
where, as usual, we have performed the integration shifting the
integration contour to the right and picking up the residues at
$z=s+n$ and $z=\nu_n+k+2$ to get an expansion in powers of $mr$
similar to Eq. (\ref{zetafun}).

As far as the second derivatives with respect to $r$ and $r'$ are
concerned, using the following identity, which can be proved using
some recursion formulas for the Bessel functions \cite{GR},
$$
[\partial_r J_\nu(\lambda r)]^2=-\frac{\nu^2}{r^2}J_\nu^2(\lambda r)
+\frac{1}{2r}\partial_r r\partial_r J_\nu^2(\lambda r)+\lambda^2
J_\nu^2(\lambda r),
$$
one can see that
\begin{eqnarray*}
\partial_r\partial_{r'}\zeta_D(s;x,x')|_{x=x'}&=&
\frac{1}{2r}\partial_r r\partial_r\zeta_D(s;x)+\frac{1}{r^2}
\partial_\theta^2
\zeta_D(s;x,x')|_{x=x'}\\
&&+\chi_D(s;x),\\
\partial^2_{r'}\zeta_D(s;x,x')|_{x=x'}&=&
-\frac{1}{2r}\partial_r\zeta_D(s;x)-\frac{1}{r^2}\partial_\theta^2
\zeta_D(s;x,x')|_{x=x'}\\
&&-\chi_D(s;x),\\
\end{eqnarray*}
where the function $\chi_D(s;x)$ is defined in Appendix {\bf B}.
If $D=4$ and $m=0$ we simply have
\begin{eqnarray}
\chi_{D=4}^{m=0}(s;x)=4\pi(s-2)\zeta_{D=6}^{m=0}(s;x),
\label{chizero}
\end{eqnarray}
while the massive case is more complicate and is studied in the
Appendix B.

Now we have all the pieces needed to compute the massive correction to
the energy-momentum tensor. For the actual calculation it is
convenient to follow \cite{BROT86} and \cite{ALO92}. We define the
renormalized stress tensor as
$$
\langle T_{\mu\nu}(x)\rangle_\alpha^R\equiv
\langle T_{\mu\nu}(x)\rangle_\alpha-
\langle T_{\mu\nu}(x)\rangle_{2\pi}
$$
which, considering only the first massive correction, turns out to be
\begin{eqnarray*}
\langle T_{\theta\theta}\rangle_\alpha^R&=&
\frac{1}{4\pi\alpha r^2}\left[2H_\alpha(0)-(6\xi-1)I_\alpha(0)+
(mr)^2\left(H_\alpha(1)+\xi\frac{\nu-1}{\nu}\right)\right],\\
\langle T_{rr}\rangle_\alpha^R&=&
\frac{-1}{4\pi\alpha r^4}
\left[2H_\alpha(0)-I_\alpha(-1)-(2\xi-1)I_\alpha(0)\right.\\
&&\left.+(mr)^2\left(H_\alpha(1)+\xi\frac{\nu-1}{\nu}\right)\right],\\
\langle T_{tt}\rangle_\alpha^R&=& \frac{-1}{8\pi\alpha
r^4}\left[I_\alpha(-1)+2(4\xi-1)I_\alpha(0)+
(mr)^2I_\alpha(0)\right],\\
\langle T_{zz}\rangle_\alpha^R
&=&\langle T_{tt}(x)\rangle_\alpha^R,
\end{eqnarray*}
The tensor can be written in the more familiar form
\begin{eqnarray}
\langle {T_\mu}^\nu(x)\rangle_\alpha^R&=&
\frac{-1}{1440\pi^2 r^4}\left [\left(\nu^4-1\right)
\mbox{diag}(1,1,-3,1)\right.\nonumber\\
&&\left.+10(6\xi-1)\left(\nu^2-1\right)
\mbox{diag}(2,-1,3,2)\right.\nonumber\\
&&+15(mr)^2(\nu-1)
(12\xi-1-\nu) \mbox{diag}(0,1,-1,0)\nonumber\\
&&\left.+15(mr)^2\left(\nu^2-1\right)
\mbox{diag}(1,0,0,1)\right].
\label{stresst}
\end{eqnarray}
In the limit $m\rightarrow 0$ the result is in agreement with that
obtained by other authors \cite{DO87a,FRSE87}.
The components $\langle T_{rr}(x)\rangle_\alpha^R$ and
$\langle T_{\theta\theta}(x)\rangle_\alpha^R$ satisfy the
equation
\beas
\frac{d}{dr}(r {T^r}_r)&=&{T^\theta}_\theta,
\enas
which follows from the conservation law $\nabla_\mu {T^\mu}_\nu=0$.
Another interesting check of the above result is the comparison with the
conformal anomaly formula, Eq. (\ref{traccia}). For a massive scalar field and general
coupling, the trace of the one-loop renormalized stress tensor is given
by an anomalous contribution $\z(s=0;x)$ and a non-anomalous
contribution which depends on the quantum state chosen \cite{birrel}:
\beas
\langle {T_\mu}^\mu(x)\rangle^R&=&
\z(s=0;x)+\left[\frac{1}{2}(1-6\xi)\Delta-m^2\right]
\langle\phi^2(x)\rangle.
\enas
Although Eq. (\ref{traccia}) was derived employing the
$\zeta$-function regularization of \cite{Zmoretti} also  for the
energy momentum tensor, it can be applied also for our mixed
$\zeta$-function and point-splitting method, provided that we employ
the expression (\ref{vacuum}) for the fluctuations in order to have a
subtraction scheme consistent with that of the stress tensor. It is
then easy to check that this relation is satisfied by the  quantities
computed above, order by order in $(mR)^2$.

It is interesting to notice that in the corrections to the stress tensor
of order $(mr)^2$  are not present the logarithmic divergences which
are instead present in the vacuum fluctuations at the same order (see
Eq. (\ref{vacuum})). The logarithmic terms appear in higher order
corrections, which can also be computed with the method we have
developed. Actually only the logarithms in the term of order $(mr)^4$
give rise to divergences, since at higher orders they are multiplied
by positive powers of $r$. Indeed, the correction of order $(mr)^4$ to
$\langle T_{\theta\theta}(x)\rangle_\alpha^R$ is
\begin{eqnarray*}
\frac{1}{4\pi\alpha r^2}\left\{-\frac{(mr)^4}{8}
\left[2\frac{\ln\nu+1}{\nu}+(2\xi-1)A(r)+4\xi-1\right]\right.\\
\left.+(mr)^{2+2\nu}\nu[2\xi+(4\xi-1)\nu]\frac{\Gamma(-\nu-1)
\Gamma(\nu+1/2)}{\sqrt{\pi}\Gamma(2\nu+1)}\right\},
\end{eqnarray*}
from which one can also compute the
correction to $\langle T_{rr}(x)\rangle_\alpha^R$
by means of the conservation law. The correction
to $\langle T_{tt}(x)\rangle_\alpha^R=
\langle T_{zz}(x)\rangle_\alpha^R$ is
\begin{eqnarray*}
-\frac{1}{8\pi\alpha r^4}\left\{\frac{(mr)^4}{4}
\left[2B(r)-(4\xi-1)(A(r)+1)-\frac{1}{\nu}\right]\right. \\
\left.+(mr)^{2+2\nu}2(4\xi-1)\nu^2\frac{\Gamma(-\nu-1)
\Gamma(\nu+1/2)}
{\sqrt{\pi}\Gamma(2\nu+1)}\right\}.
\end{eqnarray*}
In the above equations we have set
\begin{eqnarray*}
A(r)&=&2G_\alpha(2)+2\gamma+2\ln\frac{mr}{2}.\\
B(r)&=&\frac{\nu-1}{\nu}\left(\ln\frac{mr}{2}+
\gamma-\frac{1}{2}\right)+\frac{\ln\nu}{\nu}.
\end{eqnarray*}

An interesting point to be discussed is the dependence on the
parameter $\xi$ which fixes the coupling of the scalar field with the
gravity. In the introduction we said that we would consider the
minimally coupled case only, $\xi=0$: this because it has been shown
in \cite{AlKaOt96} that the idealized conical space is a good model of
the space time of a cosmic string  only for the minimally coupled
case. For  nonminimally coupled  fields quantities like $\langle
\phi^2(x)\rangle$ will depend on the details of the metric in the core
of the string even very far from the string. However, one could be
interested, e.g., in  finite-temperature fields in the Rindler space,
where the (Euclidean) manifold there is a true conical singularity and
not just an idealization of a non-singular metric. In these cases it
is interesting to consider also nonminimally coupled fields.

It is clear that in a realistic model of cosmic string all the
quantities may depend on the parameter $\xi$. Moreover, during the
process of renormalization of the theory the dependence on $\xi$ can
appear even in the case $R=0$, and actually, in an interacting theory
this parameter becomes a running coupling constant (see, e.g.,
\cite{BuOdSh92}).

Nevertheless,  in our simplified model of idealized cosmic string and
non-interacting scalar field there are quantities, such as the Green
functions and the  $\zeta$ function which do not depend on $\xi$.
Actually, from the mathematical point of view, it is not clear which
is the meaning of the field equation $[-\Delta+m^2+\xi R]\phi=0$ when
the curvature $R$ has Dirac's delta singularities, and our choice of
$\xi=0$ allowed us to avoid the problem. A possible way to  define the
problem is to smooth out the singularity, as done in
\cite{AlOt90,ALO92,FUSOL95}: as a result, one finds that when the
regularization of the singularity is removed to recover the conical
space also the dependence of the Green functions on the parameter
$\xi$ vanishes \cite{ALO92}. We can argue that also the $\zeta$
function is independent of $\xi$, since the modes used to construct
the $\zeta$ function are essentially the same as those for the Green
functions.

Therefore, we conclude that the choice of $\xi$ is irrelevant when
computing the $\zeta$ function, while, of course, it can affect other
quantities. In particular, this means that there is no contradiction
computing the energy-momentum tensor applying $D_{\mu\nu}(x,x')$ with
$\xi\neq 0$ to the Green or $\zeta$ function computed setting $\xi=0$,
as we have done above and as done by most authors.

\section{Back reaction on the background metric}
\markboth{The massive case}{Back reaction on the background metric}
\label{sex5}

In this section we make a first application of the obtained
results, employing the stress tensor computed in the previous
section to find the backreaction on the metric of the idealized
cosmic string due to the quantum fluctuations of  the massive
scalar field near the string.

In semi-classical approach to quantum gravity  the
gravitational field is kept classical while the others fields
are quantized. A possible step forward is considering
the ``backreaction'' of the quantum fields on the classical
background metric: the vacuum average of
energy-momentum tensor of the fields is computed (and
suitably renormalized) in the
background metric, and in general it involves corrections
of order $\hbar$. Then the vacuum average of the stress
tensor is inserted in the Einstein equations as source of the
gravitational field in place of the classical stress-tensor:
\begin{eqnarray}
G_{\mu\nu}=R_{\mu\nu}-\frac{1}{2}g_{\mu\nu}R=
8\pi\langle T_{\mu\nu}\rangle.
\label{einstein}
\end{eqnarray}
Then, if possible,  these equations are solved to get the
corrections of order $\hbar$ to the background metric due
to the quantum effects. Of course this approach has a limited
range of applicability, but it is commonly believed that it is
sufficiently reliable far from the Planck scale. It is however
unclear where the Planck scale physics starts being relevant.

Apart from the question of the reliability, the semi-classical
programme is not easily applicable: first, one has to able to compute
the stress tensor of the field in curved background, with the related
problems of regularize and renormalize it; second, one has to be able
to solve the quantum-corrected Einstein equations (\ref{einstein}), at
least at order $\hbar$ in the perturbations.

One of the backgrounds in which the above programme can be easily
carried on is just that of the idealized cosmic string we are
discussing in this chapter: the simple form of the metric and of the
renormalized stress tensor allowed Hiscock \cite{Hiscock87} to solve
the Einstein equations and compute the corrections to the background
metric in the case of a massless conformally coupled scalar field. More
recently, Guimar{\~a}es \cite{MEX97} has performed the same
calculation in a more general case in which the cosmic string carries
a magnetic flux that gives rise to an Aharonov-Bohm interaction with a
charged scalar field.

In the previous section we have computed the massive corrections to the
stress tensor of a scalar field in the region $mr\ll1$,
and so we can extend the work of Hiscock to include these corrections.
Notice that, since the stress tensor diverges
as $r\rightarrow 0$,  the region $mr\ll1$ is just where
one expects that the vacuum polarization of the field gives rise
to relevant backreaction effects, while at large distances from
the string the stress tensor decreases as $\exp{-2mr}$, and so
the backreaction effects should be negligible.

Following Hiscock \cite{Hiscock87}, we write the back-reacted
metric around the string as
\begin{eqnarray}
ds^2=e^{2\Phi(r)}(-dt^2+dr^2+dz^2)+e^{2\Psi(r)}d\theta^2,
\label{backmetric}
\end{eqnarray}
which is the most general metric compatible with the symmetries
of the system. The metric functions $\Phi$ and $\Psi$ are then
expanded around the background metric
\begin{eqnarray*}
\Phi(r)&=&\phi_0(r)+\phi(r),\hspace{1cm}\phi_0(r)=0,\\
\Psi(r)&=&\psi_0(r)+\psi(r),\hspace{1cm}\psi_0(r)=\ln\,r.
\end{eqnarray*}
Here the perturbations $\phi$ and $\psi$ are of order $\hbar$. In what
follows we write explicitly the Plank constant in order to stress the
role of the quantum corrections. The Einstein equations
(\ref{einstein}) for the metric (\ref{backmetric}) are then linearized
in the perturbations, and on the right hand side we insert the stress
tensor given in Eq. (\ref{stresst}). The equations for $\phi(r)$ and
$\psi(r)$ obtained in this way are then easily solved. The final
result for the perturbed metric of the string, at first order in
$\hbar$ and $(mr)^2$, is given by
\begin{eqnarray*}
ds^2&=&\left[1-\frac{\hbar A}{r^2}
+\hbar m^2B\ln mr\right](-dt^2+dr^2+dz^2)\\
&&+r^2\left[1+\frac{\hbar C}{r^2}+
\hbar m^2 D\ln mr\right]d\theta^2,
\end{eqnarray*}
where we have set
\begin{eqnarray*}
A&=&\frac{\nu^2-1}{360\pi}[\nu^2+1-10(6\xi-1)],\hspace{2cm}
B=\frac{\nu^2-1}{12\pi}[12\xi-1-\nu],\\
C&=&\frac{\nu^2-1}{72\pi}[\nu^2+1+2(6\xi-1)],\hspace{2cm}
D=\frac{\nu^2-1}{6\pi}[12\xi+1+\nu].
\end{eqnarray*}
It is convenient to introduce a system of coordinates in which
the radial coordinate is the proper distance from the string.
Therefore, we introduce the new coordinate $R$ as
\beas
R&=&r+\frac{\hbar A}{r}+\hbar m^2 B r(\ln\,mr-1),
\enas
so that $g_{rr}=1$ and the metric becomes
\begin{eqnarray*}
ds^2&=&\left[1-\frac{\hbar A}{R^2}
+\hbar m^2 B\ln mR\right](-dt^2+dz^2)+dR^2\nonumber\\
&&+R^2\left[1+\frac{\hbar (C-A)}{R^2}+
\hbar m^2(D-2B)\ln mR+2\hbar m^2B\right]d\theta^2.
\end{eqnarray*}
We see that, as a consequence of the backreaction, the geometry is no
longer locally flat. In particular, the space $(R,\theta)$ now is not
a simple flat cone, but rather a cone in which the deficit angle
depends on the coordinate $R$. Indeed, the deficit angle reads
$$
\Delta=2\pi\frac{\nu-1}{\nu}-\frac{\hbar(\nu^4-1)}{90\nu R^2}
\left[1+60(mR)^2\ln mR \right]+\frac{\hbar m^2 (\nu^2-1)}{3\nu}
[\nu+1-12\xi].
$$
A few remarks on this result. In the massless case, the correction to
the unperturbed the deficit angle is negative for any value of $R$ (if
$\nu>1$), and its absolute value decreases as $R^{-2}$ as $R$ grows
from $0$ to $\infty$ \cite{Hiscock87}. The massive correction changes
this behaviour, since now the correction to the deficit angle is
negative if $mR\stackrel{<}{\sim}0.0815$, but it is positive above,
till the approximation that $mR$ is small breaks down. It is quite
surprising that the apparently small massive correction can affect so
strongly the backreaction. Moreover, it is interesting to notice that
the $R$-dependent part is independent on $\xi$. Finally, the massive
correction introduces also a shift in the deficit angle constant with
respect to $R$, whose sign depends  on $\xi$ and $\nu$.

\section{Conclusions}
\markboth{The massive case}{Conclusions}
\label{sex6}

In this chapter we have studied the $\zeta$ function of a massive scalar
field in a cosmic string background, and we have obtained an
expression, Eq. (\ref{zetafun}), which is useful in the region near the
core of the string, $mr\ll 1$. By means of this  expression
we have  computed the massive corrections to the vacuum
fluctuations, Eq. (\ref{vacuum}), and to the energy-momentum tensor,
Eq. (\ref{stresst}), up to order $(mr)^4$, going beyond the known
results. Higher corrections are also computable.

As a first application of the results  we have
studied of the backreaction of the quantum fluctuations of the
massive scalar field on the background metric of the cosmic string,
showing that the massive corrections are not negligible.

Possible extensions of the results of this Chapter could be  the
inclusion of a magnetic flux carried by the string, which gives
Aharonov-Bohm effects, and the case of spin-$1/2$ fields.  Also the
limit $mr\gg 1$ is worth studying, for example rewriting the $\zeta$
function (\ref{zetafun}) in terms of the hypergeometric function
${}_1F_2[a;b,c;z]$ and then using its asymptotic behaviour for large
$z$. However, the vacuum fluctuations and the energy-momentum tensor
in this limit have been already obtained by other authors with
different methods \cite{ShiraHire92,GuimLin94}. The obtained results
could also be useful discussing the the scalar self-interacting theory
(see, e.g., \cite{CKV94}).

\section{Appendix A}
\label{appa}

In this Appendix we study the function $H_\alpha(s)$, which is
defined as the analytic continuation of following series:
\begin{eqnarray*}
H_\alpha(s)=\sum_{n=1}^\infty\nu_n^2
\frac{\Gamma(\nu_n-s+1)}{\Gamma(\nu_n+s)},
\hspace{1cm}\nu_n=\frac{2\pi}{\alpha}|n|.
\end{eqnarray*}
It can be studied and analytically continued proceeding exactly
as for the function $G_\alpha(s)$ studied in the Appendix of \cite{ZCV}.
Then one sees that the series converges for $\mbox{Re} s>2$ and that
the analytic continuation for $[n/2]<\mbox{Re} s<2$ (here $[n/2]$
represents the integer part of $n/2$) is given by
\begin{eqnarray*}
\sum_{j=0}^{[n/2]+1}c_j(s)\nu^{3-2s-2j}\zeta_R(2s+2j-3)
+\sum_{k=1}^\infty \nu_n^2 f_n(\nu_k,s),
\end{eqnarray*}
where $\zeta_R$ is the Riemann $\zeta$ function and the function
$f_n(\nu,s)$ is generally unknown, but vanishes for $s=1/2,0,-1/2,
-1,\dots$. The coefficients $c_j(s)$ vanish for $s=-n/2$,
($n=-1,0,1,2,\dots$) for all $j>(n+1)/2$, and the first ones
have been given in \cite{ZCV}.

The function $H_\alpha(s)$ has then a simple pole in $s=2$ and
near this pole we have
\begin{eqnarray*}
H_\alpha(s)=\frac{1}{2\nu(s-2)}+\frac{1}{\nu}(\gamma-\ln\nu)+
{\cal O}\left((\nu-2)^2\right).
\end{eqnarray*}
Moreover, it has simple poles at $s=1+\nu_n+k$, $k=1,2,\dots$, due to
the gamma function in the numerator of the terms of the series, with
obvious residue. Finally, it is possible to compute the value of the
function $H_\alpha(s)$ for some useful value of $s$:
\begin{eqnarray*}
H_\alpha(0)&=&\frac{1}{120\nu}(\nu^4-1),\nonumber\\
H_\alpha(1)&=&-\frac{1}{12\nu}(\nu^2-1).
\end{eqnarray*}

\section{Appendix B}
\label{appb}

In this Appendix we study the function $\chi_D(s;x)$, which is defined
by the analytic continuation of the following spectral representation
\begin{eqnarray}
\chi_D(s;x)&=&\frac{2 (4\pi)^{-\frac{D-2}{2}}}{\alpha\Gamma((D-2) /2)}
\times\nonumber\\
&&\times\int_0^\infty dk\, k^{D-3}\sum_{n=-\infty}^\infty\int_0^\infty
d\lambda\,\lambda^3 [\lambda^2+k^2+m^2]^{-s}J_{\nu_n}^2(\lambda r),
\nonumber\\
&&\label{chi}
\end{eqnarray}
which is the same as $\zeta_D(s;x)$ with $d\lambda\,\lambda
\rightarrow d\lambda\,\lambda^3$. Note that the massless case
is trivial, since in that case the function $\chi$ is simply related
to the $\zeta$ function, see Eq. (\ref{chizero}). In the massive case,
we can proceed in analogy to the massive $\zeta$ function, namely
employing Eq. (\ref{general}): considering $D=4$ and using Eq.
(\ref{chizero})
\begin{eqnarray*}
\chi^{m}_{D=4}(s;x)&=&\frac{1}{2\pi i \Gamma(s)}
\int_{\sigma-i\infty}^{\sigma-i\infty}
\Gamma(z)\chi^{m=0}_{D=4}(z;x)m^{2z-2s}\Gamma(s-z) dz
\nonumber\\
&=&\frac{4\pi}{2\pi i \Gamma(s)}
\int_{\sigma-i\infty}^{\sigma-i\infty}
\Gamma(z)(z-s)\zeta^{m=0}_{D=6}(z;x)m^{2z-2s}\Gamma(s-z) dz.
\end{eqnarray*}
Then we go on as usual splitting $\zeta^{m=0}_{D=6}(z;x)$
as $\zeta=\zeta_<+\zeta_>$ and performing the integrals
shifting the integration contour to the left and picking up
the residues of the poles. The final result is an expansion
in powers of $mr$:
\begin{eqnarray}
\chi^{m}_{D=4}(s;x)&=&\frac{r^{2s-6}}{4\pi\alpha\Gamma(s)}
\left\{\sum_{n=0}^\infty\frac{(-1)^n}{n!}(s+n-2)(mr)^{2n}
I_\alpha(s+n-2)
\right.\nonumber\\
&&+\sum_{n=0}^\infty\frac{(-1)^n}{n!}(n+1)(mr)^{6+2n-2s}
\frac{\Gamma(n+1/2)\Gamma(s-n-3)}{2\sqrt{\pi}\Gamma(n+1)}
\nonumber\\
&&+(mr)^{6-2s}
\sum_{n=1}^\infty\sum_{k=0}^\infty
\frac{(-1)^k}{k!}(\nu_n+k+1)(mr)^{2\nu_n+2k}\times\nonumber\\
&&\left.\times\frac{\Gamma(\nu_n+k+1/2)
\Gamma(s-\nu_k-k-3)}{2\sqrt{\pi}
\Gamma(2\nu_n+k+1)}\right\}.
\nonumber\\
&&\label{chifinal}
\end{eqnarray}

\cleardoublepage
\newpage

\chapter{Black hole entropy}

\label{BHENTROPY}
\section*{Introduction}

This Chapter is concerned with the computation of the quantum
corrections to the Bekenstein-Hawking entropy  by means of the conical
singularity method. In the first section we remind some basic
facts about the  Bekenstein-Hawking entropy and how it can be computed
by means of the conical method. Then we discuss the
quantum corrections to the tree-level entropy due to the quantum
fluctuations of the vacuum outside the horizon, which form a thermal
atmosphere. Here we show that these corrections diverge at the
horizon as a consequence of the divergence of the Tolman local
temperature at the horizon, and a discussion about the
non-ultraviolet nature of the divergence is given. In section
\ref{conicmethod}
we show that within the conical singularity method there are
two inequivalent ways to compute the quantum corrections: in the {\em
local} approach the ultraviolet divergences are regularized in the
local quantities, while in the {\em integrated} approach the
ultraviolet regularization is performed in the quantities after the
integration over the manifold.  The consequences of the two approaches
are quite different, and we argue that from the physical point of view
the local approach is more reasonable. In section \ref{renohordiv} we
discuss the possibility of the renormalization of the horizon
divergences, showing that it is not possible to perform such a
renormalization if the effective action is computed in the local
approach, in contrast to what happens in the integrated one.
Finally, we shortly discuss the case of a non-minimally coupled field.

\section{Tree-level black hole entropy}
\label{classicbhe}
\markboth{Black hole entropy}{Tree-level black hole entropy}

In this section we want to discuss the classical Bekenstein-Hawking
entropy of a black hole or of the Rindler space using the conical
singularity method. By classical we mean that we do not consider the
quantum corrections due to matter fields propagating outside the
horizon or the geometric entropy; these corrections will be discussed
in later sections. The Hilbert-Einstein action reads
\bea
W_G=-\frac{1}{16\pi G}\int d^4x\sqrt{-g}\left[R+2\Lambda\right],
\label{gravaction}
\ena
where $G$ is the Newton constant, $\Lambda$ is the cosmological
constant. We prefer to denote the action as $W$ instead of $S$ to
avoid any possible confusion with the entropy. In order to ensure a
consistent variational principle, it is also necessary to add to the
action (\ref{gravaction}) a surface term on the boundary of the volume
in which the action is evaluated \cite{GH77,hawking78}. This happens
because the above action contains second derivatives of the metric
which invalidate the variational principle in the usual form. It is
however possible to eliminate the second derivatives integrating by
parts and canceling the resulting surface term by adding an
appropriate surface term to the action. This procedure yields an
action suitable for the path integral. It is easy to show that the
correct surface term is
\bea
\frac{1}{8\pi G}\int d^3x \sqrt{h} \,[K],
\label{surfaction}
\ena
where, in the case of asymptotically flat metrics, $[K]$
 is the difference of the extrinsic curvature for the metric $g$ and
that of the flat-space metric, and $h$ is the induced metric
on the boundary.

According to Gibbons-Hawking path integral approach
\cite{GH77,hawking78}, it is possible to employ the periodic imaginary
time formalism of the path integral to discuss the gravitational
thermodynamics which arises from the gravitational action. Of course,
since we work at tree level the effective action will coincide with
the above classical action. As it is well know, nothing particular
happens when the spacetime has a globally defined time-like Killing
vector field which is not null anywhere: the entropy turns out to be
zero \cite{GH77}. Instead, when the space time has a Killing horizon,
as it is the case for black holes or the Rindler wedge, non-trivial
thermodynamical effects appear and the entropy is non-zero. This
entropy is usually one quarter of the event horizon area. Let us
illustrate this for the Schwarzschild case.

Consider the action (\ref{gravaction}) with $\Lambda=0$
and the surface term (\ref{surfaction}). The Euclidean Schwarzschild
metric reads
\bea
ds^2=\left(1-\frac{2MG}{\rho}\right) d\tau^2+
\left(1-\frac{2MG}{\rho}\right)^{-1}d\rho^2+\rho^2d\Omega^2.
\label{schwarz}
\ena
Introducing the coordinate $r=[8MG(\rho-2GM)]^{1/2}$ one finds
that the metric close to the horizon $\rho=2GM$ takes the form
\beas
ds^2=\frac{r^2}{16M^2G^2} d\tau^2+dr^2+(2GM)^2d\Omega^2.
\label{schwarz2}
\enas
In this form of the metric it is clear that there is a conical
singularity at the horizon $r=0$ unless the period of the Euclidean
time $\tau$ is $8\pi MG$. The absence of singularities in the
Euclidean manifold then fixes the temperature of the black hole at
$T_H=\b_H^{-1}=(8\pi GM)^{-1}$, the Hawking temperature. The Hawking
temperature can also be written as $T_H=\kappa/2\pi$, where $\kappa$
is the surface gravity on the horizon; we remind that the zeroth law
of black hole thermodynamics states that all parts of the event
horizon of a black hole at equilibrium have the same surface gravity.
Since the scalar curvature $R=0$ vanishes everywhere the value of the
action is entirely determined by the surface term: for the tree-level
partition function we have \cite{GH77}
\beas
\ln Z=-\frac{1}{8\pi G\hbar}\int [K]=-\frac{\b_H^2}{16\pi G\hbar},
\enas
where the bounding surface has been chosen as the surface $r=r_0$.
Now, from the standard thermodynamic formula
$S=-\b^2\pa_\b(\b^{-1}\ln Z)$, we obtain
\bea
S&=&\frac{\b_H^2}{16\pi G\hbar}\nonumber\\
&=&\frac{4\pi GM^2}{\hbar}=\frac{\cal A_H}{4G\hbar},
\label{BHentropy}
\ena
where ${\cal A_H}=4\pi(2GM)^2$ is the area of the event horizon. By
writing also the speed of light, we see that $S={\cal
A_H}/4l^2_{\mbox{\scriptsize P}}$, where $l_{\mbox{\scriptsize
P}}=(\hbar G/c^3)^{1/2}$ is the Planck length. It is worth reminding
that such relation was first derived from the thermodynamical analogy
in black hole physics \cite{beken72}.

There are three important facts about the above result to be
remarked. The first one is the problem of the statistical-mechanics
origin of the above entropy: there is not yet an explanation of the
black hole entropy in terms of number of microscopical states
corresponding to a given configuration of the black hole, and
actually it is not even clear whether such an explanation
will ever exists (for a review of the black hole entropy problem
see \cite{beken94}). At this regard, it must be said that there are very
recent progress form strings and M-theory, as we wrote in the
Introduction. The second one is related to the first and
is the presence of the Planck constant in the black hole entropy: this
is really a puzzle, since the tree level entropy is a classical
quantity and should not depend on quantum constants. The third fact to
be remarked is that the above derivation is ``{\em on-shell }'', that
is to say that in deriving the free energy $F=-\b^{-1}\ln Z$ with
respect to the temperature to get the entropy we have supposed that
the Hawking relationship between $\b$ and $M$ holds. In other words,
we have compared the free energy of two black holes with slightly
different  mass (and temperature). This also means that we have always
considered smooth manifolds. By contrast, in the so called ``{\em
off-shell }'' approach one allows $M$ and $\b$ to vary independently
and therefore a conical singularity appears in the manifold. In what
follows we show how the off-shell approach works. For a discussion
of the two approaches see \cite{ffz96}.

Later on, we will be interested in discussing one-loop quantum effects
and so now we do not consider only the plain Hilbert-Einstein action,
but we also add to it the quadratic terms in the curvature which
naturally arise in the one-loop renormalization of quantum fields
\cite{birrel}. The gravitational action then reads
\bea
W_G=\int d^4x\sqrt{-g}\left[-\frac{R+2\Lambda}{16\pi G}
+c_1R^2+c_2 R^{\mu\nu}R_{\mu\nu}+c_3R^{\mu\nu\a\b}R_{\mu\nu\a\b}
\right],
\label{gravaction2}
\ena
where $G$ is now the bare Newton constant, $\Lambda$ is the bare
cosmological constant, and $c_i$ are new bare gravitational
coupling constants. Since we work in four dimensions, the generalized
Gauss-Bonnet theorem implies that only two of these constants
are independent \cite{birrel}. Within the conical singularity method
it is not necessary to put the system in a finite volume and
to consider the surface term, since the entropy arises in a different
way. As before, we can compute the free energy $F(\b_H)=-\b_H^{-1}\ln Z$,
$\ln Z=-W_G$ on the smooth manifold $\M_{\b_H}$, but now when we
take the derivatives with respect to the temperature we vary $\b$
independently from $M$: this procedure can be considered as a canonical
approach, in the statistical-mechanics meaning, in which the temperature
is varied while the others thermodynamical quantities are kept fixed. We
remind that $M=\langle E\rangle$, where $E$ is the energy.
Therefore, for computing the entropy as
\beas
S(\b_H)=\b^2\pa_\b F(\b)|_{\b=\b_H},
\enas
it is necessary to compute the free energy for states, namely manifolds,
with temperature slightly different from the Hawking one: this
introduces a conical singularity in the manifold, just at the horizon
surface ${\cal H}$. Near the set of singular points the manifold
looks like $C_\a\times{\cal H}$, where, as usual, $C_\a$ is a two
dimensional cone and the cone angle is $\a=2\pi\b/\b_H$.

In order to compute the free energy on the conical manifold
we can employ the method developed of the smoothed singularity
\cite{FUSOL95} and summarized in section (\ref{riemangeom}).
There we have seen that the term linear in the curvature
is well-defined and causes no problem. Instead, the terms
quadratic in the curvature are mathematically ill-defined,
since they contain terms which are divergent and regularization
dependent. Nevertheless, we have already seen that these
ill-defined parts vanish as $(\b-\b_H)^2$ for small deficit
angles, and so they give contributions proportional to
$\b^{-1}(\b-\b_H)^2$ in the free energy. As a consequence,
they do not contribute to the entropy computed at $\b_H$.

Then, from the gravitational action (\ref{gravaction2}) and the
formulae (\ref{QRiem}) one obtains the following expression for the
tree-level entropy \cite{FUSOL96}
\bea
S(G,c_i)_{\b_H}=\frac{{\cal A_H}}{4G}-4\pi\int_{\cal H}
\left(2c_1R+c_2R_{\mu\nu}n^\mu_i n^\nu_i+
2c_3R_{\mu\nu\a\b}n^\mu_i n^\a_i n^\nu_j n^\b_j\right),
\nonumber\\
&&\label{BHentropy2}
\ena
where ${\cal A_H}$ is the area of the horizon ${\cal H}$, and $R$,
$R_{\mu\nu}$, and $R_{\mu\nu\a\b}$ are those of the smooth
manifold $\M_{\b_H}$.

The remarkable fact about the above result is that it differs
from the Be\-ken\-stein-Haw\-king entropy (\ref{BHentropy}) by
the contributions due to the quadratic terms in the gravitational
action. The same result can be derived using the Noether charge
method suggested by Wald \cite{wald93,jackob94}. It must be
noted that in the Rindler approximation to a large-mass
Schwarzschild black hole such terms are not present.

\section{Quantum corrections to the black hole entropy}
\label{quantumbhe}
\markboth{Black hole entropy}{Quantum corrections}

In this section we want to discuss the quantum corrections to the
entropy of a black hole or of the Rindler wedge. It was first
suggested in 1985 by 't Hooft \cite{thooft} that one should consider
the entropy of the quantum fluctuations of the matter fields living
in thermal equilibrium in
the vicinity of the horizon of a black hole as a quantum correction to
the tree level entropy. In the same paper 't Hooft introduced the
``brick wall'' model, in which all the wave functions are required to
vanish at some fixed distance $\ep$ from the horizon. Then,
employing a WKB approximation 't Hooft found that thermodynamical
quantities such as the free energy $F$ and the entropy $S$
diverge as $\ep\ra 0$, because of a ``rather alarming divergence
at the horizon of the number of energy levels a particle can
occupy''. This physically unacceptable divergence is supposed
to be cured in a theory in which the interaction of the fields
with the horizon, namely with ingoing and outgoing particles,
is taken into account. Moreover, 't Hooft noticed that if
the value of $\ep$ is chosen in such a way that the entropy
of the fields equals the Bekenstein-Hawking entropy, then
the energy contribution of the radiation is a non-negligible
fraction of the black hole mass: this fact means that
back reaction effects on the black hole geometry are not
negligible if one wants to develop a consistent theory.
The interpretation of the above choice of $\ep$ is
that in this way the black hole position is entirely
due to the quantum corrections. However, this interpretation
does not explain why it is possible to compute the entropy
at tree level, as we have seen above, and why the entropy
does not depend on the number of quantum fields considered.
Therefore, we will not follow this interpretation.

We will see that the ``horizon divergence'' is not avoidable: in some
sense, there is an analogy with the ultraviolet catastrophe in the
black-body radiation \cite{BARB94}, and this makes the horizon
divergence a very interesting subject.

Since 't Hooft seminal paper, other different models have been
proposed for computing these corrections to the entropy
and almost as many models for curing the
divergence.  We will not try here to review all these models, but
rather we would like to gather some understanding of the nature of the
horizon divergence and its relations with the usual divergences of the
quantum theory. Then, we will also discuss one of the simplest
way for curing the divergence, namely the proposal by
Susskind and Uglum of reabsorbing it in the renormalization of the Newton
constant. Among the alternative methods that we will not discuss
there are at least two worth to be mentioned.
One is that of the ``fuzzy horizon'' (see, e.g.,
\cite{fronov93,barfrozel95,sorkin97}), in which the quantum
fluctuation of the black hole radius, with a spreading of the order of
the Planck length, implements in a natural way the brick-wall cutoff.
Another idea followed by Frolov \cite{frolov95} (see also \cite{BM})
is  that the Bekenstein-Hawking entropy $S^{\mbox{\scriptsize
BH}}$ does not coincide with the statistical-mechanical
entropy $S^{\mbox{\scriptsize SM}}$, defined counting the number of
degrees of freedom, but rather with the thermodynamical entropy
$S^{\mbox{\scriptsize TD}}$, defined as the total response
of the free energy to the change of temperature,
$dF=-S^{\mbox{\scriptsize TD}}dT$. For a black hole
$S^{\mbox{\scriptsize SM}}\neq S^{\mbox{\scriptsize TD}}$,
because of the explicit dependence of the Hamiltonian of the
system on the temperature through the relation $\b=8\pi M$.
It is then shown that the one-loop contribution to
$S^{\mbox{\scriptsize TD}}$ is finite and independent of the
number of fields.

A very interesting approach for computing the black hole entropy that,
to avoid overloading,  will not be discussed in this thesis is that of
the ``entanglement entropy''
\cite{bomb86,srednicki93,DO94,kabstrass94,CW94,kabat}. In this
approach the entropy has a pure quantum-mechanical origin, since it is
due to the ignorance of an observer external to the horizon  about the
field correlations existing between points on different sides of the
horizon, so that the state of the field is described by a density
matrix rather than by a pure state. About this interesting approach we
limit to say that it gives the same results as those of the conical
singularity method we will employ, since although conceptually
different, the actual calculations are the same as those of the
conical method. In particular, even for the entanglement entropy
calculations we could follow a local on an integrated approach
with essentially the same kind of problems and discussions (see below).

The physical origin of the horizon divergences can be traced back in
the divergence of the local temperature at the horizon (see, e.g.,
\cite{BARB94}). Let us consider a system in thermal equilibrium at
asymptotic temperature $T$ in a black hole background of the form
\bea
ds^2=-\l(r)dt^2+\frac{1}{\mu(r)}^2 dr^2 +r^2d\Omega_{D-2}.
\label{BHbackground}
\ena
The functions $\l(r)$, $\mu(r)$, are such that the metric is asymptotically
the Minkowski one for large $r$, and there is a non-degenerate horizon
at $r=r_0$. According to the equivalence principle, in a static spacetime
a system in thermal equilibrium has a Tolman local temperature given by
\bea
T(x)=\frac{T}{\sqrt{-g_{00}(x)}},
\label{tolman}
\ena
so that in presence of a canonical horizon the local temperature
diverges on the horizon if $T$ is finite. It is worth noticing that it
is $T$ and not $T(x)$ that appears as $1/\b$ in the
statistical-mechanical relations. Therefore, the local density of any
extensive dimensionless quantity such as the entropy or $\b F$ is
proportional to the local temperature at the power $(D-1)$, so that
\beas
\b F,\, S\propto\int d^{D-1}x\sqrt{h}\left[T(x)\right]^{D-1},
\enas
where $h=\det g_{ij}$. Thus we see that $\b F$ or the entropy
diverge as $\rho^{-(D-2)}$, or logarithmically if $D=2$, where $\rho$
is the proper distance from the horizon,
$\rho(r)=\int_{r_0}^r\sqrt{g_{rr}} dr$.
Indeed, introducing a cutoff in the integration at a proper distance
$\epsilon$ and expanding near $r_0$, we get that
\bea
\b F,\, S\propto\frac{{\cal A_H}}{\epsilon^{D-2}}
\left(\frac{T}{T_H}\right)^{D-1},
\label{blueshift}
\ena
where $T_H$ is the Hawking temperature and ${\cal A_H}$ is the area
of the horizon. We see that the divergent entropy is proportional to the
area of the horizon as the Bekenstein-Hawking one.

Now that we have seen the origin of the horizon divergences as a
simple consequence of the divergence of the local temperature
at the horizon, we would also like to have a
quantum field theory's point of view. In particular,  we would like to
see whether it is possible to understand the horizon divergences in
terms of the usual ultraviolet or infrared divergences of the quantum
theory. Roughly speaking, in flat space an ultraviolet divergence
arises when we sum field modes with an arbitrary high energy, while an
infrared one arises summing over modes with arbitrary small energy.

Let us then consider the computation of the free energy using the
canonical approach for a neutral scalar field. For simplicity,
we consider the Lorentzian Rindler metric, but this will not affect
the final conclusion. The one-loop free energy is defined as
\beas
F(\b)=\frac{1}{\b}\sum_i\ln\left(1-e^{\b \omega_i}\right),
\enas
where the sum runs over single-particle states of the normal ordered
Hamiltonian ${\hat{H}}=\sum_i\omega_i a^\dagger a_i$. The field
equation is the usual Klein-Gordon equation
\beas
[\Box-m^2]\phi(x)=\left[ -\frac{1}{r^2}\pa_\tau^2+\frac{1}{r}\pa_r
+\pa_r^2+\pa_y^2+\pa_z^2\right]\phi(x)=0
\enas
The symmetries of the problem allows us to write solutions in the form
\beas
\phi(x)=U(r) e^{-i\omega\tau+i{\bf k}\cdot{\bf x}_\perp},
\enas
where ${\bf x}_\perp=(y,z)$. Defining $\l=\sqrt{{\bf k}^2+m^2}$,
the equation can be written as
\bea
\left[\frac{d^2}{dr^2}+\frac{1}{r}\frac{d}{dr}-\frac{(i\omega)^2}{r^2}\right]
U(r)=\l^2 U(r),
\label{eqbessel}
\ena
which is the modified Bessel equation of order $i\omega$. The most
general solution of this equation is
\beas
U(r)=A\,I_{i\omega}(\l r)+B\, K_{i\omega}(\l r).
\enas
The eigenvalues and the eigenfunctions are determined by the
boundary conditions. The boundlessness of $I_{i\omega}(x)$ as $x\ra\infty$
requires $A=0$. The normalized positive frequency modes are then
\bea
\phi_{{\bf k}\omega}(x)=\frac{\omega}{\pi}\sqrt{2\omega
\sinh (\pi\omega)}\,e^{-i\omega\tau+i{\bf k}\cdot{\bf x}_\perp}
\,K_{i\omega}(\l r).
\label{rindlermodes}
\ena
Moreover, to regulate the theory in the framework of 't Hooft
brick-wall model, we require that $\phi(x)$ vanishes at $r=\ep$, where
$\ep$ is close to zero. Then the Rindler eigenfrequencies are the
solutions of the equation
\beas
K_{i\omega}(\l \ep)=0.
\enas
Some kind of approximation is necessary to solve this equation.
For instance \cite{SU}, we can employ the WKB approximation:
define $z=\ln \l r$ and $E=-\frac{1}{2}\omega^2$, so that Eq. (\ref{eqbessel})
can be written as a time-independent Schr\"{o}dinger equation
\beas
\left[-\frac{1}{2}\frac{d^2}{dz^2}+V(z)\right]
U(z)=E\,U(z)
\enas
for a particle of unit mass moving in the potential
\beas
V(z)&=&\left\{\begin{array}{ll}\frac{1}{2}e^{2z}&\mbox{ if }z>\ln \l\ep\\
\infty &\mbox{ otherwise.}\end{array}\right.
\enas
The turning points of the classical motion occur when $V(z)=E$,
namely $z=\ln\,\l\ep$ and $z=\ln\,\omega$. The WKB quantization
condition is
\beas
2\pi n&=&2\int_{\ln\,\l\ep}^{\ln\,\omega}dz\sqrt{E-V(z)}.
\enas
This integral can be calculated and gives
\bea
n=\frac{\omega}{2\pi}\left\{\ln\frac{1+\sqrt{1-(\l\ep/\omega)^2}}
{1-\sqrt{1-(\l\ep/\omega)^2}}-2\sqrt{1-(\l\ep/\omega)^2}\right\},
\label{numberstates}
\ena
which is an implicit equation for the eigenfrequencies $\omega$.
The eigenfrequencies depend on both the quantum numbers $n$ and ${\bf k}$,
and are denoted by $\omega_n({\bf k})$.

Requiring that the square root be real gives the crucial condition
$\omega_n({\bf k})>\l\ep$. Therefore, we see that from the point of
view of the field energy levels the divergence problem seems of
{\underline{infrared}} nature: any field mode with finite frequency
at the horizon must have vanishing frequency at infinity, and so
the of number of states with arbitrary small frequency is infinite.
The cutoff $\ep$ has just the effect of cutting the sum over these
states. The same conclusion was given by Barbon \cite{BARB94}
in the black hole metric, where the divergence is also related
to the infinite volume of the optical space
(see Chapter \ref{OPTICCHAP}).

The infrared nature of the horizon divergences has been quite
overlooked by many authors, which claim that the divergences are of
ultraviolet nature.  Our opinion is that a common source of confusion
is the computation of the free energy in the conical singularity
framework by using the integrated heat kernel; we will see below that
in this approach the horizon divergences are not present, while there
are the usual ultraviolet divergences.

It is interesting to notice that the lowest positive eigenvalue goes
to zero as the cutoff is removed independently of the mass of the
field: this means that near the horizon all the fields behave as
massless \cite{BARB94}.

From the result (\ref{numberstates}) Susskind and Uglum computed
the WKB approximation of the free energy of the scalar field in the
Rindler wedge, obtaining
\bea
F(\b)=-\frac{\pi^2{\cal A_H}}{180\ep^2\b^4},
\label{helmholtz}
\ena
which is in agreement with the general result, Eq. (\ref{blueshift}),
obtained above from the equivalence principle.

\section{Quantum corrections to black hole entropy
 in the conical singularity method}
\label{conicmethod}
\markboth{Black hole entropy}{BH entropy in the conical singularity method}

In this section we want to treat in details the computation of the
the quantum corrections to a large mass Schwarzschild black
hole using the Rindler approximation and the conical singularity
method. We will not discuss here the renormalization or other
procedures to remove the horizon divergences, problem that will
be treated in a following section. So, our aim is simply
to compute the free energy and the entropy of the one-loop
quantum fluctuations of a scalar field in the Rindler wedge at
temperature $1/T$: by using the path integral with imaginary
time formalism, the problem reduces to the computation of
the one-loop effective action of the field on the conical
manifold $C_\a\times R^2$, where $\a=\b$ since for the Rindler
space the Unruh-Hawking temperature is $T_H=1/2\pi$.\footnote{
Hereafter we use $\b$ instead of $\a$ for the conical angle
to stress the connection with the temperature. Notice that $\b$
is a shortcut for $2\pi\b/\b_H$, and so it is adimensional.}
Hence, the heat kernel kernel and the $\z$ function on the cone
reviewed in the first Chapter are just the tool we need for this
computation.

We will start by considering the local approach for a minimally
coupled scalar field, and we will see that both the $\z$ function
and the heat kernel give the same result which is agreement
with the general result obtained from the Tolman temperature,
Eq. (\ref{blueshift}), and so we have horizon divergences. Then
we will also consider an integrated approach using the integrated
$\z$ function or heat kernel, showing that in this approach the
horizon divergences are not present. We will argue that the
correct approach is the local one.

\subsection{Local approach}
\label{localapproach}

In the finite temperature case, the one-loop effective action is more
properly called one-loop free energy $F_\b$, and it  can be obtained
as the integral over the manifold of the corresponding one-loop
effective Lagrangian density, $\L_\b^{\mbox{\scriptsize eff}}(x)$,
\bea
W_\b^{(1)}&=&\b F_\b=\int_\M d^4 x\sqrt{g}
\L_\b^{\mbox{\scriptsize eff}}(x).
\label{actiolagran}
\ena
From the general theory of the $\z$-function regularization, reviewed
in section \ref{shorthkz}, it is easy to see that $\L_\b^{\mbox{\scriptsize
eff}}(x)$ can be obtained from the local $\z$ function as
\beas
\L_\b^{\mbox{\scriptsize eff}}(x)&=&
-\frac{d}{ds}\z_\b(s;x|A)|_{s=0}+\frac{1}{2}\z_\b(s=0;x|A)\ln\mu^2.
\enas
In the case of a minimally coupled massless scalar field
we are considering, the local $\z$ function of the small fluctuations
operator on the flat cone $C_\b\times R^2$ ($D=4$) is given
by Eq. (\ref{coniczeta}):
\beas
\z_\b(s;x|A)&=&\frac{r^{2s-4}}{4\pi\b\Gamma(s)}I_\b(s-1).
\enas
From this $\z$ function we obtain the effective Lagrangian
density
\beas
\L_\b^{\mbox{\scriptsize eff}}(x)&=&-\frac{1}{8\pi\b r^4}I_\b(-1)\\
&=&-\frac{1}{1440\pi^2 r^4}\left[\left(\frac{2\pi}{\b}\right)^4
+10\left(\frac{2\pi}{\b}\right)^2-11\right].
\enas
The free energy of the fluctuations can then be obtained
from Eq. (\ref{actiolagran}): the integral over $r$ diverges
in $r=0$, and so we introduce a cutoff in the integration
at  a proper distance $r=\ep$ from the horizon:
\bea
F_\b&=&-\frac{{\cal A_H}}{2880\pi^2\ep^2}
\left[\left(\frac{2\pi}{\b}\right)^4
+10\left(\frac{2\pi}{\b}\right)^2-11\right],
\label{freeener}
\ena
where ${\cal A_H}$ is the area of the horizon. Notice
that $F_\b$ is positive for $\b<2\pi$. From the free energy
we compute also the entropy
\bea
S_\b&=&\b^2\pa_\b F_\b\nonumber\\
&=&\frac{{\cal A_H}}{180\b\ep^2}
\left[\left(\frac{2\pi}{\b}\right)^2+5\right].
\label{zentropy}
\ena
The latter expression, evaluated at $\b=2\pi$ gives the
one-loop quantum correction to the entropy of the
large mass Schwarzschild black hole in the Rindler
approximation.

It is important to remark that the cutoff $\ep$ we have
introduced is physically different from that of `t Hooft
\cite{thooft}, since in the brick-wall model the fields
are required to vanish at a proper distance $r=\ep$
from the horizon, while in computing the $\z$ function
the fields vanish in $r=0$. This could explain why
in $F_\b$, Eq. (\ref{freeener}), besides the term proportional
to $T^4$ there are also terms proportional to
$T^2$ which are not present in 't Hooft result or in the
WKB derivation of Susskind and Uglum \cite{SU}
(see Eq. (\ref{helmholtz})). The difference in the constant
term is of no importance, since it reflects just a
different  subtraction procedure and it does not
contribute to the entropy. We will come back to this problem
in Chapter \ref{OPTICCHAP}.

Since the two cutoffs correspond to different physical
situations, the results are different. Nevertheless,
the results contain the same essential features, namely
the presence of a horizon divergence as the cutoffs are
removed, the same correct Planckian
behavior $T^4$ of $F_\b$ and $S_\b$ at high temperatures,
and the proportionality to the horizon area. The term
proportional to $T^2$ gives a different low-temperature
behavior  and a different numerical coefficient in the entropy
at the Unruh-Hawking temperature. We then conclude that the
coefficient of the quantum corrections to the entropy depends on the
specific model considered.

Let us now briefly consider the massive case.
The $\z$ function on the cone for a massive scalar field
was computed in \cite{massive} (see Chapter \ref{massivechap}) in the
limit $mr<<1$ as a power series in $(mr)^2$. We stress that
the region $mr<<1$ is just that near the horizon, and so
the approximation is the right one for considering horizon
effects. Taking into account only the first massive
correction we have
\beas
\L_\b^{\mbox{\scriptsize eff}}(x)&=&-\frac{1}{8\pi\b r^4}
\left[I_\b(-1)-(mr)^2I_\b(0)+\O((mr)^4)\right],\\
F_\b&=&-\frac{{\cal A_H}}{16\pi\b}
\left[\frac{1}{\ep^2}-\frac{1}{R^2}\right]
I_\b(-1)+\frac{m^2{\cal A_H}}{8\pi\b}I_\b(0)\ln \frac{R}{\ep},\\
S_\b&=&\frac{\cal A_H}{180\b}
\left[\frac{1}{\ep^2}-\frac{1}{R^2}\right]
\left[\left(\frac{2\pi}{\b}\right)^2+5\right]-
\frac{m^2 {\cal A_H}}{12\b}\ln \frac{R}{\ep},
\enas
where we have considered $F_\b$ and $S_\b$ in the
region $\ep<r<R$, with $R$ such that $mR<<1$. We see that
there is an additional logarithmic divergence at the horizon.
Those written above are the only divergent terms in the
power expansion.

Let us now consider the computation of the same quantities
by means of the local heat kernel: we will show that the
heat kernel gives the same result as the $\z$ function.
This makes clear that the difference between the local
and integrated approaches is not related to the particular
regularization method employed, but just to the order in which
the regularization of the ultraviolet divergences and the
integration over the manifold are performed. This is not
so obvious considering the massless case only, since
the integrated approach cannot be applied to this case,
o r to be more exact it yields a vanishing effective action.

Formally, the effective one-loop Lagrangian density
can be computed as \cite{birrel} (see also subsection
\ref{regulrelations})
\beas
\L_\b^{\mbox{\scriptsize eff}}(x)&=&
-\frac{1}{2}\int_0^\infty \frac{dt}{t}K_t(x,x|A),
\enas
where $K_t(x,x|A)$ is the massive heat kernel on $C_\b\times R^2$,
which can be trivially obtained from any of the integral
representations of section \ref{conickernel}. The above expression is
just formal, since the the integral is divergent in $t=0$, as can be
easily seen from the asymptotic expansion: some kind of
regularization is needed. The two regularization we will
consider are the dimensional regularization and the Schwinger
proper-time regularization.

In the  dimensional regularization we add $2\varepsilon$ flat
dimensions with $\varepsilon$ chosen such that the above integral is
convergent \cite{birrel}. We consider $D=4$ and the integral
representation (\ref{HKreprE}) with $x=x'$: we then have
\bea
\L_\b^{\mbox{\scriptsize eff}}(x)
&=&-\frac{1}{2}\int_0^\infty
\frac{dt}{t} \frac{e^{-tm^2}}{(4\pi t)^{\frac{D+2\varepsilon}{2}}}
\left[1+\frac{i}{2\b}\int_\Gamma dw\,
e^{-\frac{r^2}{t}\sin^2\frac{w}{2}}
{\mbox{ctg}\,}\frac{\pi w}{\b}\right]\nonumber\\
&&\hspace{-3cm}=-\frac{m^{4+2\varepsilon}}{2(4\pi)^{2+\varepsilon}}
\left[\Gamma(-2-\varepsilon)+\frac{i}{\b}\int_\Gamma dw
\frac{{\mbox{ctg}\,}\frac{\pi w}{\b}}
{(m^2r^2\sin^2w/2)^{1+\varepsilon}}
K_{2+\varepsilon}\left(2mr|\sin^2\frac{w}{2}|\right)\right],\nonumber\\
&&\label{kerneff}
\ena
where $K_\nu(z)$ is the Mac Donald function. Only the first
term is divergent as $\varepsilon\ra 0$: in the
contour integral we can safely take the limit $\varepsilon\ra0$.
Considering the limit of small $mr$ we can use the asymptotic
expansion of $K_\nu(z)$ for small $z$ \cite{GR}
$$
K_2(z)=\frac{2}{z^2}-\frac{1}{2}+\O(z^2\ln z).
$$
Inserting this expansion in Eq. (\ref{kerneff}) we get
\bea
\L_\b^{\mbox{\scriptsize eff}}(x)&=&
-\frac{m^{4+2\varepsilon}}{2(4\pi)^{2+\varepsilon}}
\Gamma(-2-\varepsilon)\nonumber\\
&&-\frac{i}{64\pi^2\b r^4}\left[\int_\Gamma dw
\frac{{\mbox{ctg}\,}\frac{\pi w}{\b}}{\sin^4\frac{w}{2}}-
(mr)^2\int_\Gamma dw
\frac{{\mbox{ctg}\,}\frac{\pi w}{\b}}{\sin^2\frac{w}{2}}+\dots\right]
\nonumber\\
&=&-\frac{m^{4+2\varepsilon}}{2(4\pi)^{2+\varepsilon}}
\Gamma(-2-\varepsilon)
-\frac{1}{8\pi\b r^4}\left[I_\b(-1)-(mr)^2I_\b(0)+\dots\right],
\nonumber\\
&&
\label{heatdimens}
\ena
where we have used the integrals given in Eq. (\ref{C2C4}).
The first term is divergent: however, it does not depend
on $\b$ or on $r$, and so it is easily recognized as the
usual ultraviolet divergence for a massive scalar
field in the flat Euclidean spacetime. We should
expect such terms, because of the local flatness
of the cone: the ultraviolet divergences arises from
very small wavelength modes, which are insensible
to the global topology of the cone and sense only the
local flatness of the cone; just for this reason
the ultraviolet divergences are exactly those of the
flat space. The reason why the divergences appear
here and not in the $\z$ function calculation can be
traced back to the relation among different regularization procedures
discussed in subsection (\ref{regulrelations}).

The above divergences can then be renormalized away by means of a
standard renormalization procedure \cite{birrel}. As far as
the finite part is concerned, it is exactly the conical contribution
obtained by means of the $\z$ function.

Now we show that it is possible to obtain the same result
using the Schwinger proper-time regularization,
up to regularization dependent, flat-space ultraviolet
divergent terms. We consider also this regularization
because it is used by many authors working on this
subject and we want to compare their results
with ours.

In this regularization the effective Lagrangian density
reads
\beas
\L_\b^{\mbox{\scriptsize eff}}(x)&=&-\frac{1}{2}\int_\delta^\infty
\frac{dt}{t} K_t^\b(x,x|A)\nonumber\\
&=&-\frac{m^4}{32\pi^2}\left[\Gamma(-2, \delta m^2)
+\frac{i}{2\b}\int_\Gamma dw\,{\mbox{ctg}\,}\frac{\pi w}{\b}
\int_{\delta m^2}^\infty\frac{dy}{y^3}e^{-y-\frac{m^2 r^2}{y}
\sin^2\frac{w}{2}}\right]
\enas
where $\Gamma(z,x)$ is the incomplete Euler's gamma function
\cite{GR}. For $mr<<1$ and $r^2{ \stackrel{\scriptsize >}{\scriptsize \sim}}
\delta$ the integral over $y$ can be approximated as
\beas
\int_{\delta m^2}^\infty\frac{dy}{y^3}e^{-y-\frac{a^2}{y}}
&\simeq& \int_0^\infty\frac{dy}{y^3}e^{-y-\frac{a^2}{y}}
-\int_{0}^{\delta m^2}\frac{dy}{y^3}e^{-\frac{a^2}{y}},
\enas
and so, for $\delta\ra 0$ we have
\beas
\L_\b^{\mbox{\scriptsize eff}}(x)&\simeq&
-\frac{m^4}{32\pi^2}\Gamma(-2,\delta m^2)\nonumber\\
&&\hspace{-2cm}-\frac{1}{32\pi^2 r^4}\frac{i}{2\b}\int_\Gamma dw
\left[\frac{2m^2r^2}{\sin^2\frac{w}{2}}K_2(2mr|\sin\frac{w}{2}|)
+\frac{\Gamma(2,\frac{r^2}{\delta}\sin^2\frac{w}{2})}
{\sin^4\frac{w}{2}}\right] {\mbox{ctg}\,}\frac{\pi w}{\b}.
\enas
Since
\beas
\Gamma(2,\frac{a^2}{\delta})&=&
e^{-a^2/\delta}[1+\frac{a^2}{\delta}],
\enas
vanishes exponentially as $\delta\ra 0$ we can neglect it. Furthermore,
\beas
\Gamma(-2,\delta m^2)&=&\frac{1}{2}\Gamma(0,\delta m^2)+
\frac{e^{-\delta m^2}}{2 m^4 \delta^2}[1-\delta m^2]\\
&=&\frac{1}{2}\left[\frac{1}{m^4\delta^2}-\frac{2}{m^2\delta}
-\gamma-\ln \delta m^2+\O(\delta)\right].
\enas
Finally, expanding the Mac Donald function as done for the
dimensional regularization, we get
\bea
\L_\b^{\mbox{\scriptsize eff}}(x)&=&
-\frac{m^4}{64\pi^2}\left[\frac{1}{m^4\delta^2}-\frac{2}{m^2\delta}
-\gamma-\ln \delta m^2+\O(\delta)\right]\nonumber\\
&&-\frac{1}{8\pi\b r^4}
\left[I_\b(-1)-(mr)^2I_\b(0)+\dots\right].
\label{proptimresult}
\ena
For this expression the same considerations as for the
dimensional regularization hold.

The conclusion of this section is that at local level the
$\z$-function and the heat-kernel regularization procedures for the
conical singularity method yield the same result, up to regularization
dependent flat-space ultraviolet divergences, and these results are in
substantial agreement with the predictions based on the Tolman
temperature and the WKB approximation. Another important conclusion is
that the conical part, namely the temperature dependent part, is not
ultraviolet divergent: like in ordinary finite temperature theory the
ultraviolet properties are determined by the local geometry and are
not sensible to the choice of the quantum state. This conclusion
disagrees with those of other authors (see, e.g., \cite{fursaev95})
which employ the integrated approach. Indeed, in the following section
we will show that in the integrated approach the ultraviolet
divergences acquire a non-trivial dependence on the temperature.

\subsection{Integrated approach}
\label{globalapproach}

In this section we consider the integrated approach for
computing the quantum corrections to the black hole
entropy, namely we compute the one-loop effective action on the
cone starting from the integrated heat kernel or $\z$ function.
As a result, we will not obtain horizon divergences in the free
energy or in the entropy. This rather surprising result is then
explained showing that the regularization procedures for the
ultraviolet divergences and the horizon divergences do not
commute.

Let us start by considering the integrated heat kernel on
$C_\b\times R^2$ for a massive, minimally coupled scalar
field: from  Eq. (\ref{HKtrace}) we obtain
\bea
K_t^\b(A)=e^{-tm^2}\left[\frac{{\cal A_H}V(C_\b)}{16\pi^2 t^2}
+\frac{{\cal A_H}}{48\pi t}
\left(\frac{2\pi}{\b}-\frac{\b}{2\pi}\right)\right].
\label{HKtraceD4}
\ena
By means of a Mellin transform we can compute the corresponding
integrated $\z$ function:
\beas
\z_\b(s|A\mu^{-2})&=&\frac{1}{\Gamma(s)}\int_0^\infty
dt\,t^{s-1}K_t^\b(A)\\
&&\hspace{-1.7cm}=\frac{{\cal A_H}V(C_\b)m^4}{16\pi^2(s-1)(s-2)}
\left(\frac{m}{\mu}\right)^{-2s}
+\frac{{\cal A_H}m^2}{48\pi (s-1)}
\left(\frac{m}{\mu}\right)^{-2s}
\left(\frac{2\pi}{\b}-\frac{\b}{2\pi}\right).
\enas
From this expression we can then compute the one-loop
effective action, which reads
\bea
W_\b[\z]&=&-\frac{1}{2}\frac{d}{ds}\z_\b(s|A\mu^{-2})|_{s=0}\nonumber\\
&=&\frac{{\cal A_H}V(C_\b)m^4}{16\pi^2}
\left(\ln\frac{m}{\mu}-\frac{3}{4}\right)
-\frac{{\cal A_H}m^2}{48\pi}
\left(\ln\frac{m}{\mu}-\frac{1}{2}\right)
\left(\frac{2\pi}{\b}-\frac{\b}{2\pi}\right).\nonumber\\
&&\label{WZF}
\ena
We see that the effective action is finite: there are neither
ultraviolet divergences nor horizon divergences. While the absence of
ultraviolet divergences is expected from the general theory of the
$\z$-function regularization, the absence of the horizon divergences
is quite surprising when considering the results of the previous
sections. We also note that as $m\ra 0$ the effective action vanishes:
also this fact sounds strange, since we know that near the horizon
all fields behave as massless, and so we expect that presence
the mass changes the effective action only by a small
contribution.

Before commenting further this result, let us see what happens
employing another regularization: comparing different regularizations
of the effective action is an useful exercise, since it allows us to
understand more clearly which terms are regularization dependent
and which are not. Let us consider, for instance, the Schwinger
proper-time regularization: an easy calculation shows that
\bea
W_\b[PT]&=&-\frac{1}{2}\int_\delta^\infty\frac{dt}{t}
K_t^\b(A)\nonumber\\
&=&-\frac{{\cal A_H}V(C_\b)m^4}{32\pi^2}
\left[\frac{1}{m^4\delta^2}-\frac{2}{ m^2\delta}-\ln m^2\delta
+\frac{3}{2}-\gamma+\O(\delta)\right]\nonumber\\
&&-\frac{{\cal A_H}m^2}{96\pi}
\left(\frac{2\pi}{\b}-\frac{\b}{2\pi}\right)
\left[\frac{1}{m^2\delta}+\ln m^2\delta+\gamma-1+
\O(\delta)\right].\nonumber\\
&&\label{WPT}
\ena
Within this regularization we get the usual ultraviolet divergences in
the flat-space contribution integrated over the manifold (first row),
but the surprising fact is that we also get divergences in the conical
part. As a consequence, there is a divergent contribution to the
entropy of the system:
\bea
S_\b[PT]&=&\b^2\pa_\b \frac{W_\b[PT]}{\b}\nonumber\\
&=&\frac{{\cal A_H}m^2}{24\b}
\left[\frac{1}{m^2\delta}+\ln m^2\delta+\gamma-1+
\O(\delta)\right],
\label{ptentropy}
\ena
which survives also in the massless limit giving
$S_\b={\cal A_H}/{24\b\delta}$. This last expression
is remarkably similar to the expression (\ref{zentropy})
obtained by means of the local approach: both are proportional
to the area ${\cal A_H}$ of the horizon and diverge as the cutoffs
are removed (we remind that $\delta$ has the dimensions of a length
squared, such as $\ep^2$). There are, however, important differences:
in Eq. (\ref{zentropy}) we have terms that depend on the temperature
as $T^3$ and $T$, while in Eq. (\ref{ptentropy}) there is only the
term proportional to $T$ (with a coefficient that is $2/3$ of that
in (\ref{zentropy})).  Furthermore, we have introduced $\delta$ as
an ultraviolet cutoff, and now we find that it plays a r\^{o}le
similar to that of the horizon cutoff $\ep$.

In order to understand better what is going on here, we study
how is related the above result obtained by means of
the proper-time regularization with those of different regularizations.
We have already computed the $\z$ function result: in order to
compare them, we have to decouple $\delta$ and $\mu^2$ in
Eq. (\ref{WPT}) by introducing the arbitrary mass scale
$\mu$:
\beas
W_\b[PT]&=& W_\b[\z]+\frac{{\cal A_H}V(C_\b)}{16\pi^2}
\left[-\frac{1}{2\delta^2}+\frac{m^2}{\delta }+\frac{m^4}{2}
\ln \mu^2\delta+\frac{\gamma m^4}{2}\right]\nonumber\\
&&-\frac{{\cal A_H}}{48\pi}
\left(\frac{2\pi}{\b}-\frac{\b}{2\pi}\right)
\left[\frac{1}{2\delta}+\frac{m^2}{2}\ln \mu^2\delta
+\frac{\gamma m^2}{2}\right].
\enas
As expected, the difference among the effective actions
computed by means of the two regularizations are divergent
terms in the proper-time cutoff $\delta$. Notice that the constant
term can be reabsorbed in a redefinition of $\mu$ and $\delta$.

For sake of completeness, we give also the result of dimensional
regularization:
\beas
W_\b[dim]&=&\left[\frac{{\cal A_H}V(C_\b)m^4}
{32\pi^2}-\frac{{\cal A_H}m^2}{48\pi}
\left(\frac{2\pi}{\b}-\frac{\b}{2\pi}\right)\right]
\left[\frac{1}{\varepsilon}+\ln\frac{m}{\mu}+
\frac{\gamma-1}{2}\right]\nonumber\\
&=&W_\b[\z]+\left[\frac{{\cal A_H}V(C_\b)m^4}
{32\pi^2}-\frac{{\cal A_H}m^2}{48\pi}
\left(\frac{2\pi}{\b}-\frac{\b}{2\pi}\right)\right]
\left[\frac{1}{\varepsilon}+
\frac{\gamma}{2}\right]
\enas
As for $W_\b[\z]$, $W_\b[dim]$ vanishes in the limit
$m\ra 0$, while $W_\b[PT]$ has a finite massless limit.

From the above examples, it is clear that the effective
action computed from the integrated heat kernel
 (\ref{HKtraceD4}) is finite up to regularization
dependent terms, and actually vanishes in the massless limit.
Furthermore, the regularization dependent terms come
just in the form expected in ordinary field theory, namely
they are proportional to the Seeley-DeWitt coefficients
\beas
A_0&=&{\cal A_H}V(C_\b),\\
A_1&=&\frac{\pi{\cal A_H}}{3}
\left(\frac{2\pi}{\b}-\frac{\b}{2\pi}\right).
\enas
As a consequence, these divergent terms can be renormalized
away by redefining the bare cosmological constant $\Lambda_B$
and the bare Newton constant $G_B$ in the gravitational action:
for example, in the proper-time regularization the bare constants
in the bare gravitational Lagrangian density
\beas
\L_G(x)&=&\frac{1}{16\pi G_B}\left[R+2\Lambda_B\right]
\enas
are renormalized to
\beas
G&=&\frac{G_B}{1+16\pi G_B A},\\
\Lambda&=&\frac{\Lambda_B+8\pi G_B A}{1+16\pi G_B B},
\enas
where
\beas
A&=&-\frac{1}{16\pi^2}
\left[-\frac{1}{2\delta^2}+\frac{m^2}{\delta }+\frac{m^4}{2}
\ln \mu^2\delta+\frac{\gamma m^4}{2}\right],\\
B&=&\frac{1}{16\pi^2}
\left[\frac{1}{2\delta}+\frac{m^2}{2}
\ln \mu^2\delta+\frac{\gamma m^2}{2}\right].
\enas

Actually, as a trivial consequence of the fact that the complete
integrated heat kernel (\ref{HKtraceD4}) coincides with the sum
of the first two terms of the asymptotic, the whole effective
action $W_\b$ computed out of it can be renormalized in the
redefinition of the bare gravitational coupling constants. As a
consequence, even taking into account the quantum corrections,
the black hole entropy has the usual form
\beas
S_{BH}&=&\frac{{\cal A_H}}{4G},
\enas
provided that $G$ is the renormalized Newton constant
\cite{SU,delamy95,fursaev95,FUSOL96,LW}.

After having renormalized the divergences, the effective action is
finite as, for instance, that given in Eq. (\ref{WZF}). There are,
however, two questions which remain to be answered. The first one is
why we get ultraviolet divergences in the temperature dependent part
of the free energy, and so an ultraviolet divergent contribution to
the entropy, while it is well known (see, e.g. \cite{AAA95}) that in
ordinary quantum field theory, even in presence of boundaries, the
ultraviolet divergences are temperature-independent? The second
question is where have the horizon divergences gone and, in general,
why the integrated approach gives results different from those of the
local approach?

The answer to the first question has to be found in the peculiar
nature of the conical geometry. First of all, we remind that in the
previous section we have seen that in the local quantities the
ultraviolet divergences affect only the vacuum contribution part.
Therefore, something strange must happen when passing form local to
integrated quantities. This problem is probably related to the fact
that the integrated Seeley-DeWitt coefficient $A_1$ is not a locally
computable geometric invariant, because it contains terms proportional
to $\b^{-1}$ (see section  \ref{conickernel}): it is just this
singular behaviour that introduces the temperature dependence in the
ultraviolet divergences. From the physical point of view, the terms in
question arise from particle loops which encircles the conical
singularity at $r=0$. These kind of loops are sensitive to the deficit
angle of the cone and so they give the non-trivial dependence on the
temperature \cite{SU,LW}. Instead, the particle loops which do not
intersect the singularity sense only the locally flat space, and so
integrating their contributions to the effective action  over the
manifold yields a result proportional to $\b$, which do not contribute
to the entropy. From this point of view the non-trivial ultraviolet
divergences in $W_\b$ computed from the integrated heat kernel arise
from very small loops which encircle the singularity. It would be of
great interest to understand the r\^{o}le of these loops in the local
quantities.

Let us now consider the second question. Dimensional reasons
imply that, in the massless case and $D=4$, the only possible
dependence of the effective Lagrangian density on the proper
distance from the horizon is $r^{-4}$. It is therefore unavoidable
that the integration over the manifold to obtain the effective action
yields also divergences in $r=0$. How is then possible that
the effective $W_\b$ computed from the integrated heat kernel
(\ref{HKtraceD4}) does not show any horizon divergences?

The reason of the disappearance of the horizon divergences
can be explained by considering the discussion about the
integrated $\z$ function done in section \ref{globalzeta}.
There we have shown that, in order to obtain a integrated $\z$
function which agrees with the integrated heat kernel
(\ref{HKtraceD4})  we have to integrate the local
$\z$ function {\underline{before}} making any analytic continuation
to the physical region of the complex $s$ plane. In other
words, it is as if, in order to integrate $\L(r)=a/r^4$,
we inserted a regularizing function $r^{2s}$ with $2s>2$,
so that the integration converges in $r=0$,
$$
\int_\ep^R dr\, r^{2s-3}=\frac{R^{2s-2}}{2s-2}-
\frac{\ep^{2s-2}}{2s-2},
$$
and {\underline{before}} continuing $s$ to the region near $s=0$
we took the limit $\ep\ra 0$. It is clear that this procedure kills
the horizon divergences. It is worth remarking that this is just
the procedure which gives an integrated $\z$ function which agrees
with the integrated heat kernel (\ref{HKtraceD4}) and
actually it gives a vanishing effective action in the limit $R\ra\infty$,
just as $W_\b[\z]$ in the massless limit after renormalization.

Can the above procedure be considered correct? At first sight it is
quite similar to the dimensional or the $\z$ function regularizations,
and so, if we can regulate ultraviolet divergences in that way, why
shouldn't we do the same with horizon divergences? However, it must be
stressed the fact that, as we have seen in the discussion about the
different regularization procedures, the regularization procedures of
the ultraviolet divergences are physically well founded: they allow to
isolate the divergent terms and these have just the correct form to be
renormalized by redefining the bare coupling constants in the
gravitational action. It can happen that in some regularization the
divergent terms are not even present, but nevertheless we know that
the divergent terms canceled by the regularization could be, in other
procedures, renormalized away.

No such physical ground exists for the above regularization of the
horizon divergences, not being supported by a corresponding
renormalization procedure, and so it can be regarded as nothing more
than a trick for sweeping the dirt under the rug.

Not being physically founded, the regularization of the horizon
divergences yields results which are not correct, namely
results for $F_\b$ and $S_\b$ which do not contain the horizon
divergences which were predicted on the very general ground of the
behaviour of the local temperature  and confirmed by the WKB
approximation.

It must be stressed again that the apparent horizon divergences that
appear in $W_\b[PT]$ or $W_\b[dim]$ should be interpreted more
properly as usual ultraviolet divergences that can be renormalized
into the gravitational action, as we have seen above. Nevertheless,
the unusual dependence on the temperature of these divergences has
generated much confusion in literature, since many authors have
considered them as horizon divergences \cite{kabat,fursaev95,FUSOL96,ffz97}.
The confusion has been increased by the fact that in two dimensions
the effective action obtained from the integrated heat kernel plus
proper-time regularization coincides with that obtained in from the
local approach. In a paper by Solodukhin \cite{SOL95} it is
particularly clear how, in two dimensions, the ultraviolet divergent
part and the ultraviolet finite but horizon divergent part arise in
the effective action (notice  that the coefficient of the  ultraviolet
divergence is correct only for small deficit angles). It would be
interesting to understand the reason behind this singular coincidence.

Anyway, in other dimensions the temperature dependence and the
coefficients are different: in particular, the ultraviolet divergences
in the free energy have the same dependence on the temperature in any
dimension \cite{EMP95},  proportional to $\b^{-1}[(2\pi/\b)-(\b/2\pi)]$,
while the horizon divergences in $D$ dimensions have a leading term
proportional to $\b^{-D}$. It is important to notice that, as it has
been shown by Elizalde and Romeo \cite{elirom96}, it is possible
to obtain the correct dependence on the temperature of the
free energy even starting from the integrated heat kernel
(\ref{HKtraceD4}) by means of a careful application of the
relation between heat kernel and $\z$ function. The drawback
of this approach is a more difficult comparison of the results
because the cutoffs needed in the approach of \cite{elirom96} are
different from those we are using here. Anyway, since the results
of \cite{elirom96} are in agreement with those of the WKB
approximation of \cite{SU}, they also support the local
approach discussed here.

Another point of view of the above problem is the following.
Let us consider the local $\z$ function: the analytic continuation
in $s$ is the procedure that regularizes the ultraviolet divergences
of the theory, and so if we integrate over $r$ before continuing in $s$
we regularize the horizon divergences before the ultraviolet ones.
Instead, if we regularize the ultraviolet divergences in the local
$\z$ function by analytic continuation, then there is no way to
avoid the horizon divergences. Therefore, the two regularization
procedures do not commute (see also \cite{ffz97}).

There are strong arguments in favor of the second procedure.
Indeed, the analytic continuation in the local $\z$ function allows
us to define a regularized theory at local level which agrees
with any other local method \cite{DO78,DC79,FRSE87,DO94}.
The local approach is also in agreement with the results
of the optical approach that will be discussed in Chapter
\ref{OPTICCHAP}.
Therefore it allows us to compute fundamental quantities
such as the propagator, the energy-momentum tensor,
$\langle\phi^2\rangle$ and so on. It would be hard to accept
a theory in which it is not possible to compute such quantities.
Furthermore, the local approach has a smooth massless limit, while it
is not clear how to treat the massless case in the integrated
approach. Therefore, the author's opinion is that the analytic
continuation in the local $\z$ function must be performed before the
integration. Only after having defined and regularized the local
quantities we can integrate them to obtain the integrated ones.

We have seen that in the local $\z$ function approach the origin
of the non-commutativity of  ultraviolet and horizon
regularization is mathematically clear.  It remains to be
explained from the point of view of the integrated heat
kernel and its  physical  origin. In particular, it is not obvious
why one cannot start from the integrated heat kernel
to compute the effective action. Nevertheless,
since we have seen that both at local and integrated level
all the approaches agree within a certain regularization
procedure, we think that a similar explanation should
exist.

The conclusion of this long discussion is that the integrated approach
does not yield physically reasonable results, and the quantities on
the cone should be computed starting from the local one. We have given
the explanation of this phenomena within the $\z$ function
regularization, but it would be of great interest to understand it
also from different points of view.

\section{Renormalization of horizon divergences}
\markboth{Black hole entropy}{Renormalization of horizon divergences}
\label{renohordiv}

We have seen in the previous sections that the one-loop
quantum corrections to the black hole entropy are
divergent as the horizon cutoff  is removed.
Susskind and Uglum \cite{SU} have suggested that
these divergences have the correct form to be
absorbed the Bekenstein-Hawking formula as a
renormalization of the Newton constant so that
the entropy remains
\bea
S={\cal A_H}/4G
\label{ancoraBH}
\ena
even after the quantum corrections, provided that $G$ is renormalized
Newton constant. Quantum effects could also introduce corrections
coming from quadratic terms in the curvature, as we have seen in
section \ref{classicbhe}.

The purpose of this section is to investigate the hypothesis
of Susskind and Uglum. Other authors have considered
this issue \cite{BAEMP95,delamy95,fursaev95,FUSOL96,LW},
but, according to the results of the previous section,
it seems to us that in most cases
\cite{delamy95,fursaev95,FUSOL96,LW} only the
ultraviolet divergences appearing in the integrated
approach have been renormalized, essentially in the same
way as we have done above. In their approach the horizon
divergences are not present, and so the possibility of
their renormalization remains to be discussed.

Discussing the integrated approach in the previous section we have seen
that the whole one-loop effective action can be renormalized away in a
redefinition of the bare gravitational coupling constants, so that
the Bekenstein-Hawking formula (\ref{ancoraBH}) remains unchanged.
Although we were unsatisfied with the integrated approach, and we regard
the local approach as giving more reasonable results,
this behaviour is very interesting. In particular it could
explain why the entropy is always given by Eq. (\ref{ancoraBH})
regardless the number and the nature of the quantum fields.
Actually, one could make a  further step and regard the whole
black hole entropy as a low-energy consequence of the
quantum corrections (see, e.g., \cite{ffz97}).
Therefore, we would like to see if the idea by Susskind and Uglum
works also within the local approach.

Let us start from the two dimensional case. In two dimensions
the renormalized effective action for a massless scalar field
reads
\beas
W_\b^{D=2}=-\frac{1}{12}\left(\frac{2\pi}{\b}-\frac{\b}{2\pi}\right)
\ln\frac{R}{\ep}
\enas
where $R$ is a volume cutoff. We see that the logarithmic
horizon divergence is proportional to the Seeley-DeWitt
coefficient (see Eq. (\ref{HKtrace}))
$$
A_1=\frac{1}{6}\int_{C_\b}d^2x\sqrt{g}\,R=
\frac{\pi}{3}\left(\frac{2\pi}{\b}-\frac{\b}{2\pi}\right)
$$
and therefore we can renormalize this divergence
by redefining the Newton constant, in agreement with
Susskind and Uglum hypothesis. The renormalization
is unambiguous because of the geometrical meaning
of the coefficient of the divergence.

Let us now consider the less trivial four dimensional case:
from Eq. (\ref{freeener}) we have
\beas
W_\b&=&-\frac{{\cal A_H}}{16\pi \ep^2}I_\b(-1)\nonumber\\
&=&-\frac{{\cal A_H}}{480\pi^2 \ep^2}\frac{\pi}{3}
\left(\frac{2\pi}{\b}-\frac{\b}{2\pi}\right)
\left[\left(\frac{2\pi}{\b}\right)^2+11\right].
\enas
We see that we can na\"{\i}vely consider also this
horizon divergence as proportional to $A_1$,
but the factor in the square brackets could have
some non-trivial geometrical meaning and so invalidate the
reasoning. Indeed, if we consider the singular heat kernel
expansion on the cone and the coefficients given by
Eq. (\ref{singcoeff}), we see that the divergence seems
rather proportional to the local coefficient $a_2(x)$.
If this were the case, the divergence could be renormalized
by redefining the coupling constants of the terms quadratic
in the curvature in the gravitational action (\ref{gravaction2}).
Unfortunately, the corresponding integrated coefficient
$A_2$ vanishes with the usual choice for the test
function $f$.

Therefore, the conclusion of this short section
is that it seems not possible to renormalize the
horizon divergences in the way proposed by
Susskind and Uglum. This, at least, if for renormalization
we mean the standard procedure for which the divergent terms
in the effective action of the matter fields which are proportional
to geometrical quantities such as the Seeley-DeWitt coefficients
are canceled by redefining the coefficients of corresponding
geometrical terms in the gravitational action (see \cite{birrel}):
for the horizon divergences it was not possible to identify
such geometrical meaning and so no consistent renormalization
is possible. Our opinion is that the argument of Susskind and Uglum
fails because is based on an expansion in powers of $(2\pi-\b)$:
at first order, also quantities which are not proportional to
$a_1(x)$ seem to be proportional to it.
For example, suppose we had found a divergence proportional to
$a_2(x)$: in the standard procedure should renormalize the coupling
constants of the quadratic terms in (\ref{gravaction2}). Instead, in
the argument by Susskind and Uglum, by expanding in powers of
$(2\pi-\b)$ we would have found a term proportional to $a_1(x)$ which
renormalizes the bare Newton constant, plus higher order terms which do
not contribute to the entropy. This example makes clear the arbitrariness
of the renormalization suggested in \cite{SU}. Finally, it has to be noted
that, although the renormalization of the (horizon) divergences works
in the integrated approach, it also raises a serious problem,
since it is easy to see that in order to carry out the renormalization
one must introduce an infinite negative bare entropy, which has
no statistical-mechanical origin \cite{frofur98}.

\section{Non-minimal coupling}
\markboth{Black hole entropy}{Non-minimal coupling}
\label{nomincap}

In this section we want to briefly discuss the case of non-minimally
coupled massless scalar fields, which has been covered in some recent
papers \cite{SOL97,hotta,moretticonic,Zmoretti}. In the case of
non-minimally coupling, the equation for the modes reads
\beas
A_\xi\phi_{n\l}(x)&=&\l^2\phi_{n\l}(x),\\
A_\xi&=&-\Delta_\a+\xi R.
\enas
Since on the cone the scalar curvature behaves such as a
$\delta$-function at $r=0$, it is not even clear how to give
a mathematical meaning to the above equation, let alone
solving it.

There is, however, an alternative way to discuss
the case $\xi\neq 0$ \cite{hotta,moretticonic} and
works as follows. Let us write the partition
function as
\beas
Z_\b[\xi]&=&\int \D\phi\, e^{-S_\xi[\phi]}=
\int \D\phi\, e^{-S_{\xi=0}[\phi]-\frac{\xi}{2}\int d^4x\,\sqrt{g}
R\phi^2}\\
&=&Z_\b[\xi=0]\,e^{-\frac{\xi}{2}\int d^4x\,\sqrt{g}R
\langle\phi^2\rangle_{\xi=0}}
\enas
The value of the vacuum fluctuations
is given by Eq. (\ref{vacuum}) in the limit $m\ra 0$:
\beas
\langle\phi^2\rangle_{\xi=0}&=&\frac{1}{48\pi^2 r^2}
\left[\left(\frac{2\pi}{\b}\right)^2-1\right].
\enas
According to the discussion in section \ref{riemangeom}, we
can represent the scalar curvature of the flat cone as
$R=2\left(\frac{2\pi}{\b}-1\right)\delta(r)$, where $\delta(r)$
is such that $\int_0^\infty \delta(r) \,r\,dr=1$. It follows that
the free energy can be written as
\beas
F_b[\xi]&=&F_\b[\xi=0]+\frac{\xi {\cal A_H}}{48\pi^2}
\left[\left(\frac{2\pi}{\b}\right)^2-1\right]
\left(\frac{2\pi}{\b}-1\right)
\int_0^\infty\frac{dr}{r}\delta(r).
\enas
The last integral is ill-defined, but, according to the
horizon regularization we employ in this Chapter we
could think it as $1/2\ep^2$. However, this is not
even necessary if we are interested in computing
the entropy on-shell,
$\b=2\pi$: the only thing that matters is the fact
that it does not depend on the temperature, so that
we immediately obtain
\beas
S(\xi)|_{\b=2\pi}&=&S(\xi=0)|_{\b=2\pi}.
\enas
Therefore, the quantum corrections to the black-hole entropy in the
Rindler approximation due to scalar fields do not depend on the value
of the coupling \cite{moretticonic}. We stress that this result
holds only at the Unruh-Hawking temperature.

From the physical point of view the above conclusion is very
reasonable. In fact, the Lorentzian Rindler space is flat everywhere
and so the physical quantities should not depend on $\xi$. In the
Euclidean formulation only the state with $\b=2\pi$ corresponds to a
state in the Lorentzian formulation, and so only for that value we can
expect an independence on $\xi$.

\cleardoublepage
\newpage

\chapter{Thermal partition function of photons and
gravitons in a Rindler wedge}

\label{PHOTONS}

\section*{Introduction}
\markboth{Photons and gravitons}{Introduction}

Most of the work on the quantum corrections to the black hole entropy
and discussed in the previous Chapter is carried on using the
scalar field. In literature, results for higher spins have
been obtained translating earlier results obtained for the closely
related cosmic string background \cite{DO94}. In a recent and interesting
paper \cite{kabat} Kabat investigated the corrections
to the black hole entropy coming form scalar, spinor and vector fields
by explicitly writing the field modes in the Euclidean Rindler space
and then what we have called the integrated approach. In
the vector field case  he has obtained an unexpected ``surface''
term, which corresponds to particle paths beginning and ending at the
horizon. This term gives a negative contribution to the entropy of the
system and, in fact, is large enough to make the total entropy
negative at the equilibrium temperature. Kabat argued that this term
corresponds to the low-energy limit of string  processes which couple
open strings with both ends attached to the horizon and closed strings
propagating outside the horizon diagrams and discussed by Susskind and
Uglum \cite{SU} as responsible for black hole entropy within string
theory.

 In this Chapter, based on the paper \cite{ielmo}, we apply the local method
of \cite{ZCV} (see section \ref{Coniczeta}) to the case of the Maxwell
field and the graviton field. As a result, in the case of the photon
field we confirm that  there is a `surface term' which would give a
negative contribution to the entropy, as obtained by Kabat in
\cite{kabat}. However, beside getting a different temperature
dependence, we show that it depends on the gauge-fixing parameter and
so we discuss how it is possible to discard it. In this way we also
avoid embarrassing negative entropies. In the case of the graviton we
get similar surface terms and show that one can get consistent
physical results by discarding them. We also discuss the appearance of
similar terms in more general manifolds. After discarding the surface
terms we get the reasonable result that the effective action and all
the thermodynamical quantities are just twice those of the minimally
coupled scalar field: this is in agreement with the results of the
point-splitting method \cite{FRSE87,ALO92}, the heat kernel method
\cite{LW,BS,FM}, and, apart from the surface terms, also with Kabat
\cite{kabat}.

We remind that the Rindler wedge is  a globally hyperbolic manifold
defined by the inequality $x>|t|$, in the usual set of rectangular
coordinates $(t,x,y,z)$ of Minkowski space-time. In this wedge we can
define a new set of static coordinates by setting $t=r\sinh\tau$ and
$x=r\cosh\tau$, with $0<r<\infty$ and $-\infty<\tau<\infty$. Then the
Minkowski metric takes the form of the Rindler metric:
\begin{eqnarray}ds^2=-r^2 d\tau^2+dr^2+dy^2+d^2z.
\label{Rindler}
\end{eqnarray}
One can see that lines of constants $r,y$, and $z$ are trajectories of
uniformly accelerated particles, with proper acceleration $a=r^{-1}$.

As we said above, the  importance of the Rindler metric is mainly due
to the fact that it can be seen as an approximation of the metric of a
large mass Schwarz\-schild black hole outside the event horizon. Indeed,
consider the Schwarzschild metric, which describes an uncharged,
nonrotating black hole of mass $M$
\begin{eqnarray}
ds^2&=&-\left(1-\frac{2GM}{R}\right ) dT^2+\left(1-\frac{2GM}{R}\right
)^{-1} dR^2+R^2 d\Omega_2, \nonumber\\
&&d\Omega_2=d\theta^2+\sin\theta\,
d\varphi^2, \nonumber
\end{eqnarray}
where $M$ is the mass of the black hole. In the region outside the
event horizon, namely, ${2GM<R<\infty}$, we can define new
coordinates $\tau$ and $r$ by
\begin{eqnarray*}
\tau&=&\frac{T}{4GM},\\
r&=&\sqrt{8GM(R-2GM)},
\end{eqnarray*}
and so the metric takes the form
\begin{eqnarray*}
ds^2&=&-r^2\left(1+\frac{r^2}{16G^2 M^2}\right )^{-1}d\tau^2+
\left(1+\frac{r^2}{16G^2 M^2}\right )dr^2\\
& &+4G^2 M^2\left(1+\frac{r^2}{16G^2 M^2}\right )^2d\Omega_2.
\end{eqnarray*}
If we take the large mass limit, the last term becomes the metric of a
spherical surface with very large radius that can be approximated by a
flat metric $dy^2+dz^2$. Then, in this limit, the metric becomes the
Rindler one, Eq. (\ref{Rindler}). Actually, even if we do not
consider the large mass limit, the approximation should become better
and better as we approach the event horizon, $r=0$.

The Rindler metric is also related with the study of the cosmic string
background: the metric around an infinitely long, static, straight and
with zero thickness cosmic string can be written as
\begin{eqnarray}
ds^2=-dt^2+dz^2+dr^2+r^2d\varphi,\hspace{1cm}0\leq\varphi\leq\alpha,
\nonumber
\end{eqnarray}
where the polar angle deficit $2\pi-\alpha$ is related to the mass per
unit length of string $\mu$ by $2\pi-\alpha=8\pi G\mu$. Since the
metric is ultrastatic, we can perform a Wick rotation, $t \rightarrow
it$, and the metric becomes equal to the Euclidean Rindler metric.
Therefore, we can identify the thermal partition function of a field
at temperature $\alpha^{-1}$ in the Rindler wedge with the
zero-temperature Euclidean-generating functional of the same field in
a cosmic string background.

The rest of this Chapter is organized as follows. In section \ref{sectwo} we
compute the one-loop effective action for the electromagnetic field on
the manifold $C_\beta\times\mbox{R}^2$ using the $\zeta$-function
regularization. We use this result to compute the quantum correction
to the black hole entropy in the framework of conical singularity
method. In section \ref{secthree} we formulate a general conjecture on the
appearance of Kabat-like surface terms in the case of integer spin and
general manifolds. In section \ref{secfour} the conjecture is checked in the case
of the graviton. Section \ref{secfive} is devoted to the discussion of the
results.

\section{Effective action for the photon field}
\markboth{Photons and gravitons}{Effective action for the photon field}
\label{sectwo}

In a curved space-time with Lorentz signature the action of the
electromagnetic field is  $S=\int{\cal L}(x) \sqrt{-g}d^4x$,
where the Lagrangian scalar density\footnote{We
adopt the convention that the indices $a,b,\dots=\tau,r,y,z$ are for
the whole manifold, the greek indices are for the pure cone,
$a,b,\dots=\tau,r$, and the indices $i,j,\dots=y,z$ are for the
transverse flat directions.} is \cite{birrel}
\begin{eqnarray}
{\cal L}_{\mbox{\scriptsize em}}(x)&=&-\frac{1}{4}
F_{ab}F^{ab},\nonumber\\
F_{ab}&=&\nabla_a A_b-\nabla_b A_a=
\partial_aA_b-\partial_a A_b.
\label{em1}
\end{eqnarray}
We need also the gauge-fixing term and the contribution of the
ghosts:
\begin{eqnarray}
{\cal L}_G&=&-\frac{1}{2\alpha}(\nabla^a A_a)^2,\nonumber\\
{\cal L}_{\mbox{\scriptsize ghost}}&=&\frac{1}{\sqrt{\alpha}}
g^{ab}\partial_a c\partial_b c^{\ast},
\label{em2}
\end{eqnarray}
where $c$ and $c^\ast$ are anticommuting scalar fields. The dependence
on the gauge-fixing parameter $\alpha$ is relevant only in presence of
a scale anomaly. SInce this is not the case here, we shall ignore it in the
rest of this paper.

We are interested  in the finite temperature theory and so we change
$\tau\rightarrow
i\tau$ and identify $\tau$ and $\tau+\beta$. The metric of the
Rindler space-time turns to Euclidean signature,
$ds^2=r^2d\tau^2+dr^2+dy^2+dz^2$, and the vector D'Alembertian
operator $\Box$ becomes the vector Laplace-Beltrami operator $\Delta$.
In the following this operator will be simply called Laplacian. The
one-loop effective action for this theory will then be given by the
following determinants:
\begin{eqnarray}
\ln Z_\beta&=&-\frac{1}{2}\ln\det \mu^{-2}\left (g^{ab}(-\Delta)-R^{ab}+
(1-\frac{1}{\alpha})\nabla^{a}\nabla^{b}\right )+\ln
Z_{\beta,\mbox{\scriptsize ghosts}},\nonumber\\
&&
\label{em10}
\end{eqnarray}
where $\mu^2$ is the renormalization scale and the effective action of
the ghosts is minus twice the effective action of a scalar massless
field, which is well known \cite{cheeger83}\cite{ZCV}. It is important to note
that the determinant has to be evaluated on the whole set of
eigenfunctions, not only on the physical ones \cite{DV}.

We work on the manifold $C_\beta\times \mbox{R}^2$,
where $C_\beta$ is
the cone with angular deficit equal to $2\pi-\beta$. This manifold is
flat everywhere but on the tip of the cone, where the curvature has a
$\delta$-function singularity. Nevertheless,  the modes we use vanish
on the tip, and so we can consider $R_{ab}=0$. Note also that, due to
the flatness, the covariant derivatives commute. Hence, we  are left
with the problem of computing the determinant of the operator
$[g^{ab}(-\Delta)+ (1-\frac{1}{\alpha})\nabla^{a}\nabla^{b}]$ acting
on vectors. In order to define this determinant we use the
$\zeta$-function regularization: first, suppose we have a complete set
of eigenfunctions of the operator, indicated as $A_a^{(i,n\lambda{\bf
k })}(x)$, with eigenvalue $\nu_i^2(n\lambda{\bf k})$. Here, ${\bf
k}=(k_y,k_z)$, $a=\tau,r,y,z$ and $i=1,\dots,4$ is the polarization
index. In this notation we have taken into account the triviality of
the transverse dimension and the fact that we have a discrete index
$n$ since the $\tau$ coordinate is compact and we impose periodic
boundary conditions. Then we can define the local, diagonal heat
kernel as
\begin{eqnarray}
K^{(i)}(t;x)=\sum_n\int d\mu(\lambda)\,d^2{\bf k}\;
e^{-t\nu_i^2} g^{ab} A_a^{(i)}(x)A_b^{(i)\ast}(x),
\label{em12}
\end{eqnarray}
where $d\mu(\lambda)$ is an appropriate integration measure. The
corresponding local spin-traced $\zeta$ function can be obtained
through a Mellin transform:
\begin{eqnarray}
\zeta(s;x)=\frac{1}{\Gamma(s)}\int_0^\infty dt\, t^{s-1}
\sum_i K^{(i)}(t;x).
\label{em16}
\end{eqnarray}
Alternatively, we can define the local $\zeta$ function as the inverse
power of the kernel of the above differential operator: the spectral
representation gives directly
\begin{eqnarray}
\zeta(s;x)=\sum_i\sum_n\int d\mu(\lambda)\,d^2{\bf k}\;
[\nu_i^2(n\lambda{\bf k})]^{-s} g^{ab} A_a^{(i)}(x)A_b^{(i)\ast}(x)
\label{emZETA}
\end{eqnarray}
In general, both the Mellin transform and the inverse power of the
operator require analytic continuation arguments to be defined at the
physical values of $s$.

We can also define a global $\zeta$ function by tracing over the space
indices:
\begin{eqnarray}
\zeta(s)= \int d^4x \sqrt{g} \, \zeta(s;x).
\label{em17}
\end{eqnarray}
This last step is delicate: in general, the operation of tracing over
the space indices requires the introduction of a smearing function,
since the manifold is noncompact and there can be nonintegrable
singularities in the local $\zeta$ function, and a particular choice of
the smearing function could sweep away important information. This is
one of the reasons why we prefer to work with a local formalism as
long as possible. Once we have computed and analytically continued the
$\zeta$ function, we can write the effective lagrangian density and the
effective action as
\begin{eqnarray}
{\cal L}_\beta(x)&=&-\frac{1}{2}
\zeta'(s=0;x)+\frac{1}{2}\zeta(s=0;x)\ln \mu^2,\nonumber\\
\ln Z_{\beta}&=&\int d^4 x \sqrt{g}\,{\cal L}_\beta(x).
\label{em18}
\end{eqnarray}
Of course, to the above expression we have to add the contribution of
the ghosts, which is minus two times the effective lagrangian density
of a scalar field.

A suitable set of normalized eigenfunctions of the operator
$[g^{ab}(-\Delta)+(1-\frac{1}{\alpha})\nabla^a\nabla^b]$ (equivalent to
Kabat's set \cite{kabat} if $\alpha=1$) is the following: setting
$k=|{\bf k}|$
\begin{eqnarray}
A_a^{(I,n\lambda{\bf k})}&=&\frac{1}{k}\epsilon_{ij}\partial^j\phi=
\frac{1}{k}(0,0, ik_z\phi,-ik_y\phi)\nonumber,\\
A_a^{(II,n\lambda{\bf
k})}&=&\frac{\sqrt{g}}{\lambda}\epsilon_{\mu\nu}\nabla^\nu\phi=
\frac{1}{\lambda}(r\partial_r\phi,-\frac{1}{r}\partial_\tau\phi,0,0),
\nonumber\\
A_a^{(III,n\lambda{\bf k})}&=&\frac{1}{\sqrt{\lambda^2+{\bf k}^2}}
(\frac{k}{\lambda}\nabla_\mu-\frac{\lambda}{k}\partial_i)\phi
\nonumber\\
&=&\frac{1}{\sqrt{\lambda^2+{\bf k}^2}}
(\frac{k}{\lambda}\partial_\tau\phi,\frac{k}{\lambda}\partial_r\phi,
-\frac{\lambda}{k}\partial_y\phi,-\frac{\lambda}{k}\partial_z\phi),
\nonumber\\
A_a^{(IV,n\lambda{\bf k})}&=&\frac{1}{\sqrt{\lambda^2+{\bf k}^2}}
\nabla_a\phi=\frac{1}{\sqrt{\lambda^2+{\bf k}^2}}
(\partial_\tau\phi,\partial_r\phi,\partial_y\phi,\partial_z\phi),
\label{em20b}
\end{eqnarray}
where $\sqrt{g}\epsilon_{\mu\nu}$ is the Levi-Civita pseudotensor on
the cone, $\epsilon_{ij}$ is the Levi-Civita pseudo-tensor on R$^2$ in
Cartesian coordinates, and $\phi=\phi_{n\lambda{\bf k}}(x)$ is the
complete set of normalized eigenfunctions of the  Friedrichs
self-adjoint extension of the scalar Laplacian on $C_\beta\times
\mbox{R}^2$\cite{KS91} (see the discussion at the beginning of
section \ref{conickernel}):
\begin{eqnarray}
\phi_{n\lambda{\bf k}}(x)&=&\frac{1}{2\pi\sqrt{\beta}}e^{ik_y y+ik_z z}
e^{i\frac{2\pi n}{\beta}\tau} J_{\nu_n}(\lambda r),\nonumber\\
&&\hspace{5mm}
n=0,\pm 1, \dots;\,\,\lambda\in \mbox{R}^+;\,\, k_y, k_z\in \mbox{R}
\nonumber\\
\Delta\phi_{n\lambda{\bf k}}(x)&=&-(\lambda^2+{\bf
k}^2)\phi_{n\lambda{\bf k}}(x).
\label{em21}
\end{eqnarray}
Here, $J_{\nu_n}$ is the Bessel function of first kind and
$\nu_n=\frac{2\pi|n|}{\beta}$. Using the relation
(\ref{besselnormal}) one can check that the  modes
(\ref{em20b}) are normalized according
to
\begin{eqnarray}
(A^{(i',n'\lambda'{\bf k}')},A^{(i,n\lambda{\bf k})})&\equiv&\int
d^4x\,\sqrt{g}\,g^{ab} A_a^{(i',n'\lambda'{\bf
k}')\ast}A_b^{(i,n\lambda{\bf k})}\nonumber \\
&=&\delta_{i'i}\delta_{n'n}\delta^{(2)}({\bf k}-{\bf
k}')\frac{1}{\lambda}\delta(\lambda-\lambda'),\nonumber
\end{eqnarray}
The first three eigenfunctions (\ref{em20b}) satisfy $\nabla^a A_a=0$
and have eigenvalue $\lambda^2+{\bf k}^2$, while $A_a^{(IV)}$ is a pure
gauge and has eigenvalue $\frac{1}{\alpha}(\lambda^2+{\bf k}^2)$.

Using these eigenfunctions, we can compute the diagonal
$\zeta$ function using the spectral representation (\ref{emZETA}):
after the integration over $d{\bf k}$, the contributions of the modes
to the diagonal $\zeta$ function are
\begin{eqnarray}
\zeta^{(I)}(s;x)&=&\zeta^{\mbox{\scriptsize scalar}}(s;x),\nonumber\\
\zeta^{(II)}(s;x)&=&\frac{\Gamma(s-1)}{4\pi\beta\Gamma(s)}\sum_n
\int_0^\infty
d\lambda\,\lambda^{1-2s}[\frac{\nu_n^2}{r^2}J_{\nu_n}^2(\lambda
r)+(\partial_r J_{\nu_n}(\lambda r))^2],\nonumber\\
\zeta^{(III)}(s;x)&=&\!\!\frac{s-1}{s}\zeta^{\mbox{\scriptsize
scalar}}(s;x)\nonumber\\
&&+\frac{\Gamma(s-1)}{4\pi\beta\Gamma(s+1)}\sum_n\int_0^\infty
d\lambda\lambda^{1-2s}[\frac{\nu_n^2}{r^2}J_{\nu_n}^2(\lambda
r)+(\partial_r J_{\nu_n}(\lambda
r))^2],\nonumber\\
\zeta^{(IV)}(s;x)&=&\frac{\alpha^s}{s}\zeta^{\mbox{\scriptsize
scalar}}(s;x)\nonumber\\
&&+\frac{\alpha^s\Gamma(s)}{4\pi\beta\Gamma(s+1)}\sum_n\int_0^\infty
d\lambda\,\lambda^{1-2s}[\frac{\nu_n^2}{r^2}J_{\nu_n}^2(\lambda
r)+(\partial_r J_{\nu_n}(\lambda r))^2],\nonumber
\end{eqnarray}
where the spectral representation of the local $\zeta$ function of a
minimally coupled scalar field on $C_\beta\times \mbox{R}^2$ is
\begin{eqnarray}
\zeta^{\mbox{\scriptsize scalar}}(s;x)=
\frac{\Gamma(s-1)}{4\pi\beta\Gamma(s)}\sum_{n=-\infty}^{\infty}
\int_0^\infty
d\lambda\,\lambda^{3-2s} J_{\nu_n}^2(\lambda r).\nonumber
\end{eqnarray}
Now, looking for a way close to that followed by Kabat \cite{kabat},
we use the following identity, which can be proved using some
recursion formulas for the Bessel functions \cite{GR},
\begin{eqnarray}
2\left [ \frac{\nu_n^2}{r^2}J_{\nu_n}^2(\lambda r)+[\partial_r
J_{\nu_n}(\lambda r)]^2 \right ]= 2\lambda^2 J_{\nu_n}^2(\lambda
r)+\frac{1}{r}\partial_r r\partial_r J_{\nu_n}^2(\lambda r),
\label{em25}
\end{eqnarray}
and so the spin-traced local $\zeta$ function becomes
\begin{eqnarray}
\zeta(s;x)&=&(1+\frac{s-1}{s}+\frac{\alpha^s}{s})
\zeta^{\mbox{\scriptsize
scalar}}(s;x)  \nonumber \\
& &+\frac{s+1+\alpha^s(s-1)}{2s}
\frac{\Gamma(s-1)}{4\pi\beta\Gamma(s)}\times\nonumber\\
&&\times
\sum_n\int_0^\infty \,\lambda^{1-2s}[2\lambda^2
J_{\nu_n}^2(\lambda r)+\frac{1}{r}\partial_r
r\partial_r J_{\nu_n}^2(\lambda r)] \:,\nonumber
\end{eqnarray}
namely
\begin{eqnarray}
\zeta(s;x) &= &(3+\alpha^s)\zeta^{\mbox{\scriptsize scalar}}(s;x)+
\frac{s+1+\alpha^s(s-1)}{2s}\zeta^{\mbox{\scriptsize V}}(s;x),
\label{ema0}
\end{eqnarray}
where we have set
\begin{eqnarray}
\zeta^{\mbox{\scriptsize V}}(s;x)=\frac{1}{r}\partial_r r\partial_r
\frac{\Gamma(s-1)}{4\pi\beta\Gamma(s)}\sum_{n=-\infty}^{\infty}
\int_0^\infty d\lambda\,\lambda^{1-2s}J_{\nu_n}(\lambda r)^2.
\label{em101}
\end{eqnarray}
Notice that the term $\zeta^{\mbox{\scriptsize V}}(s;x)$ arises from
the ``conical'' components of the field, i.e. $A_\tau$ and $A_r$. In
particular its source is the second term in the right-hand side of Eq.
(\ref{em25}) only. This terms will produce the Kabat ``surface term''
as we will see shortly.

We have taken $\frac{1}{r}\partial_r r\partial_r$, which is in fact
the Laplacian $\Delta$, outside the integral and the series, but this
is a safe shortcut: indeed, one could first let $\Delta$ act on the
Bessel function using $\partial_r J_\nu(\lambda r)=\lambda
J_{\nu-1}(\lambda r)-\frac{\nu}{r}J_\nu(\lambda r)$, go through some
tedious calculations and get the same result as Eq. (\ref{em27}).

So far, the expressions for $\zeta^{\mbox{\scriptsize scalar}}$ and
$\zeta^{\mbox{\scriptsize V}}$ are just formal, since one can easily
see that there is no value of $s$ for which they converge.  The
correct way to compute $\zeta^{\mbox{\scriptsize scalar}}$ in this
background has been recently given by Zerbini, Cognola and Vanzo
\cite{ZCV}, following an earlier work of Cheeger \cite{cheeger83}, and the
result in four dimensions is (see section \ref{Coniczeta})
\begin{eqnarray}
\zeta^{\mbox{\scriptsize scalar}}(s;x)=
\frac{r^{2s-4}}{4\pi\beta\Gamma(s)}
I_\beta(s-1), \nonumber
\end{eqnarray}
where the function $I_\b(s)$ is defined in (\ref{Ifunction}).
We remind that it is analytic in the whole complex plane but
in $s=1$, where it has a simple pole with residue
$\frac{1}{2}(\frac{\beta}{2\pi}-1)$. Following the same procedure used
in \cite{ZCV} to obtain the above result, we can compute the
contribution to the $\zeta$ function coming from
$\zeta^{\mbox{\scriptsize V}}(s;x)$. The essential step to give a
sense to Eq. (\ref{em101}) is the separation of the small eigenvalue
$\nu_0$ from the others \cite{cheeger83}: define
\begin{eqnarray}
\zeta_<^{\mbox{\scriptsize
V}}(s;x)&=&\Delta\frac{\Gamma(s-1)}{4\pi\beta \Gamma(s)}
\int_0^\infty
d\lambda\,\lambda^{1-2s}J_0^2(\lambda r), \nonumber \\
\zeta_>^{\mbox{\scriptsize V}}(s;x)&=&2\Delta
\frac{\Gamma(s-1)}{4\pi\beta \Gamma(s)}\sum_{n=1}^{\infty}
\int_0^\infty d\lambda\,\lambda^{1-2s}J_{\nu_n}^2(\lambda r).
\nonumber
\end{eqnarray}
The integrals over $\lambda$ can be computed \cite{GR}: for
$\frac{1}{2}<\mbox{Re} s<1+\nu$
\begin{eqnarray}
\int_0^\infty  d\lambda\,\lambda^{1-2s} J^2_\nu(\lambda
r)=r^{2s-2}\frac{\Gamma(s-\frac{1}{2}) \Gamma(\nu-s+1)}{2\sqrt{\pi}
\Gamma(s)\Gamma(\nu+s)}. \nonumber
\end{eqnarray}
Therefore, in the strip $\frac{1}{2}<\mbox{Re} s<1$ we get
\begin{eqnarray}
\zeta_<^{\mbox{\scriptsize
V}}(s;x)=-\Delta\frac{r^{2s-2}\Gamma(s-1)}{4\pi\Gamma(s)^2}
\frac{\Gamma(s-\frac{1}{2})}{\sqrt{\pi}}G_{2\pi}(s), \nonumber
\end{eqnarray}
while
\begin{eqnarray}
\zeta_>^{\mbox{\scriptsize
V}}(s;x)=\Delta\frac{r^{2s-2}\Gamma(s-1)}{4\pi\Gamma(s)^2}
\frac{\Gamma(s-\frac{1}{2})}{\sqrt{\pi}}G_{\beta}(s), \nonumber
\end{eqnarray}
which is valid in the strip $1<\mbox{Re} s<1+\nu_1$, since the series
defining $G_\beta(s)$ converges for $s>1$. Both expressions can now be
analytically continued the whole complex plane and then summed, so
we can write
\begin{eqnarray}
\zeta^{\mbox{\scriptsize
V}}(s;x)&=&\Delta\frac{r^{2s-2}\Gamma(s-1)}{4\pi\Gamma(s)^2}
I_\beta(s)\nonumber\\
&=&\frac{(s-1)r^{2s-4}}{\pi\beta\Gamma(s)}I_\beta(s).
\label{em27}
\end{eqnarray}
This result could be obtained directly from Eq. (\ref{em101}), noting
that
\begin{eqnarray}
\zeta^{\mbox{\scriptsize V}}(s;x)=\Delta[\frac{s}{s-1}
\zeta^{\mbox{\scriptsize scalar}}(s+1;x)]\:.\nonumber
\end{eqnarray}
 Note also that
$\zeta^{\mbox{\scriptsize V}}(s;x)|_{\beta=2\pi}=0$ and
$\zeta^{\mbox{\scriptsize V}}(s=0;x)=0$.

Now we can write the final result for the local $\zeta$ function of the
electromagnetic field: after adding the contribution of the ghosts,
which is just $-2\zeta_\beta^{\mbox{\scriptsize scalar}}(s;x)$, we get
\begin{eqnarray}
\zeta^{\mbox{\scriptsize e.m.}}(s;x)&=&
(1+\alpha^s)\zeta^{\mbox{\scriptsize scalar}}(s;x)+
\frac{s+1+\alpha^s(s-1)}{2s}\zeta^{\mbox{\scriptsize
V}}(s;x)\nonumber\\
&=&(1+\alpha^s)\frac{r^{2s-4}}{4\pi\beta\Gamma(s)}
I_\beta(s-1)\nonumber\\
&&+\frac{s+1+\alpha^s(s-1)}{2s}
\frac{(s-1)r^{2s-4}}{\pi\beta\Gamma(s)}I_\beta(s).
\label{em28}
\end{eqnarray}
{From} this expression we can easily see that
$\zeta^{\mbox{\scriptsize e.m.}} (s;x)|_{s=0}=0$ and
\begin{eqnarray}
{\zeta^{\mbox{\scriptsize e.m.}}}'(s;x)|_{s=0}=
\frac{1}{2\pi\beta
r^4}I_\beta(-1)-(1-\frac{1}{2}\ln\alpha)\frac{1}{\pi\beta
r^4}I_\beta(0),
\label{em29}
\end{eqnarray}
Therefore, the one-loop effective Lagrangian density for the
electromagnetic field on $C_\beta\times \mbox{R}^2$ is
\begin{eqnarray}
{\cal L}_\beta^{\mbox{\scriptsize e.m.}}(x)&=&2{\cal
L}_\beta^{\mbox{\scriptsize
scalar}}(x) -\frac{(1-\frac{1}{2}\ln\alpha)}{2\pi\beta
r^4}I_\beta(0)\nonumber\\ &=&\frac{1}{4\pi\beta
r^4}I_\beta(-1)-\frac{(1-\frac{1}{2}\ln\alpha)}{2\pi\beta
r^4}I_\beta(0).
\label{em30}
\end{eqnarray}
Since $I_{2\pi}(s)=0$,  we can notice that both terms of the effective
Lagrangian density vanish when the conical singularity disappears,
$\beta=2\pi$.

A few remarks  on this result. First, no surprise that in in the
effective Lagrangian density we get a contribution which is  twice
that of a scalar field. More surprising is the second term: after the
integration over the spatial variables, it gives rise to what Kabat
\cite{kabat} calls ``surface'' term and interprets as a low-energy relic
of stringy effects foreseen by Susskind and Uglum \cite{SU}. This term
would give a negative contribution to the entropy of the system, at
least for  for $\alpha<e^2$, and actually also the total correction to
the entropy at the black hole temperature $\beta=2\pi$ would be
negative for $\alpha<e^{6/5}$, which is clearly a nonsense if we want
to give a state counting interpretation to the entropy. However, in
the four-dimensional case we get that it is not gauge invariant, in
contrast with Kabat's result.

With this regard, it is interesting to note that in two dimensions, i.e.,
on $C_\beta$, the result is indeed independent on the gauge-fixing
parameter: using the modes of the e.m. field on $C_\beta$ given by Kabat
\cite{kabat} and following the same procedure as above, before adding the
contribution of ghosts we get
\begin{eqnarray}
\zeta_{d=2}^{\mbox{\scriptsize e.m.}}(s;x)=(1+\alpha^s)
\left [\zeta_{d=2}^{\mbox{\scriptsize scalar}}(s;x)+
\zeta_{d=2}^{\mbox{\scriptsize V}}(s;x)\right ], \nonumber
\end{eqnarray}
where
\begin{eqnarray}
\zeta_{d=2}^{\mbox{\scriptsize
scalar}}(s;x)&=&\frac{r^{2s-2}}{\beta\Gamma(s)} I_\beta(s),\nonumber
\\
\zeta_{d=2}^{\mbox{\scriptsize
V}}(s;x)&=&\Delta\frac{r^{2s}}{2\beta\Gamma(s+1)} I_\beta(s+1),
\nonumber \end{eqnarray}
and so, adding the contribution of the ghosts we have
\begin{eqnarray}
{\cal L}^{\mbox{\scriptsize e.m.}}(x)=\frac{1}{2\pi\beta
r^2}(2\pi-\beta),\nonumber
\end{eqnarray}
which is gauge independent and, after the integration over the
manifold, gives exactly the result of Kabat.

Coming back to the four-dimensional case, we argue that a natural
(albeit not the only possible, see the final discussion)
procedure to restore the gauge invariance is simply to drop the Kabat
term, namely, the last term in Eq. (\ref{em30}), obtaining the
reasonable result ${\cal L}^{\mbox{\scriptsize e.m.}}(x) = 2{\cal
L}^{\mbox{\scriptsize scalar}}(x)$.

First of all, notice that the gauge invariance must  hold for the
integrated quantities as the effective action, namely the logarithm of
the integrated effective Lagrangian. In fact, the ghost procedure,
which takes into account the gauge invariance,  works on integrated
quantities. However, in our case, the integration of the Kabat term
produces a divergent {\em gauge-dependent}  result, and thus it seems
reasonable to discard such a local term. With this regard, it is
important to note that Kabat obtains a gauge-independent result
because, within his regularization procedure, he has the freedom to
choose an independent cutoff parameter for each mode. Instead, in our
procedure we have only one cutoff parameter $\epsilon$, to which we
give a precise physical meaning, namely the minimal distance from the
horizon.

A more general discussion might be the following. It is worth one's
while stressing that, dealing with {\em smooth compact} manifold,
local quantities as local heat kernel and local zeta-functions are
intrinsically ill defined due to the possibility of adding to them a
total covariant derivative with vanishing integral. In such a case,
the previous global quantities are well-defined, and one can
satisfactorily employ these latter instead of local quantities in
order to avoid the ill-definiteness problem.
Notice also that the gauge dependent Kabat surface term formally looks
such as a Laplacian and thus it should disappear after a global
integration, provided regularity conditions on the manifold are
satisfied, producing gauge-independent integrated quantities. However,
this is not the case for the present situation, where the background
is a noncompact manifold with a conical singularity, and the
integrated quantities diverge  requiring a regularization procedure.
We stress that the use of local quantities is preferred on the
physical ground, because they lead us to the correct temperature
dependency as we will see shortly.

Therefore, in our case  the local quantities remain ill defined  and
require a further regularization procedure in order to fix the
possible added total derivative term before we integrate. Furthermore,
the integrated quantities are divergent, so we expect we to have to
take into account also total derivative terms with a divergent
integral.  In our case this further regularization procedure consists
just in discarding the Kabat term. Notice that this procedure produces
gauge-independent local quantities.

In the next Chapter we will see that the results of the optical method
confirm this procedure.

Once we have dropped the Kabat's term, we can compute thermodynamical
quantities like internal energy and entropy: we need the effective
action and so we have to introduce a smearing function $\varphi(x)$ in
order to define the trace: $\ln Z_\beta=\int d^4x \sqrt{g}{\cal
L}_\beta(x)\varphi(x)$. Actually, since ${\cal L}_\beta$ does not
depend on the transverse coordinates $y$ and $z$, the integration on
these coordinates simply yields the infinite area of the Rindler
horizon, that we indicate as $A_\perp$. This divergence has clear
physical meaning. The integration over $\tau$ has no problem, while a
convenient smearing function for  the integration over $r$ is
$\varphi(r)=\theta(r-\epsilon)$, and so the effective action becomes
\begin{eqnarray}
\ln Z_\beta(\epsilon)=\frac{A_\perp}{8\pi\epsilon^2}I_\beta(-1).
\label{em31}
\end{eqnarray}
For $\epsilon \rightarrow 0$ we have a divergence that can be seen as
a ``horizon'' divergence \cite{thooft}, since as $r\rightarrow 0$ we
approach the horizon of the Rindler wedge (see Chapter \ref{BHENTROPY}
for a discussion on this point).

{From} Eq. (\ref{em31}) we can compute the free energy,
$F_\beta=-\frac{1}{\beta}\ln Z_\beta$, which at high temperature,
$\beta\rightarrow 0$, has a leading behavior
$-2\pi^2A_\perp/180\epsilon^2\beta^4$, in perfect
agreement with the statistical mechanics result of  Susskind and Uglum
\cite{SU}. Instead, Kabat \cite{kabat} obtains a leading behavior
$-2A_\perp/8\epsilon^2\beta^2$, where the behavior
$\beta^{-2}$ independent of the dimension of the space-time, is
typical of the global approach, as we have seen in
section \ref{conicmethod}. Of particular interest
for the black hole physics is the entropy of the system:
\begin{eqnarray}
S_\beta=\beta^2\partial_\beta
F_\beta=\frac{A_\perp}{90\beta\epsilon^2} \left
[\left(\frac{2\pi}{\beta}\right)^2+5\right].
\label{em32}
\end{eqnarray}
This equation gives, in Rindler space approximation, the one-loop
quantum correction to the black hole entropy coming form the
electromagnetic field propagating in the region outside the horizon.
It shows the well known horizon divergence \cite{thooft} (see also
\cite{BM} for a recent review on this topic): unless we
suppose the existence of a natural effective cutoff at the Planck scale
due to an (unknown) quantum gravity theory or back-reaction horizon
fluctuations etc.,\footnote{However, such a cutoff should depend on the
field spin value to produce the correct entropy factor in front of the
horizon area. See \cite{YO}.} we get a divergent entropy which is
physically unsatisfactory and contrasts with the finite
thermodynamical Bekenstein-Hawking entropy. However, this problem is
not peculiar to the photon field, as it occurs for scalar and
spinorial fields as well.

We can note that, if we took into account the surface term which we
have previously dropped,  we would obtain the unphysical, because being
gauge dependent, expression
\begin{eqnarray}
S_\beta(\alpha)=\beta^2\partial_\beta F_\beta=\frac{A_\perp}{90\beta\epsilon^2}
\left [\left(\frac{2\pi}{\beta}\right)^2+5\right] - (1-\frac{1}{2}\ln
\alpha)\frac{A_\perp}{6\beta\epsilon^2}.
\nonumber
\end{eqnarray}
As anticipated above, this expression for the entropy is negative when
the singularity is absent, $\beta = 2\pi$, and $\ln\alpha <
\frac{6}{5}$. Moreover, for $\ln\alpha<\frac{4}{3}$, $S_\beta(\alpha)$
shows a further zero of the entropy  corresponding to an inconsistent
(gauge-depending) {\em finite} temperature  {\em pure} quantum state
of the field.

With regard to the energy-momentum tensor, the very simple
relation between the $\z$ function of the Maxwell field
and that of the scalar field allows us compute the
energy-momentum tensor in a way very similar to
that employed in section \ref{enermomtens} for the scalar
case, obtaining the very simple relation
\beas
\langle T_{\mu\nu}^{\mbox{\scriptsize e.m.}}(x)\rangle
&=&2\langle T_{\mu\nu}^{\mbox{\scriptsize scalar}}(x)\rangle
(\xi=0),
\enas
where $\langle T_{\mu\nu}^{\mbox{\scriptsize scalar}}(x)\rangle$
is that given in Eq. (\ref{stresst}) with $m=0$ after an analytic
continuation to the Rindler space (see Eq. (\ref{tensor})).
This result agrees with that obtained by means of the point-splitting
procedure \cite{DC79,BROT85,FRSE87,DO94}.

\section{A general conjecture}
\markboth{Photons and gravitons}{A general conjecture}
\label{secthree}

Let us focus our attention back on Kabat's surface term in the
effective lagrangian, Eq. (\ref{em30}): is it an accident which
appears in our manifold and in the vector case only, or conversely,
is it a more general phenomenon?

We can grasp some insight by studying either  the local
$\zeta$ function, as it appears in Eq. (\ref{emZETA}), or the local
heat kernel of Eq. (\ref{em12}) and passing to the local
zeta-functions through Eq. (\ref{em16}). In fact the Kabat term
already  comes out in the heat kernel and then it remains
substantially unchanged passing to the local $\zeta$ function through
Eq. (\ref{em16}).   The components of the modes $II$, $III$ and $IV$
contain (covariant) derivatives in both the conical and ${\mbox R}^2$
indices. Using trivial (covariant) derivative rules and reminding that
$\nabla_\mu \nabla^\mu \phi = -\lambda^2 \phi$ and $\partial_i
\partial^i \phi = -{\bf k}^2 \phi$ we may transform scalar products of
(covariant) derivatives  appearing in the integrand of Eq.
(\ref{em12}) into a  covariant divergence of a vector plus a simple
scalar term. Summing over the modes, these parts produce respectively
the Kabat surface term and the `twice scalar' part of the effective
Lagrangian in Eq. (\ref{em30})  (the mode $I$ gives a contribution to
this latter part only). This is the general mechanism which produces
Kabat's term. Let us  illustrate this in more detail. Dealing with the
modes $IV$ we find
\begin{eqnarray}
g^{ab} A^{(IV)\ast}_a A^{(IV)}_b
&=& \frac{1}{\lambda^2+{\bf k}^2}
\nabla_a \phi^\ast \nabla^a \phi \nonumber \\
&=&\frac{1}{\lambda^2+{\bf k}^2} \left[  \nabla_a (\phi^\ast
\nabla^a\phi ) - \phi^\ast \nabla_a \nabla^a \phi
\right] \nonumber\\
&=&\frac{1}{\lambda^2+{\bf k}^2} \left[ \nabla_a
(\phi^\ast \nabla^a \phi )  + (\lambda^2+ {\bf k}^2) \phi^\ast  \phi
\right]\:. \label{ema1}
\end{eqnarray}
Thus, using the particular form of our modes we get
\begin{eqnarray}
g^{ab} A^{(IV)\ast}_a A^{(IV)}_b
&=&\frac{1}{2(\lambda^2+{\bf k}^2)}\nonumber
\Delta
J_{\nu_n}^2 +  J^2_{\nu_n}  \:. \nonumber
\end{eqnarray}
The modes $III$  contribute to the local heat kernel and to the
effective Lagrangian in the same way. The modes $II$ require a  little
different care: we have
\begin{eqnarray}
g^{ab} A^{(II)\ast}_a A^{(II)}_b
&=&\frac{1}{\lambda^2} g^{\mu\nu}\epsilon_{\mu\sigma}
\epsilon_{\nu\rho} \nabla^\sigma \phi^\ast \nabla^\rho \phi \nonumber\\
&=&\frac{1}{\lambda^2} \left[  \nabla^\sigma ( g^{\mu\nu}
\epsilon_{\mu\sigma} \epsilon_{\nu\rho} \phi^\ast \nabla^\rho \phi
) - g^{\mu\nu}\epsilon_{\nu\rho} \epsilon_{\mu\sigma} \phi^\ast
\nabla^\sigma\nabla^\rho \phi \right] \nonumber\\
&=&\frac{1}{\lambda^2} \left[  \nabla^\sigma ( g_{\sigma\rho}
\phi^\ast \nabla^\rho \phi ) -\phi^\ast  g_{\rho\sigma} \nabla^\rho
\nabla^\sigma \phi \right] \nonumber\\
&=&\frac{1}{\lambda^2} \left[  \nabla_\mu (\phi^\ast
\nabla^\mu \phi)  + \lambda^2 \phi^\ast  \phi \right]\nonumber\\
&=&\frac{1}{\lambda^2} \left[  \nabla_a (\phi^\ast
\nabla^a \phi)  + \lambda^2 \phi^\ast  \phi \right]\:.
\label{ema2}
\end{eqnarray}
And thus, reminding the particular form of our modes
\begin{eqnarray}
g^{ab} A^{(II)\ast}_a A^{(II)}_b
&=&\frac{1}{2\lambda^2} \nonumber
\Delta
J_{\nu_n}^2 +  J^2_{\nu_n} \nonumber \:.
\end{eqnarray}
The contribution to the effective Lagrangian is similar to the
previous ones. In both the examined cases, using the specific form of
scalar eigenfunctions, we have obtained the right-hand side of
Eq. (\ref{em25}) except for some factors  which will be arranged
summing over all the modes in the final result.
The term  $\nabla_a (\phi^\ast \nabla^a \phi)$ $(= \frac{1}{2}\Delta
J^2_{\nu_n})$ contributes only to the second term of the right-hand
side of  Eq. (\ref{ema0}), namely it contributes only to the Kabat
surface term in the effective Lagrangian in Eq. (\ref{em30}). Moreover,
the  term $\lambda^2 \phi^\ast \phi$ $(= \lambda^2 J^2_{\nu_n})$
contributes only to the remaining term in the right hand side of
Eq. (\ref{ema0}) and thus to the  ``twice scalar'' part of the same
effective Lagrangian only.

We  further remark that the previously employed  covariant derivative
identities are exactly the same which one has to use in order to check
the correct normalization of the modes.\footnote{In this case the
indices $(n\lambda {\bf k})$ which appear in the modes $A_a$ and
$A^\ast_a$ are generally different.} However, in that case the
surface terms are dropped after the formal integration in the  spatial
variables, because they do not contribute, in a distributional sense,
to the overall normalization. Conversely, following the local zeta
function method they produce Kabat-like terms.

More generally speaking, following the previous outline, one
can avoid specifying the form of the scalar eigenfunction and the use
of Eq. (\ref{em25}), remaining on a more general ground.\footnote{It is
clear from our discussion that the Kabat term gets contributions from
each mode $II,III,IV$ not depending on the corresponding eigenvalue.
This term does not coincide  with the surface term recently suggested
by  Fursaev and  Miele \cite{FM} dealing with  compact
manifolds, because this latter involves zero modes only.} This means
that we can consider a more general manifold which is topologically
${\cal M}\times \mbox{R}^2$ with the natural product metric, where
${\cal M}$ is any, maybe curved, two-dimensional manifold. The photon
effective action can be written  as
\begin{eqnarray}
\ln Z = -\frac{1}{2} \ln \det \mu^{-2} \left(
-\Delta_1 + (1- \frac{1}{\alpha}) d_0 \delta_0
\right) + \ln Z_{\mbox{\scriptsize ghost}}
\label{ema3}\:,
\end{eqnarray}
where $\Delta_1 = d_0\delta_0 +\delta_1 d_1$ is the  Hodge Laplacian
for 1-forms ($\delta_n \equiv d_n^\dagger $ with respect to the Hodge
scalar product.)
The eigenfunctions of the operator appearing in the above equation can
still be written as in Eq. (\ref{em20b}). Now, $\phi = \frac{1}{2\pi}
e^{ik_y y+ i k_z z} {\bf J}_{n,\lambda}(x^\mu)$ where ${\bf
J}_{n,\lambda}(x^\mu)$ is an eigenfunction of (the Friedrichs
extension of) the 0-forms Hodge Laplacian\footnote{Remind the Hodge
Laplacian coincides with minus the Laplace-Beltrami operator for
0-forms. This generally  does not happen for n-forms when $n>0$ in
curved manifolds.}
$\Delta^{\cal M}_0$ on ${\cal M}$, with eigenvalue $-\lambda^2$.
Employing a bit of $n$-forms algebra, one can obtain in our manifold
the same eigenvalues found in the manifold ${\cal C}_\beta \times
\mbox{R}^2$. Furthermore, once again $\delta_0 A^{(y)} = 0$,  namely $
\nabla^a A_a^{(y)}= 0 $, in case $y= I, II, III$. Then, using Eq.s
(\ref{ema1}) and (\ref{ema2}) and the definition in Eq.
(\ref{emZETA}), we get, before we take into account the ghosts
contribution,
\begin{eqnarray}
\zeta_{{\cal M}\times \mbox{\scriptsize R}^2}(s;x) = (3+\alpha^s)
\zeta^{\mbox{\scriptsize scalar}}_{{\cal M}\times \mbox{\scriptsize
R}^2}(s;x) + \frac{s+1+\alpha^s(s-1)}{2s} \zeta^{V}_{{\cal M}\times
\mbox{\scriptsize R}^2}(s;x) \:, \nonumber
\end{eqnarray}
where the surface term reads
\begin{eqnarray}
\zeta^{V}_{{\cal M}\times \mbox{\scriptsize R}^2}(s;x)
=\frac{\Gamma(s-1)}{4\pi \Gamma(s)} \nabla_a \sum_n \int d\lambda
\lambda \,{\bf J}^\ast \nabla^a {\bf J}\:.\nonumber
\end{eqnarray}
Notice that, if the manifold is regular and compact, this surface term
automatically disappears after we integrate over the spatial
variables. Instead, if the manifold ${\cal M}$ has conical
singularities or boundaries, then this term could survive the
integration.  We can further  suppose  that ${\cal M}$ contains a
Killing vector $\partial_\tau$ with compact orbits in such a manner
that we can define a temperature $1/\beta$ and interpret the effective
action as the logarithm of the photon partition function. Employing
coordinates $r, \tau$ on ${\cal M}$, we can decompose ${\bf
J}_{n,\lambda}(r,\tau)$ as ${\bf J}_{n,\lambda}(r,\tau) =
{\beta}^{-1/2}e^{-2\pi n i\tau/\beta} {\cal J}_{n,\lambda}(r)
$, ${\cal J}_{n,\lambda}(r)$ being real. The surface term reads, in
this case,
\begin{eqnarray}
\zeta^{V}_{{\cal M}\times \mbox{\scriptsize R}^2}(s;x)
=\frac{\Gamma(s-1)}{4\pi \beta \Gamma(s)} \Delta_0 \sum_n
\int d\lambda \lambda\, {\cal J}_{n,\lambda}(r)^2 \:. \nonumber
\end{eqnarray}
Equation (\ref{ema3}) holds in very general manifolds, also dropping the
requirement of a metric which is Cartesian product of the flat
$\mbox{R}^2$ metric and any other metric.

One can simply prove that, if $\phi$ is an eigenfunction of $\Delta_0$
with eigenvalue $-\nu^2$ on such a general manifold, $A = d_0 \phi$
will be an eigenfunction of the vector operator $-\Delta_1 +
(1-\frac{1}{\alpha})d_0\delta_0$ with gauge-dependent eigenvalue
$-\nu^2/\alpha$. Employing the rule in Eq. (\ref{ema1}) with $\nu^2$
in place of $\lambda^2 + {\bf k}^2$, we expect that this latter
eigenfunction should produce a (gauge-dependent) surface term into the
local zeta function.

Dealing with spin $s\geq 1$ and massless fields, because of  the
simple equation of motion form (in Feynman-like gauges at least), we
expect to find out some normal modes obtained as covariant derivatives
of the scalar field modes opportunely rearranged. Hence, barring
miraculous cancellations, the corresponding local heat kernel, local
$\zeta$ function and effective Lagrangian, should contain Kabat-like
surface terms, due to the previous mechanism. We will check this for
the graviton in the next section.\footnote{We also tried  to study the
photon case employing  a so-called  `physical gauge' as $A_z=0$. The
use of the $\zeta$-function regularization in this case is problematic
due to a remaining gauge ambiguity arising  whenever one tries to deal
with a path integral nonformal approach in axial gauges.
Nevertheless, through the same mechanism, the Kabat term seems to
survive  in this case as well. }

\section{The graviton $\zeta$ function}
\markboth{Photons and gravitons}{The graviton $\zeta$ function}
\label{secfour}

In this section we shall compute the local $\zeta$ function in the case
of a linearized graviton propagating in the Rindler wedge. We will see
that Kabat-like surface terms indeed appear, as we suggested in
the previous section. Moreover, we will find out that consistent
results arise by discarding all those terms.

Following the same procedure  used in \cite{HA93,GHP77}, which employs the
harmonic gauge, we decompose the linearized field of a graviton into
its symmetric traceless part $h_{ab}$ and its trace part $h$. Choosing
an opportune normalization factor of the fields and dropping boundary
terms, the Euclidean action (containing also the gauge-fixing part)
looks such as:
\begin{eqnarray}
S_E[h_{ab},h] = \frac{1}{32 \pi G}\int d^4x \sqrt{g}\left\{
\frac{1}{2} g^{aa'}g^{bb'} h_{ab}\nabla_c \nabla^c
h_{a'b'} +\frac{1}{4} h \nabla_d\nabla^d h\right\},
\label{lagranggrav}
\end{eqnarray}
where $g$, $g^{ab}$, and covariant derivatives are referred to the
background metric, namely, the Euclidean Rindler metric. That metric is
also used to raise and lower indices. Notice that curvature tensor
terms (see \cite{GHP77}) do not appear in the above action and this is
due to the flatness of the manifold. It is necessary to point out that
we changed the sign of  the trace field Lagrangian as this  appeared
after we performed a ``simple'' Wick rotation toward the imaginary time
on the Lorentzian Lagrangian. In fact, in order to obtain an Euclidean
Lagrangian producing a formally finite functional
integral\footnote{Remind that this functional integral contains the
exponential $\exp {(- S_E)}$}, it is also necessary to rotate the
scalar field $h$  into imaginary values during the Wick rotation. This
adjusts the sign in front of the corresponding Lagrangian
\cite{HA93,GHP77}. We can write, as far as the effective action is
concerned:
\begin{eqnarray}
\ln Z_{\mbox{\scriptsize gravitons}}
&=& -\frac{1}{2}\ln \det \mu^{-2} \left[ -g^{aa'}g^{bb'} \nabla_c \nabla^c
\right]\nonumber\\
&&-\frac{1}{2}\ln \det \mu^{-2}\left[ -\nabla_d \nabla^d
\right] + \ln Z_{\mbox{\scriptsize grv. ghosts}}
\label{zetagrav}.
\end{eqnarray}
The first determinant has to be evaluated in the $L^2$ space of
traceless symmetric tensorial field. Unessential factors in front of
the operators can be dropped into an overall added constant and thus
omitted. Furthermore, the ghost contribution has been taken into
account through the last term of the previous equation. A usual
procedure\footnote{This result holds also for local quantities.}
leads us to \cite{HA93,GHP77}
\begin{eqnarray}
\ln Z_{\mbox{\scriptsize grv. ghosts}}
= - 2 \ln Z_{\mbox{\scriptsize vector}}.
\nonumber
\end{eqnarray}
The  partition function in  $\ln Z_{\mbox{\scriptsize vector}}$ is the
partition function obtained quantizing the massless Klein-Gordon
vector field. Hence,  this also coincides with the photon partition
function evaluated in the Feynman gauge, namely $\alpha=1$ in Eq.
(\ref{em10}), {\em without} taking into account the photon ghost
contribution. Thus, from the effective graviton ghost action, two
vector $\alpha=1$ Kabat's surface terms (with the sign changed) arise.
In order to compute the above functional determinants, we have to look
for normalized modes of a self-adjoint extension of the  tensorial
Laplace-Beltrami operator $\Delta_T = g^{aa'}g^{bb'} \nabla_c
\nabla^c$ in the space of symmetric traceless tensors  and the scalar
Laplace-Beltrami operator $\Delta_S = \nabla_d \nabla^d$. Obviously,
the eigenfunctions of $\Delta_S$ can be chosen as $h_{n\lambda {\bf
k}} = \phi_{n\lambda {\bf k}}(x) $, where, as before, $\phi =
\phi_{n\lambda k}(x)$ indicates the generic eigenfunction of the
scalar Laplacian, Eq. (\ref{em21}).

In the tensorial case, we find the following nine classes of
sym\-metric trace\-less eigen\-functions:\footnote{All the components
of each eigenfunction class which do not appear in the following list
are understood to vanish.}
\begin{eqnarray}
h^{(1)}_{n\lambda {\bf k}}&:&
\:\:\:\: \frac{\sqrt{2}}{\lambda^2}\nabla_\mu \nabla_\nu
\phi + \frac{1}{\sqrt{2}} g_{\mu\nu}
\phi = h^{(1)}_{\mu\nu}=h^{(1)}_{\nu\mu}\:;\nonumber \\
h^{(2)}_{n\lambda {\bf k}} &:&
\:\:\:\: \frac{\sqrt{g}}{\sqrt{2}\lambda^2}\left\{
\epsilon_{\mu\sigma}\nabla^\sigma \nabla_\nu\phi +
\epsilon_{\nu\sigma}\nabla^\sigma \nabla_\mu\phi \right\}
= h^{(2)}_{\mu\nu}=h^{(2)}_{\nu\mu}\:;\nonumber \\
h^{(3)}_{n\lambda {\bf k}}&:&\:\:\:\:
\frac{1}{\sqrt{2}k\lambda} \partial_i \nabla_\mu \phi = h^{(3)}_{i\mu}
= h^{(3)}_{\mu i} \:;\nonumber \\
h^{(4)}_{n\lambda {\bf k}}&:& \:\:\:\:
\frac{\sqrt{g}}{\sqrt{2} k\lambda}  \epsilon_{\mu\nu}\partial_i
\nabla^\nu \phi = h^{(4)}_{i\mu}= h^{(4)}_{\mu i} \:;\nonumber \\
h^{(5)}_{n\lambda {\bf k}} &:& \:\:\:\:
\frac{\sqrt{g}}{\sqrt{2} k\lambda}  \epsilon_{\mu\nu}\epsilon_{ij}
\partial^j \nabla^\nu\phi = h^{(5)}_{i\mu}
= h^{(5)}_{\mu i} \:;\nonumber \\
h^{(6)}_{n\lambda {\bf k}} &:&\:\:\:\:
\frac{1}{\sqrt{2} k\lambda} \epsilon_{ij}
\partial^j \nabla_\mu\phi = h^{(6)}_{i\mu}
= h^{(6)}_{\mu i} \:;\nonumber \\
h^{(7)}_{n\lambda {\bf k}} &:&
\:\:\:\: \frac{\sqrt{2}}{ {\bf k}^2}\partial_i \partial_j
\phi + \frac{1}{\sqrt{2}} \delta_{ij} \phi =
h^{(7)}_{ij}=h^{(7)}_{ji}\:;\nonumber \\
h^{(8)}_{n\lambda {\bf k}} &:&
\:\:\:\: \frac{1}{\sqrt{2}{\bf k}^2}\left\{
\epsilon_{ik}\partial^k\partial_j\phi +\epsilon_{jk}
\partial^k\partial_i\phi\right\} = h^{(8)}_{ij}=h^{(8)}_{ji}\:;\nonumber \\
h^{(9)}_{n\lambda {\bf k}} &:&
\:\:\:\: \frac{1}{2} g_{\mu\nu}\phi -\frac{1}{2}\delta_{ij}\phi =
h^{(9)}_{ab}
\:.\nonumber
\end{eqnarray}
Here, $\sqrt{g}\epsilon_{\mu\nu}$ indicates the antisymmetric
Levi-Civita pseudotensor on the cone and $\epsilon_{ij}$ the
antisymmetric Levi-Civita pseudotensor on ${\mbox{R}}^2$ in Cartesian
coordinates. The previous modes satisfy
\begin{eqnarray}
\Delta_T h^{(y)}_{n\lambda {\bf k}} = -(\lambda^2 +
{\bf k}^2) h^{(y)}_{n\lambda {\bf k}},
\hspace{1cm} y= 1,2,... ,9,
\end{eqnarray}
and
\begin{eqnarray}
\Delta_S h_{n\lambda {\bf k}} = -(\lambda^2 + {\bf k}^2)
h_{n\lambda {\bf k}} \:.
\end{eqnarray}
Finally, the normalization relations are ($ y,y'= 1,2....,9$)
\begin{eqnarray}
\int d^4x \sqrt{g} \: g^{aa'}g^{bb'} \: h^{(y)\ast}_{n\lambda {\bf k}}
(x)_{ab}
h^{(y')}_{n'\lambda' k'_t}(x)_{a'b'} =
\delta^{yy'}\delta_{nn'}
\delta^{(2)}({\bf k}- {\bf k}')
\frac{\delta (\lambda-\lambda')}{\lambda}\nonumber
\end{eqnarray}
and
\begin{eqnarray}
\int d^4x \sqrt{g} h^{\ast}_{n\lambda {\bf k}}(x)
h_{n'\lambda' k'}(x) = \delta_{nn'}
\delta^{(2)}({\bf k}- {\bf k}')
\frac{1}{\lambda}
\delta (\lambda-\lambda')\:. \nonumber
\end{eqnarray}
Using Eq. (\ref{emZETA}), we can write the local $\zeta$ function as
\begin{eqnarray}
\zeta^{\mbox{\scriptsize Gravitons}}(s;x)&=& \sum_{y=1}^9
\zeta^{(y)}(s;x) + \zeta^{\mbox{\scriptsize scalar}}(s;x)\nonumber\\
&=&\hspace{-2cm}\sum_{y=1}^9
\sum_{n=-\infty}^{\infty}\int_0^{\infty}d\lambda\,\lambda
\int_{\mbox{R}^2} d^2{\bf k}\, \nu_{n}^{-2s}
g^{aa'}(x) g^{bb'}(x) h^{\ast(y)}(x)_{ab}\,h^{(y)}(x)_{a'b'}\nonumber\\
& &\hspace{-2cm}+ \sum_{n=-\infty}^{\infty}
\int_0^{\infty}d\lambda\,\lambda
\int_{\mbox{R}^2} d^2{\bf k}\,\nu_{n}^{-2s} h^\ast(x) h(x)
\label{gravheat}.
\end{eqnarray}
The latter term takes into account the graviton trace part
contribution to local $\zeta$ function. Obviously, this is exactly the
scalar local $\zeta$ function. Let us rather  consider the former term
and, in particular, the contribution due to $h^{(1)}$. Following the
sketch of the previous section, we can rearrange this term
transforming the product of the covariant derivatives into a scalar
term added to several covariant divergences of vector and tensor
fields:
\begin{eqnarray}
\zeta^{(1)}(t;x)&=& \frac{\Gamma(s-1)}{4\pi \beta \Gamma(s)}
\sum_n
\int_0^{\infty}d\lambda\,\lambda^{3-2s} \,
\phi^{\ast} \phi\nonumber\\
& &+ 4 \frac{\Gamma(s-1)}{4\pi \beta \Gamma(s)}\sum_n
\int_0^{\infty}d\lambda\,\lambda^{1-2s}\,
\nabla_a (\phi^\ast \nabla^a \phi)+ \nonumber\\
& &+ 2 \frac{\Gamma(s-1)}{4\pi \beta \Gamma(s)}
\sum_n
\int_0^{\infty}d\lambda\,\lambda^{-2s} \,
\nabla^a \nabla_b [\nabla_a \phi^\ast
\nabla^b \phi]\:.\nonumber
\end{eqnarray}
Using different notation, we finally find
\begin{eqnarray}
\zeta^{(1)}(s;x)= \zeta^{\mbox{\scriptsize scalar}}(s;x)+ 2
\zeta^{V}(s;x)+  2 \zeta^{W}(s;x)
\end{eqnarray}
where we defined
\begin{eqnarray}
\zeta^{W}(s;x) &=& \frac{\Gamma(s-1)}{4\pi \beta \Gamma(s)} \sum_n
\int_0^{\infty}d\lambda\,\lambda^{-2s} \, \nabla^a \nabla_b
(\nabla_a \phi^\ast \nabla^b \phi )\nonumber\\
&=&\frac{\Gamma(s-1)}{4\pi \beta \Gamma(s)} \sum_n
\int_0^{\infty}d\lambda\,\lambda^{-2s} \nonumber\\
& &\times\left[ \frac{1}{r}\partial_{r} r \partial_{r} (\partial_r
J_{\nu_n})^2 + \frac{1}{r}\partial_{r}(\partial_r J_{\nu_n})^2 -
\frac{\nu_n^2}{r} \partial_r \frac{J_{\nu_n}(\lambda r)^2}{r^2}\right]
\:.
\end{eqnarray}
Thus, we see that in the local $\zeta$ function the $(\alpha=1)$-Kabat
surface term $\zeta^{V}(s;x)$ reappears, together with a new surface
term $\zeta^{W}(s;x)$. The contribution of $h^{(2)}$ is similar to
the previous one  and it reads
\begin{eqnarray}
\zeta^{2}(s;x)=
\zeta^{\mbox{\scriptsize scalar}}(s;x)+ \zeta^{V}(s;x)+\zeta^{W}(s;x)+
\zeta^{U}(s;x) \nonumber
\end{eqnarray}
where, provided $\theta^{ab} =\epsilon^{ab}$ when $a,b = \mu, \nu $
and $\theta^{ab} = 0 $ otherwise,
\begin{eqnarray}
\zeta^{U}(s;x)&=&
\frac{\Gamma(s-1)}{4\pi \beta \Gamma(s)} \sum_n
\int_0^{\infty}d\lambda\,\lambda^{-2s} \, \nabla_a \nabla_b \left[
g\:\theta^{ac}\theta^{bd} \nabla_c \phi^\ast
\nabla_d \phi     \right]\nonumber \\
&=& \frac{\Gamma(s-1)}{4\pi \beta \Gamma(s)}
\sum_n
\int_0^{\infty}d\lambda\,\lambda^{-2s} \,
\partial_r \left[
\frac{J_{\nu_n}^2}{r}-(\partial_r J_{\nu_n})^2 \right]. \nonumber
\end{eqnarray}
The contributions of the remaining terms are much more trivial.
In fact, a few of algebra leads us to
\begin{eqnarray}
\zeta^{(3)}(s;x)&
=&\zeta^{(4)}(s;x)=\zeta^{(5)}(s;x)=\zeta^{(6)}(s;x)\nonumber\\
&=& \zeta^{\mbox{\scriptsize scalar}}(s;x)+ \frac{1}{2} \zeta^{V}(s;x),
 \nonumber
\end{eqnarray}
and
\begin{eqnarray}
\zeta^{(7)}(s;x)=\zeta^{(8)}(s;x)=\zeta^{(9)}(s;x)=
\zeta^{\mbox{\scriptsize scalar}}(s;x).
\nonumber
\end{eqnarray}
Finally, we have already noted above that the contribution of the trace
terms $h$ is
exactly $\zeta^{\mbox{\scriptsize scalar}}(s;x)$.
Then, taking into account the contribution of the
ghost Lagrangian, which amounts
to $-8 \zeta^{\mbox{\scriptsize scalar}}(s;x) -
2\zeta^{V}(s;x) $, we
get the final expression of spin-traced graviton $\zeta$ function:
\begin{eqnarray}
\zeta^{\mbox{\scriptsize Gravitons}}(s;x)=
2 \zeta^{\mbox{\scriptsize scalar}}(s;x)+
3 \zeta^{V}(s;x)+
3 \zeta^{W}(s;x) +\zeta^{U}(s;x).
\end{eqnarray}
Dropping the last three surface terms we obtain the reasonable
result which agrees with the counting of the true graviton degrees of
freedom: ${\cal L}^{\mbox{\scriptsize Graviton}}(x) = 2 {\cal L}^{
\mbox{\scriptsize scalar}}(x)$. Hence, all the thermodynamical
quantities coincides with those of the previously computed photon
fields.

\section{Discussion}
\markboth{Photons and gravitons}{Discussion}
\label{secfive}

In this Chapter we have computed the effective action of the photon and
graviton fields in the conical background $C_\beta\times\mbox{R}^2$,
and our main result is that it is just what one expects from counting
the number of degrees of freedom, i.e. twice that of the massless
scalar effective action. Moreover, we have got the correct Planckian
temperature dependence of the thermodynamical quantities.

To get this apparently trivial result, we had to deal with unwanted
terms arising from the presence of the conical singularity. We
discussed how the appearance of those surface terms is quite a general
phenomenon dealing with general manifolds in the case of fields with
integer non-zero spin. The presence of conical singularities needs
some further regularization procedure. In particular, this is
necessary while studying the photon field in order to restore the
gauge invariance of the integrated quantities. It could be interesting
to develop an analogue research in the case of gravitons in any
covariant gauge.

In the general case our proposal is the simplest one, namely, to
discard all the surface terms. However, we think that, away from our
local $\zeta$-function  approach,  this should not be the only possible
treatment of surface terms. In fact, comparing our results with
Kabat's it arises that, except for the two-dimensional case,  the
necessary treatment of surface terms strongly depends on the general
approach used to define and calculate the effective Lagrangian.
Moreover, it also depends on the regularization procedure used to
define the integrated quantities.

In our local $\zeta$-function approach, the meaning of the only cutoff
as the minimal distance from the horizon leads ourselves towards the
simple procedure of discarding the surface terms in order to restore
the gauge invariance. In Kabat's treatment, the meaning of the
employed cutoff is not so strict and permits one to make safe the
gauge invariance  and take on the surface terms as well. This is due
to fine-tuning of mode depending cutoffs which contain a further
gauge-fixing parameter dependence.

In our approach, when integrating the surface term it is not possible
to use an $\alpha$-dependent cutoff different from that used for
the rest and such that it cancels the $\alpha$ dependence in the
integrated quantity: in fact, no real function  $\epsilon(\alpha)$ can
absorb the factor $(1-\frac{1}{2}\ln \alpha)$, appearing in the
integrated surface term, for all the values of $\alpha$.

In any case, we think that any procedure which does not discard the
surface terms must be able to explain why the consequent result is not
in agreement with what one expects from counting the number of degrees
of freedom and  to deal with the apparently unphysical corrections to
the thermodynamical quantities arising from those terms. Maybe this is
possible in an effective low-energy string theory which does not
coincide with the ordinary quantum field theory. In the next
Chapter we will see that in the optical approach the
gauge-dependent terms vanish automatically, and the results
confirm the procedure adopted in this Chapter.

\cleardoublepage
\newpage

\chapter{Optical approach for the thermal partition
function of photons}

\label{OPTICCHAP}
\section*{Introduction}
\markboth{Optical approach to the cone}{Introduction}

In this Chapter we will discuss an approach to finite temperature
theory in the Rindler space quite different from those employed in the
previous Chapters, the so called optical approach
\cite{GIPER78,BROT85,GIDD94,BARB94,EMP95,CVZ95,BAEMP95,dAO}. Actually,
this is a very general method for discussing a finite temperature
theory in a static spacetime.\footnote{ We remind that a spacetime is
called {\em static} if it admits a metric $g_{\mu\nu}$ such that
$\pa_0 g_{\mu\nu}=0$ and $g_{0i}=0$. Moreover, if $g_{00}=1$ it is
called {\em ultrastatic}.} In this approach, instead of computing the
partition function directly in the static metric, one performs a
conformal transformation in such a way that the metric of the
resulting manifold is ultrastatic. Then, one can compute the relevant
quantities in this ``optical manifold'' using a suitable
regularization method, such the dimensional one or the $\zeta$
function, and taking into account how the various quantities transform
under conformal transformations. This method is particularly advantageous
in the Euclidean Rindler case, since this manifold has a conical
singularity that, as we have seen, can be quite tricky to deal with,
whereas the related optical manifold has no singularity. However,
there is more in this method than the mathematical content. In fact,
it has been shown \cite{GIDD94,BARB94,BAEMP95,dAO} that the canonical
partition function of a quantum field in a curved background with a
static metric is not directly related to the Euclidean path integral
with periodic imaginary time in the static manifold, but rather it is
equal to Euclidean path integral in the related optical manifold. In
particular, in \cite{BARB94} it is shown that the statistical counting
of states leads naturally to a formulation in the optical manifold. We
can also notice that, as far as we know, the equivalence of the direct
periodic imaginary time path integral formalism to the canonical
formalism for computing finite temperature effects has been proved in
ultrastatic manifolds only \cite{ALLEN86}. Let us start reviewing how
the relation among the canonical theory and the optical manifold
arises.

Let us suppose that we want compute the one-loop free energy of a
neutral Bose field in the canonical ensemble. Then, by using standard
Bose oscillator algebra we obtain
\bea
F_\b&=&-\frac{1}{\b}\ln\Tr e^{-\b H}\\
&=&\frac{1}{\b}\sum_i\ln\left(1-e^{-\b\omega_i}\right)+F_0,
\label{canofreeen}
\ena
where $\b F_0=\frac{1}{2}\sum_i \omega_i$ is the vacuum energy.
The Hamiltonian $H=\sum_i \omega_i a^\dagger_i a_i$ is defined with
respect to the Killing vector $\pa_t$. Let us then consider the
following identity \cite{ramond}:
\beas
\frac{d}{d\omega^2}
\sum_{n=-\infty}^{\infty} \ln (a^2n^2+\omega^2)&=&
\sum_n\frac{1}{a^2n^2+\omega^2}\\
&=&\frac{\pi}{\omega a}\mbox{cth}\frac{\pi \omega}{a},
\enas
where we have used the identity \cite{GR}
\beas
\mbox{cth}\pi x&=&\frac{1}{\pi x}+\frac{2x}{\pi}
\sum_{n=1}^\infty\frac{1}{x^2+n^2}.
\enas
By integrating the above relation we obtain that, up
to an unimportant arbitrary constant,
\beas
\ln \prod_{n=-\infty}^\infty
\left(\frac{4\pi^2 n^2}{\b^2}+\omega^2\right)&=&\b\omega+
2\ln \left(1-e^{-\b\omega}\right).
\enas
Therefore, we can rewrite the canonical free energy
(\ref{canofreeen}) as
\bea
F_\b&=&\frac{1}{2\b}\ln\prod_{n=-\infty}^\infty
\prod_i \left(\frac{4\pi^2 n^2}{\b^2}+\omega^2\right).
\label{canofreeen2}
\ena
Now, from the general theory we expect that the one-loop
free energy is, up to the vacuum contribution,
\bea
F_\b=-\frac{1}{\b}\ln Z=\frac{1}{2\b}\ln\det L,
\label{pathfreeen}
\ena
where $L$ is the small fluctuations operator of the
theory. By comparing equations (\ref{canofreeen2}) and
(\ref{pathfreeen}), we see that, if we interpret
(\ref{canofreeen2}) in the sense of $\z$-function regularization,
the quantities  $\frac{4\pi^2 n^2}{\b^2}+\omega^2$ must be
just the eigenvalues of the operator $L$ \cite{BARB94} but,
in the case of a finite temperature theory in a static
spacetime, the usual small fluctuations operator
$$
A=-\Delta+m^2+\xi R,
$$
has not the eigenvalues in that form. The only reasonable
operator of the theory which has the eigenvalues in the above
form is the operator we obtain from $A$ by making a
conformal transformation from the static metric $g_{\mu\nu}$
to the related ultrastatic one $g'_{\mu\nu}$:
\begin{eqnarray}
g_{\mu\nu}\rightarrow
g'_{\mu\nu}=\Omega^{2}(x)g_{\mu\nu},
\label{conformaltr}
\end{eqnarray}
so that
\begin{eqnarray}
ds^2 \rightarrow ds'^2= \Omega^{2} ds^2 \:.
\label{conformal0}
\end{eqnarray}
Choosing $\Omega^{-2} = g_{00}$, $ds'^2$ becomes the related ultrastatic
optical metric. In the case of the Euclidean Rindler space, this
conformal factor becomes singular just on the conical singularities,
which are pushed away to the infinity\footnote{The points at $r=0$ of
the optical manifold $S^1\times H^3$ are infinitely far from the
points of the manifold with $r>0$ taking the distance  as the affine
parameter  along geodesics. Strictly speaking, the former  points do
not belong to the manifold at all.} and the optical manifold is free
from singularities. Under such a transformation, the neutral scalar
field $\phi$ transforms into $\phi' = \Omega^{\frac{2-D}{2}}
\phi$ and the Euclidean action with coupling factor $\xi$ transforms
into the following more complicated action \cite{birrel}:
\begin{eqnarray}
S'[\phi']&=&  \frac{1}{2}\int d^4x \sqrt{g'}\: \phi' \left\{
-\nabla'_\mu  \nabla'^\mu + \xi_D R'+
\Omega^{-2}\left[m^2+(\xi-\xi_D) R\right]\right\}\phi',\nonumber\\
&& \label{i1.5'}
\end{eqnarray}
where $\xi_D=(D-2)/4(D-1)$ is the conformal invariant factor.
Therefore, the small fluctuations operator in the optical
space is
\beas
A'&=&-\Delta'+ \xi_D R'+
\Omega^{-2}\left[m^2+(\xi-\xi_D) R\right],
\enas
and the Euclidean path integral which, according with the
above discussion, leads to the canonical free energy
(\ref{canofreeen}) is
\beas
Z'[\b]&=&\int \D\phi' \, e^{-S'[\phi']},
\enas
where the integration measure is \cite{BARB94,BAEMP95,dAO}
\beas
\D\phi'&=&\prod_{x}d\phi'\,{g'}^{1/4}.
\enas
This expression has to be compared with that usually
employed to compute the partition function
\beas
Z[\b]&=&\int \D\phi\, e^{-S[\phi]},\\
S[\phi]&=&\frac{1}{2}\int d^4 x\sqrt{g}\,\phi A\phi,
\enas
with the functional measure  $\D\phi=\prod_{x}d\phi\,g^{1/4}$.

Summarizing, we have obtained that the canonical definition of the
free energy of a field in a static spacetime is not strictly
equivalent to the Euclidean theory with periodic imaginary time on the
static manifold,  but rather to the theory in the related ultrastatic
manifold, the optical space. This result was derived in
\cite{GIDD94,BARB94,BAEMP95} using a method similar to that
employed above, and in \cite{dAO} by means a completely different
method: starting from the phase-space formulation of the
path integral, namely in terms of the field and its conjugate momentum,
one derives an expression for the partition function in
which the functional measure is not the correct one; this
problem can be cured by making a conformal transformation
to the optical metric, in which the functional integral has the
correct form. Actually, this result is implicit in the canonical
quantization of a field at finite temperature, as is shown
in \cite{fullingreport,fullingbook}.

Notwithstanding this, most of the work on finite temperature theory is
carried out  using the path integral in the original static metric:
therefore, it is important to understand the relation of the two
definitions of partition function, $Z_\b$ and $Z_b'$. It is a
classical result \cite{POL81,DS89} that under conformal transformation
the functional integration  measure in the path integral is not
invariant and the transformation yields a  functional Jacobian. Since
it does not depend on $\phi'$ it can be carried outside the integral:
\begin{eqnarray}
 Z_\beta &=& \int {\cal D} \phi' \: J[g,g',\beta] \: e^{-S'[\phi']} =
J[g,g',\beta] Z'_\beta.
 \label{foundamental}
\end{eqnarray}
We stress that is $Z'_\beta$ which is equivalent to the canonical
partition function. The Jacobian $J[g,g',\beta]$ can be written as the
exponential of a Liouville-type action (see \cite{dAO} and references
therein), but, when the involved manifolds are regular, it is possible to
prove that such action is simply proportional to $\b$:
\begin{eqnarray}
J[g,g',\beta] = \exp (-\beta E_0) \:.
\label{jaco}
\end{eqnarray}
where  $E_0$ does not depend on $\beta$. This is substantially due
the fact that the Jacobian is the exponential of a spacetime
integral of local geometric quantities \cite{BAEMP95,CVZ95,dAO}.
Since these quantities are built out of the metric tensor which
is independent on the time coordinate and on $\b$, it follows
that the only dependence on $\beta$ in the exponent is due
to the  integration over the manifold and the Jacobian takes
the form (\ref{jaco}). The case of manifolds with conical singularities
is more subtle, since the standard formulas of Riemannian
geometry fail in this case and the above reasoning does
not hold. This case will be discussed in Appendix B,
where we will see that it is not yet clear whether
the relation (\ref{jaco})  holds also in presence of conical
singularities or not.

Since the Jacobian has the form (\ref{jaco}), then the free energy
computed in the optical metric, $F'_\b=-\frac{1}{\b}\ln Z_\b'$, and
that computed in the original metric, $F_\b=-\frac{1}{\b}\ln Z_\b$,
differ only for the renormalized zero-temperature energy, which does
not affect thermodynamical quantities such as the entropy. This
justifies the use of $Z_\b$ in most cases.

\section{Optical approach to the Rindler case}
\markboth{Optical approach to the cone}{The Rindler case}
\label{secdue}

Let us now discuss the case of the finite temperature theory in the
Rindler space, comparing the results of the optical approach with
those obtained in the previous Chapters  in the static metric.

As usual, the static manifold  is the Euclidean Rindler manifold ${\cal
C}_\beta\times R^2$ with an imaginary time period $\beta$ and
Euclidean Rindler  metric reads
\begin{eqnarray}
ds^2 = r^2d\tau^2 + dr^2 + dy^2 +dz^2
 \label{rindler}\:,
\end{eqnarray}
where $\tau \in [0 , \beta]$, $r\in R^+$, ${\bf x}= (y,z) \in R^2$,
and we have the the well-known conical singularity at $r=0$ when
$\beta\neq 2\pi$.

In the case $\xi=0$ and $m=0$, the partition function $Z_\b$  has been
computed in Chapter \ref{BHENTROPY} by a local $\zeta$-function or
heat-kernel approach  obtaining a Minkowski
renormalized free energy $F_\beta^{\scriptsize \mbox{sub}} = F_\beta
-U_{\beta=2\pi}$ and a renormalized internal energy\footnote{As it is
well known, the $(\beta=2\pi)$-thermal Rindler state locally coincides
with the Minkowski vacuum and, in renormalizing, we suppose that this
state does not carry energy density. Notice that such a Minkowski
subtraction procedure does not affect the entropy computed through
$F_\beta$.} $U_\beta^{\scriptsize\mbox{sub}} = \partial_\beta \beta
F_\beta - (\partial_\beta \beta F_\beta)|_{\beta=2\pi} $ which read
\begin{eqnarray}
F_\beta^{\scriptsize \mbox{sub}} &=&-\frac{{\cal A_H}}{2880 \pi^2
\epsilon^2} \left[  \left( \frac{2\pi}{\beta}\right)^4 +10  \left(
\frac{2\pi}{\beta}\right)^2 +13 \right] \label{f},\nonumber\\
U_\beta^{\scriptsize\mbox{sub}} &=& \frac{{\cal A_H}}{2880 \pi^2
\epsilon^2} \left[3\left( \frac{2\pi}{\beta}\right)^4
+10  \left(\frac{2\pi}{\beta}\right)^2 -13\right]
\label{FU},
\end{eqnarray}
where ${\cal A_H}$ is the (infinite) event horizon  area and $\epsilon$ a
short-distance cutoff representing the minimal distance from the
horizon \cite{thooft}.

It is worthwhile noticing that the Lorentz section of the Rindler
space is flat and hence, as far as the real time theory is concerned,
we find a complete independence on the parameter $\xi$. However, in
calculating the partition function, one has to deal with the Euclidean
section of the Rindler manifold and, considering it as a integral
kernel, the curvature $R$ takes  Dirac's $\delta$ behaviour at $r=0$
\cite{SOL95,FUSOL95}, thus the value of the parameter $\xi$ could be
important. The previous results have been carried out in the case
$\xi=0$ in the sense that the eigenfunctions employed in computing the
$\zeta$ functions properly satisfy the eigenvalue equation with no $R$
term.

In the case $\xi\neq 0$ the problems are due to the fact that the
equation for the eigenfunctions contains a Dirac $\delta$, and so it is
not mathematically clear how to treat it. In the case of a cosmic
string, the Dirac $\delta$ represents a limit case, maybe unphysical, of
the problem in which the string has a finite thickness, which is
mathematically well defined since no Dirac $\delta$ appears. In the case
of the Rindler space there is no such way out, and the only way to
avoid the problem is to consider the case $\xi=0$ (see also the discussion
at the end of section \ref{enermomtens}).

Now consider the optical approach.
If the massless field in the conical space $C_\b\times R^2$
is conformally coupled field, $\xi=\xi_4$,  we see that
the transformed action (\ref{i1.5'}) is that of a conformally coupled field in the
optical manifold $S^1\times H^3$, with metric
\begin{eqnarray}
ds'^2&=&d\tau^2+r^{-2}\left[dr^2+dy^2+dz^2\right],
\label{opticalmetric}
\end{eqnarray}
whose constant scalar curvature is $R'=-(D-1)(D-2)=-6$. In the other
cases, we have to keep a term proportional to $R$ which has a Dirac
$\delta$ behaviour at $r=0$ and thus we have an ill-defined
operator.\footnote{One possible way to get rid of this term is to
define the action in the Lorentzian manifold, where $R=0$, perform the
conformal transformation to the optical manifold, and only then use
the transformed action to write to partition function with the
periodic imaginary time formalism \cite{EMP95}. This procedure gives a
result independent on the parameter $\xi$ by nature: the coupling in
the optical manifold is always conformal. However, in our opinion this
procedure seems too {\em ad hoc}.} In the former case, namely when
$\xi=\xi_4$, the direct computation of $F'_\beta$ can be performed,
for instance, employing the $\zeta$-function approach \cite{CVZ95}
(see also \cite{elirom96} and the Appendix A in this Chapter). We
report here the well-known final result only:
\begin{eqnarray}
F'_\beta = - \frac{{\cal A_H}}{2880\pi^2\epsilon^2}\left( \frac{2\pi}{\beta}
 \right)^4\:.
\label{f'}
\end{eqnarray}
This result is in perfect agreement with the WKB result
 (\ref{helmholtz}) by Susskind and Uglum \cite{SU}.

A direct comparison of the optical result (\ref{f'}) with the
static manifold result (\ref{FU}) is not possible, since the
former has been obtained supposing a conformal coupling
in the static manifold and the latter in the case of minimal coupling.
However, we have seen in section \ref{nomincap} that the entropy
on-shell computed in the static manifold is the same for any value of
the coupling and its value is $S_\xi[\mbox{static}]|_{\b=2\pi}={\cal
A_H}/60\pi\ep^2$; on the other hand, the on-shell entropy computed
with the optical approach is
$S_{\xi=1/6}[\mbox{optical}]|_{\b=2\pi}={\cal A_H}/360\pi\ep^2$. It is
therefore clear that the results obtained in the static manifold and
in the optical are different in the numerical coefficients, although
they show the same essential features (proportionality to the horizon
area, Planckian behaviour at high temperatures and presence of the
horizon divergence).

Before commenting on this result, let us consider how it is
possible to compute the internal energy from the energy-momentum
tensor. The point is that it is necessary to distinguish between the
{\em energy} and the {\em canonical energy} \cite{frofur97}. The
energy $E$ of a system is defined in terms of the energy-momentum
tensor:
\bea
E&=&\int_{\cal B} T_{\mu\nu}\z^\mu d\sigma^\nu=
-\int_{\cal B}T_0^0\sqrt{-g}\,d^3x,
\label{theenergy}
\ena
where ${\cal B}$ is a space-like hypersurface orthogonal to
the Killing vector $\z^\mu$ and $d\sigma^\nu$ is the
future directed vector of the volume element on ${\cal B}$.
On the other hand, the canonical energy $H$ coincides
with the Hamiltonian of the system. In terms of the
Hamiltonian density ${\cal H}$ it can be written as
\bea
H&=&\int_{\cal B} {\cal H} \,\sqrt{-g}d^3x
\label{thehamiltonian}
\ena
where
\bea
{\cal H}&=&\frac{1}{2}\left[-g^{00}(\nabla_0\phi)^2+
g^{ij}\nabla_i\phi\,\nabla_j\phi+(m^2+\xi R)\phi^2\right].
\label{theHdensity}
\ena
By comparing the above expression for the Hamiltonian
density and the definition of the energy-momentum tensor
\cite{birrel}
\bea
T_{\mu\nu}&=&\nabla_\mu\phi\nabla_\nu\phi-\frac{1}{2}
\left(\nabla_\a\phi \nabla^\a\phi+m^2\phi^2\right)\nonumber\\
&&+\xi \left[\left(R_{\mu\nu}-\frac{1}{2}g_{\mu\nu}R\right)\phi^2
+g_{\mu\nu}\nabla_\a\nabla^\a\phi^2-\nabla_\mu\nabla_\nu\phi^2\right],
\label{theemtensor}
\ena
one sees that  the following relation holds
\bea
{\cal H}&=&-T_0^0+\xi\left[R_0^0\phi^2+g^{ij}\nabla_i\nabla_j\phi^2\right],
\label{therelation}
\ena
and so in general the energy and the canonical energy are different.
Quantistically, we can interpret this relation as
\bea
\langle{\cal H}\rangle
&=&-\langle T_0^0\rangle
+\xi\left[R_0^0\langle\phi^2\rangle+
g^{ij}\nabla_i\nabla_j\langle\phi^2\rangle\right].
\label{theQrelation}
\ena
It is important to notice that it is the canonical
energy density ${\cal H}$ which corresponds, up to a
subtraction, to the energy density that one derives from
the one-loop effective action density.

Let us then apply the relation (\ref{theQrelation}) to the case
of the Rindler space: in this static manifold the
energy-momentum tensor can be obtained for any value of the
coupling by continuing into the Rindler space the
cosmic string result (\ref{stresst}): setting $m=0$ we have
\begin{eqnarray}
\langle T^\mu_\nu\rangle_\beta(\xi) &=&
\frac{1}{1440 \pi^2 r^4} \left[\left(
\left(\frac{2\pi}{\beta}\right)^4 -1\right) \mbox{diag}(-3,1,1,1)
\right.\nonumber\\
&&\left. +10(6\xi-1)\left( \left(\frac{2\pi}{\beta}\right)^2 -1\right)
\mbox{diag}(3,-1,2,2) \right].
\label{tensor}
\end{eqnarray}
We remind that the above tensor coincides with the point-splitting
result. Reminding also  that in this case $R_0^0=0$ and
that $\langle\phi^2\rangle$ is given by (\ref{vacuum}) for $m=0$,
an easy calculation shows that for any value of $\xi$
\bea
\langle{\cal H}\rangle
&=&-\langle T_0^0\rangle_\xi+\frac{\xi}{8\pi^2 r^4}
\left[\left(\frac{2\pi}{\b}\right)^2-1\right]\nonumber\\
&=&\frac{1}{480\pi^2 r^4}\left[\left( \frac{2\pi}{\beta}\right)^4
+10\left(\frac{2\pi}{\beta}\right)^2 -11\right].
\label{canoene}
\ena
By comparing the corresponding internal energy $U=\int \langle{\cal
H}\rangle\sqrt{-g}d^{3}x$ with that computed from the effective
action, Eq. (\ref{FU}), we see that the results do not agree.

Summarizing, we have seen that in the conformally coupled
case the free energy of a scalar field in the Rindler case
can be easily computed in the optical manifold. According to the
discussion in the previous section, the obtained result (\ref{f'})
should correspond to the canonical free energy and we have
seen that it agrees with the result of the WKB approximation.

On the other hand, we have seen that the optical result is different
from that of the static manifold, and the difference is the term
proportional to $T^2$, which is not present in the former case and is
present in the latter. Furthermore, in the computation in the static
manifold we have came across a contradiction by comparing the internal
energy computed directly from the effective action and that computed
from the energy-momentum tensor. Also in this case the difference is
in the term proportional to $T^2$.
We can restate this negative result saying that quantity
$Z_\b$ computed in the physical static manifold
cannot be considered the partition function of the system
\cite{Zmoretti}, since a fundamental relation such as
\beas
-\pa_\b\ln Z_\b&=&\int\langle{\cal H}\rangle\sqrt{-g}d^3x
\enas
does not hold.

There could be several explanation for the above negative result.
First of all, it is worth noticing that no such
contradictions appear in the results of the optical method, and
in the next sections we will see that the situation is the same
also for the Maxwell field, namely the optical result is free from
contradictions, while the direct computation suffers from the
same problems as the scalar case. Therefore, the simplest explanation
could be that the equivalence of the formulations in the
static and in the optical manifold breaks down because of
anomalous temperature-dependent terms in the Jacobian
due to the conical singularity. This, of course, could explain
also the difference between the entropies computed in the
two methods.

However, there could be other explanations. For instance, the
contradictions found here could be an indication of the fact that
the usual statistical-mechanical relations have to be modified
because of the explicit dependence on the temperature of the
Hamiltonian in the black-hole or Rindler background, as
argued by Frolov \cite{frolov95}. A more radical conclusion
could be the inconsistency of the off-shell approach
to the black-hole entropy, as partially  argued in \cite{moiel}.
In any case, this issue requires further investigations.

\section{Optical approaches in the case of photons}
\markboth{Optical approach to the cone}{Optical approaches for photons}
\label{sectre}

In Chapter \ref{PHOTONS} we have discussed the computation of
the one-loop partition function of photons gas in a Rindler
wedge  (see also \cite{ielmo})  generalizing the local $\z$-function
procedure in \cite{ZCV}. This  procedures works in the static Rindler
manifold.
The
found Minkowski renormalized free energy  amounts to $2 F^{\scriptsize
\mbox{sub}}_\beta + (2 - \ln \alpha) F_\beta^{\scriptsize
\mbox{surface}}$, where $F_\beta^{\scriptsize \mbox{sub}}$ is the
scalar free energy  previously discussed, Eq. (\ref{FU}), and the
$F_\beta^{\scriptsize \mbox{surface}}$ is a ``surface'' term which
arises integrating a total derivative and has the form ${\cal A_H}
[(2\pi/\beta)^2-1]/(24\pi^2\epsilon^2)$ (see \cite{kabat} and
\cite{ielmo} for more comments), finally $\alpha$ is the gauge-fixing
parameter. Notice that also this anomalous gauge-dependent term
involves a $\beta^{-2}$ dependence. We suggested  dropping this latter
gauge-dependent term as the simplest procedure to remove the
unphysical gauge dependence. Anyway, we stressed that other procedures
could also be possible.

Another interesting feature is that the free energy of the photons
suffers of the same discrepancy encountered in the previous section
for the scalar field:  the free energy computed
from the path integral in the static manifold and the free
energy obtained by integrating the component
$T_{00}$ of the energy-momentum tensor are different.
Indeed, for the renormalized energy density
we have the following  very simple relation \cite{DC79,FRSE87}:
\begin{eqnarray}
\langle T^0_0\rangle_\beta^{\scriptsize \mbox{phot. p.-s.}} =
2 \langle T^0_0\rangle_\beta^{\scriptsize \mbox{p.-s.}}(\xi = 0) \:,
\label{tensor2}
\end{eqnarray}
where on the right the stress tensor is that of a massless scalar
field. It is worth while noticing that $\xi = 0$ takes place on the
right-hand side instead of $\xi= 1/6$. Hence, the energy density of
the electromagnetic field does not amount to twice that of a
conformally coupled scalar field, as one could naively expect
considering that the electromagnetic field is conformal invariant in
four dimensions. As far as the internal energy is concerned, we find
the same unforeseen relation.

In the (Lorentzian) Rindler space the scalar curvature $R$ is zero
everywhere and the parameter $\xi$ remains as a relic of the fact that
$T_{\mu\nu}$ is obtained by varying the metric $g_{\mu\nu}$ in the
field Lagrangian \cite{birrel}.\footnote{It is worthwhile noticing
that one has to consider the theory within the curved space time in
order to discuss on the physics in the flat spacetime. Anyhow, the
extension of the theory to a curved spacetime is not unique  and this
involves some subtleties regarding also the regularization procedure.
The choice between different regularization procedures should be made
on the basis of what  is the physics that one is trying to describe.
Obviously, the general hope is that, at the end of the complete
renormalization procedure involving matter fields and gravity, all
these different regularization approaches give rise to equivalent
physical results.  See \cite{BAEMP95} and \cite{LW} for a discussion
on these topics.} Employing the general expression of
$T_{\mu\nu}(\xi)$ \cite{FRSE87,birrel} in terms of the Hadamard
function, one finds that, in the case $R=0$, the global conserved
quantities as total energy should not depend on the value of $\xi$.
This is  because the contributions to those quantities due to $\xi$
are discarded into boundary surface integrals which generally vanish.
However, this is not the case dealing with the Rindler wedge because
such integrals diverge therein.\footnote{Similar problems appear
working in subregions of the Minkowski space in presence of boundary
conditions \cite{birrel}.}

Therefore, the unexpected relation  (\ref{tensor2}) for the
photon field should not affect the integrated quantities in  more
``regular'' theories,  restoring the naively expected relation
between the considered quantities.

In the Introduction we have stressed the importance of the optical
method in the scalar case: therefore, now we go to investigate how it
is possible to apply optical-manifold method to the Maxwell
field. As a by-product,  we will see that in the Rindler case the
unphysical gauge-fixing dependence automatically disappears in the
optical approach, confirming the procedure adopted in \cite{ielmo}.

There are two possible ways to implement
the optical method. The simplest one consists of defining the partition
function directly as a functional integral on the optical manifold.
However, there is another more complicated possibility: it consists of
starting with a functional integral in the initial static manifold,
performing the conformal transformation and finally dropping the
functional Jacobian. This is, in fact, the simplest generalization of
the results obtained in the scalar case. Both  methods produce the
same final functional integral in the simpler conformally coupled
scalar case, but in the case of the Maxwell field the two procedures
do not seem to be equivalent, as we shall see,  due to the presence of
gauge-fixing and ghost terms.

\subsection{Optical approaches in the case of  general static
manifolds}

Let us start reviewing the formalism we use dealing with the photon
field. The complete action for the electromagnetic field in any
covariant gauge and on a general Euclidean manifold, endowed with a
metric $ds^2 = g_{\mu\nu} dx^\mu dx^\nu$, which we shall suppose
{\em static} and where $\partial_0$ is the global (Euclidean) timelike
Killing vector with closed orbits of period $\beta$. Using the in
Hodge de Rham formalism  we have
\begin{eqnarray}
S^{\scriptsize \mbox{em}}&=&
\int d^4x\left[\frac{1}{4}  \langle F,F\rangle + \frac{1}{2\alpha}
\langle A,d\delta A\rangle\right] + S_{\scriptsize
\mbox{ghost}}(\alpha)\nonumber\\
&=& \frac{1}{2}\int d^4x\left[\langle  A,\Delta A\rangle -
\left(1-\frac{1}{\alpha}\right)\langle A,d\delta A\rangle\right]+
S_{\scriptsize \mbox{ghost}}(\alpha)
\label{1}\:.
\end{eqnarray}
In order to maintain the gauge invariance of the theory, it is
important to keep the dependence on the gauge-fixing parameter of the
ghost action, as one obtains by varying the gauge-fixing condition
$\frac{1}{\sqrt{\alpha}}\delta A=0$ \cite{NvN}:
\begin{eqnarray}
S_{\scriptsize \mbox{ghost}}(\alpha) = -\frac{1}{\sqrt{\alpha}}\int
d^4x \sqrt{g}\: \overline{c}\Delta c,
\label{ghost}
\end{eqnarray}
where $\Delta$ is the Hodge-de Rham Laplacian for 0-forms and $c$,
$\overline{c}$ are anticommuting scalar fields.
Usually, the dependence on the gauge-fixing parameter is absorbed
rescaling the ghost fields, but in the presence of a scale anomaly this
rescaling gives rise to a nontrivial contribution, which is essential
to maintain the gauge invariance of the theory. This is just the case
here: in fact, the contribution of the action (\ref{ghost}) to the
one-loop effective action is proportional to that of a minimally
coupled scalar field, which has a scale anomaly in four dimensions.

Some comments on  the formalism in Eq. (\ref{1}) are in order.
$F\equiv \partial_\mu A_\nu -\partial_\nu A_\mu =
\nabla_\mu A_\nu -\nabla_\nu A_\mu $ is the 2-form representing the
photon strength field, $\nabla_\mu$ being the covariant derivative;
the brackets stand for the $p$-forms Hodge local product:
\begin{eqnarray}
\langle G,H\rangle = G \wedge * H =\sqrt{g}\,
g^{\mu_1\nu_1}...g^{\mu_p\nu_p} G_{\mu_1... \mu_p} H_{\nu_1...\nu_p}
\nonumber
\end{eqnarray}
For future reference we also define the internal product
\begin{eqnarray}
G \cdot H = g^{\mu_1\nu_1}...g^{\mu_p\nu_p} G_{\mu_1... \mu_p}
H_{\nu_1...\nu_p}  \nonumber\:.
\end{eqnarray}
We remind the reader
that $\delta = (-1)^{N(p+1)+1}*d*$ is the formal adjoint of
the operator $d$ with respect to the scalar product of $p$-forms
induced by the integration of the previous Hodge local product;
finally, $\Delta = d \delta + \delta d$ is the Hodge-de Rham Laplacian
of the $p$-forms. In order to perform calculations through the usual
covariant derivative formalism the following relations for $0$-forms
and $1$-forms are quite useful:
\begin{eqnarray}
\Delta \phi &=& -\nabla_\mu \nabla^\mu \phi \:, \nonumber\\
\delta A
&=& -\nabla_\mu A^\mu \nonumber\:,\\ (\Delta A)_\mu
&=& - \nabla_\nu
\nabla^\nu A_\mu + R^\nu_\mu A_\nu\:.\nonumber
\end{eqnarray}

The second line of Eq. (\ref{1}) represents the complete photon action
now expressed in terms of the vector field $A_\mu$  and the ghost
fields only and it is the one usually employed in order to compute the
partition function of the photon field by means of a functional
integral. The partition function of photon at the temperature
$T=1/\beta$ is then formally expressed by
\begin{eqnarray}
Z_\beta &=& \int {\cal D}A \: \exp
- \frac{1}{2}\int d^4x \,\left[\langle  A,\Delta A\rangle -
\left(1- \frac{1}{\alpha}\right) \langle A,d\delta A\rangle\right]
\nonumber \\
&& \times \int {\cal D}c {\cal D}\overline{c}
\exp  -S_{\scriptsize \mbox{ghost}}(\alpha) \nonumber\:.
\end{eqnarray}

In order to compute this partition function, we want to pass to the
related optical manifold, and so we consider a conformal
transformation, Eq. (\ref{conformal0}), with $\Omega^2 = g_{00}$.
Notice that, since we work in four dimensions, the $p$-forms $A$ and
$F$ have a vanishing mass dimension and thus they must be conformally
invariant, namely $A=A'$ and $F=F'$. Furthermore the following
identity arises:
\begin{eqnarray}
\langle F,F\rangle' &=& \langle F,F\rangle \label{2}\:.
\end{eqnarray}

\subsection{First general approach}

As we said above, the way to proceed is twofold. As a first way, we
can suppose to have performed the conformal transformation {\em
before} we start with the field theory. This means that we define the
partition function of photons in the static manifold as a path
integral directly in the optical manifold. In such a case the
expression of the
partition function is defined by
\begin{eqnarray}
Z^{(1)}_\beta &=& \int {\cal D}A \: \exp
- \frac{1}{2}\int d^4x\,\left[\langle  A,\Delta' A\rangle' -
\left(1- \frac{1}{\alpha}\right)\langle A,d\delta' A\rangle'\right]
\nonumber \\
&& \times \int {\cal D}c' {\cal D}\overline{c}' \exp
-S'^{(1)}_{\scriptsize \mbox{ghost}}(\alpha)
 \label{firstcase}\:,
\end{eqnarray}
where
\begin{eqnarray}
S'^{(1)}_{\scriptsize \mbox{ghost}}(\alpha) = -\frac{1}{\sqrt{\alpha}}
\int d^4x \sqrt{g'}\: \overline{c}'\Delta' c' \:,
\nonumber
\end{eqnarray}
and where the primed metric and variables appearing in the previous
functional integral are the optical ones. In other words, for the
one-loop Euclidean effective action $-\ln Z^{(1)}_\beta$ we have
\begin{eqnarray}
\ln Z^{(1)}_\beta = -\frac{1}{2} \ln \det \mu^{-2}\left[\Delta' -
\left(1-\frac{1}{\alpha}\right)d\delta'\right]
+\ln Z^{(1)}_{\beta, \scriptsize \mbox{ghost}}(\alpha)
\label{zeta1'}\:.
\end{eqnarray}
Here $\mu$ is an arbitrary renormalization scale necessary on a
dimensional ground in the above formula and  denoting the presence of
a scale anomaly if it does not  disappear from the final formulae.

For future reference we  note that the effective action of the ghosts,
except for the $\alpha$ dependent factor, amounts trivially to minus
twice the Euclidean effective action of an uncharged massless scalar
field with the Euclidean action minimally coupled with the
gravitation. Therefore its contribution to the one-loop effective
action can be written immediately from the $\zeta$ function of a
minimally coupled scalar field,
$\zeta^{\scriptsize\mbox{m.c.s.}}(s;x)$, in the same background,
taking the $\alpha$-dependence into account:
\begin{eqnarray}
&&\hspace{-7mm}\ln Z^{(1)}_{\beta, \scriptsize \mbox{ghost}}(\alpha)=
-\int d^4x\sqrt{g'}\left[\frac{d}{ds}\zeta^{\scriptsize
\mbox{m.c.s.}}(s;x)|_{s=0}+\zeta^{\scriptsize
\mbox{m.c.s.}}(s;x)|_{s=0}\ln \sqrt{\alpha}\mu^2\right].
\nonumber\\
&&\label{fantom1}
\end{eqnarray}

\subsection{Second general approach}

As a second way,  we can suppose to define the partition function
directly in the static manifold, adding also the gauge-fixing
term and the ghost Lagrangian to the pure electromagnetic action,
and only {\em after} perform the conformal transformation to the
optical metric.
In this way we have to find how all the pieces in the path integral
transform under the conformal transformation. In particular, the
operator $\Delta - (1-\alpha^{-1}) d\delta$  transforms into another
operator $\Lambda_\alpha$, which we are going to write shortly. As
regards the functional Jacobian which arises from the functional
measure, a direct
generalization of the discussion made in the scalar case
tells us that it has to be ignored if we are interested in computing
the thermal partition function. However, we would have to take it into
account if we were computing, for example, the zero-temperature
effective action in a cosmic string background.

Hence, employing this second procedure, we shall assume  the photon
partition function to be defined by
\begin{eqnarray}
Z^{(2)}_\beta &=&  \int {\cal D}A' \: \exp -\left\{\frac{1}{2}\int
d^4x \langle  A',\Lambda_\alpha A'\rangle'\right\}
 \int {\cal D}c' {\cal D}\overline{c'} \exp  -S'^{(2)}_{\scriptsize
\mbox{ghost}}(\alpha)\nonumber\:,
\end{eqnarray}
In other words, for the Euclidean effective action $-\ln
Z^{(2)}_\beta$ we have
\begin{eqnarray}
\ln Z^{(2)}_\beta =- \frac{1}{2} \ln \det (\mu^{-2} \Lambda_\alpha )
+\ln Z^{(2)}_{\beta, \scriptsize \mbox{ghost}}(\alpha)
\label{zeta2'}\:.
\end{eqnarray}
The form of $S'^{(2)}_{\scriptsize \mbox{ghost}}(\alpha)$
is that of Eq. (\ref{ghost}) after a conformal
transformation:
\begin{eqnarray}
S'^{(2)}_{\scriptsize \mbox{ghost}}(\alpha) = -
\frac{1}{\sqrt{\alpha}}\int d^4x \sqrt{g'} \: \overline{c}'
\left[\Delta' +\frac{1}{6}( R'-\Omega^{-2} R)\right]   c'
\label{5}\:,
\end{eqnarray}
where $c'=\Omega c $, $\overline{c}'=\Omega\overline{c}$. For future
reference, we note that this effective action of the ghosts amounts
trivially to minus twice the Euclidean effective action of an
uncharged massless scalar field $\varphi$ with the Euclidean action
($\Delta' = - \nabla'_\mu\nabla'^\mu$) endowed by an
$\alpha-$depending overall factor
\begin{eqnarray}
S^{(2)}(\alpha) =\frac{1}{\sqrt{\alpha}} \int d^4x \sqrt{g'} \:
\frac{1}{2}\varphi \left[\Delta' +\frac{1}{6}( R'-\Omega^{-2}
R)\right] \varphi
\label{6'}\:.
\label{ghost2}
\end{eqnarray}
When the static manifold is flat, $R=0$, the contribution of the
ghosts to the effective action can be written in terms of the $\zeta$
function of a conformally coupled scalar field:
\begin{eqnarray}
\ln Z^{(2)}_{\beta, \scriptsize \mbox{ghost}}(\alpha)&=&
-\int d^4x\sqrt{g'}\left [\frac{d}{ds}\zeta^{\scriptsize
\mbox{c.c.s.}}(s;x)|_{s=0}+\zeta^{\scriptsize
\mbox{c.c.s.}}(s;x)|_{s=0}\ln \sqrt{\alpha}\mu^2\right].
\nonumber\\
&&\label{fantom2}
\end{eqnarray}

Now, let us find the explicit form of the operator $\Lambda_\alpha$.
The following identity holds:
\begin{eqnarray}
\delta A &=& \frac{1}{\Omega} (\delta' A - \eta\cdot A) \label{2.5}\:,
\end{eqnarray}
provided  the $1$-form $\eta$ be defined as
\begin{eqnarray}
\eta = d (\ln \Omega) \equiv  \partial_\mu \ln \Omega \label{3}\:.
\end{eqnarray}
Taking into account that $\delta  = d^\dagger$ and employing Eq.s
(\ref{1}), (\ref{2}) and (\ref{3}) we get the identity
\begin{eqnarray}
S^{\scriptsize \mbox{em}}&=& \frac{1}{2}\int d^4x \left[\langle  A,
\Delta A\rangle - \left(1-\frac{1}{\alpha}\right) \langle  A, d\delta
A\rangle\right]+ S_{\scriptsize \mbox{ghost}}(\alpha)
\nonumber \\
&=& \frac{1}{2}\int d^4x\left[\langle  A, \Delta' A\rangle' -
\left(1-\frac{1}{\alpha}\right) \langle  A, d\delta' A\rangle'
+\frac{1}{\alpha}\langle A,\eta \eta\cdot A\rangle' \right.
\nonumber\\
&&\left. -\frac{1}{\alpha}\langle A, (\eta\delta' + d \eta \cdot)
A\rangle' \right]+ S'^{(2)}_{\scriptsize \mbox{ghost}}(\alpha)\:.
\label{4}
\end{eqnarray}
Looking at the first line of Eq. (\ref{4}) we find the explicit form
of the operator $\Lambda_\alpha$
\begin{eqnarray}
\Lambda_\alpha = \Delta' -\left(1-\frac{1}{\alpha}\right)d\delta' +
\frac{1}{\alpha} \eta\eta\cdot - \frac{1}{\alpha}(\eta \delta' + d
\eta \cdot)
\label{omega} \:.
\end{eqnarray}
Notice that the use of such an operator is equivalent to employing an
unusual gauge-fixing term in the initial photon Lagrangian which reads
\begin{eqnarray}
\frac{1}{\alpha} \langle  A , ( d -\eta)(\delta'-\eta\cdot)\:
A\rangle\:. \label{fixing}
\end{eqnarray}

\section{The case of the Rindler space}
\markboth{Optical approach to the cone}{The case of the Rindler space}
\label{secquattro}

Let us check the physical results arising from Eq.s (\ref{zeta1'}) and
(\ref{zeta2'}) in the case of the Rindler space.
Setting $\Omega^2=r^2$ in Eq. (\ref{conformal0}), the related
ultrastatic optical metric reads
\begin{eqnarray}
ds'^2 = d\tau^2 + r^{-2}(dr^2+dy^2+dz^2)
\label{coformalrindler}\:.
\end{eqnarray}
Obviously, this is the natural metric of $S^1\times H^3$ which does
not contain conical singularities. We remind one that $R'^\mu_\nu
=-2\mbox{ diag}(0,1,1,1)$ and $R' = -6$. As for the $1$-form $\eta$
necessary to define the operator $\Lambda_\alpha$, we get
\begin{eqnarray}
\eta_\mu = \frac{2}{r}\delta^r_\mu \label{etar}\:.
\end{eqnarray}

We want to employ a local $\zeta$-function
\cite{camporesi,libro,report} regularization technique and hence we
define the determinant of an (at least) symmetric operator $L$
through
\begin{eqnarray}
-\frac{1}{2} \ln\det (\mu^{-2} L) = \frac{1}{2}\int d^4x \sqrt{g'}\:
\left[ \zeta'(s=0;x) + \zeta(s=0;x) \ln \mu^2\right],
\label{determinant}
\end{eqnarray}
where the {\em local} $\zeta$ function of the operator $L$ is defined,
as usual, by means of the analytic continuation in the variable $s\in
C$ of the spectral representation of the complex power of the operator
$L$:
\begin{eqnarray}
\zeta(s;x) = \sum_n  \lambda_n^{-s}
 A_n(x) \cdot A_n^\ast(x) \label{zetafunction}\:.
\end{eqnarray}
Above, $A_n(x)$ is a $1$-form eigenfunction of a suitable
self-adjoint extension of the operator $L$ and $\lambda_n$ is its
eigenvalue. The index $n$ stands for all the quantum numbers, discrete
or continuous, needed to specify the spectrum. The set of these modes
is supposed complete and (Dirac, Kroneker) $\delta$ normalized. We
will make also use of the following notation  for the $1$-forms on
$S^1\times H^3$:
\begin{eqnarray}
A \equiv (a | B ) \:, \nonumber
\end{eqnarray}
where $a$ indicates a $1$-form on $S^1$ and $B$ a $1$-form on $H^3$.
All the operations between forms which appear after ``$|$'' are
referred to the manifold $H^3$ and  its metrical structure only. Latin
indices $a,b,c,d,...$ are referred to the coordinates $r,y,z$ on $H^3$
only.

A suitable set of eigenfunctions of the operator $\Delta' -
(1-\alpha^{-1})d\delta'$ as well as $\Lambda_\alpha$ as  can be
constructed using the following complete and normalized set of
eigenfunction of the scalar Hodge de Rham Laplacian on $S^1\times
H^3$:
\begin{eqnarray}
\phi^{({\bf k},n,\omega)}(\tau, r,{\bf x}) = \frac{e^{i{\bf k}{\bf x}}\:
e^{i\nu_n \tau}}
{2\pi^2 \sqrt{\beta}} \sqrt{2\omega \sinh (\pi\omega) }\: r
K_{i\omega}(kr) \label{phi}\:,
\end{eqnarray}
where $\nu_n = \frac{2\pi n}{\beta}$, $n\in Z$, $\omega\in R^+$, ${\bf
k}= (k_y,k_z)\in R^2$, $k= |{\bf k}|$ and all the previous
eigenfunctions have eigenvalue $(\nu^2+ \omega^2 +1)$.
$K_{i\omega}(x)$ is the usual MacDonald function with an imaginary
index. The normalization reads
\begin{eqnarray}
\int d^4x \sqrt{g'}\: \phi^{({\bf k},n,\omega)\ast}\phi^{({\bf
k}',n',\omega')} = \delta^{nn'}\:\delta^2({\bf k}-{\bf
k}')\delta(\omega-\omega')\:.\nonumber
\end{eqnarray}

In the following, we report some relations which are very useful in
checking the results which we shall report shortly. It is convenient
to define the $1$-form $\xi = - d(1/r) = \eta/2r$ on $H^3$. On $H^3$
we have:
 $\delta \eta =4$, $\Delta \eta = 0$, $d \eta= 0$, $d\xi =0$,
 $\Delta \xi = - 3\xi $, $\nabla^{a} \xi_b= - \delta^a_b/r$.
 Furthermore remind that, if $f$ is a 0-form and
$\omega$ an 1-form:
\begin{eqnarray}
[\Delta (f \omega)]_a = f  [\Delta \omega]_a +
 \omega_a \Delta f - 2 (\nabla_b f) \nabla^b \omega_a\:.
\label{formula1}
\end{eqnarray}
Finally, on a $3$-manifold the following relation holds
\begin{eqnarray}
 \Delta * (\omega\wedge\omega') & = & *[(\Delta \omega)\wedge
\omega'] +
*[\omega \wedge \Delta \omega'] +R * (\omega\wedge\omega')
- * [(R \omega)\wedge \omega']\nonumber \\
&& - * (\omega \wedge R \omega') - 2 * (\nabla_a \omega \wedge
\nabla^{a}\omega'),\nonumber\\
&&  \label{formula2}
\end{eqnarray}
where obviously $[* (\nabla_a \omega \wedge \nabla^{a}\omega')]_e :
=\sqrt{g} \epsilon_{ebc} (\nabla_d \omega^{b})\nabla^d \omega'^c$,
$\omega$ and $\omega'$ are $1$-forms and the Ricci tensor acts on
$1$-forms trivially as $(R\omega)_a = R_a^b\omega_b$.

\subsection{First optical approach}

Let us now consider the first optical approach, in which we define the
path integral directly in the optical manifold, see Eq.
(\ref{firstcase}). Starting from the scalar eigenfunctions, one can
obtain the following set of eigenfunctions of the operator $\Delta -
(1- \alpha^{-1})d\delta$ on $S^1\times H^3$.
\begin{eqnarray}
A^{(1)} &=& \frac{\sqrt{\omega^2+1}}{|\nu|\sqrt{\omega^2+ \nu^2 +1}}
(\:\partial_\tau \phi \:|\: d\phi \:) \nonumber\\
&=&\frac{\sqrt{\omega^2+1}}
{|\nu|\sqrt{\omega^2+ \nu^2 +1}} (\partial_\tau \phi, \partial_r \phi,
\partial_y \phi, \partial_z \phi)
\nonumber\\
A^{(2)} &=& \frac{1}{\sqrt{\omega^2+ \nu^2 +1}} (\:\partial_\tau \phi
\:|\: \frac{-\nu^2}{\omega^2+1} d \phi\:)
\nonumber\\
A^{(3)} &=& \frac{1}{k} (\:0\:|\: *d (\xi \phi) \:) =\frac{1}{k} (0,0,
\partial_z \frac{\phi}{r}, -\partial_y \frac{\phi}{r})
\nonumber \\
A^{(4)} &=& \frac{1}{k\omega}(\:0\:|\: \delta d (\xi \phi)\:)
=\frac{r}{k\omega} (0, \frac{k^2}{r}\phi, \partial_r\partial_y
\frac{\phi}{r}, \partial_r \partial_z \frac{\phi}{r})
\nonumber.
\end{eqnarray}
The last three modes are transverse, $\delta A =0$, whereas the  first
one is a pure gauge mode. From a little Hodge algebra,  the following
normalization relations can be proved:
\begin{eqnarray}
\int d^4x \langle  A^{(J,\omega,n,{\bf k})*}, A^{(J',\omega',n',{\bf
k}')}\rangle =\delta^{JJ'} \delta^{nn'}\:\delta^2({\bf k}-{\bf
k}')\:\delta(\omega-\omega').
\end{eqnarray}
As far as the eigenvalues are concerned, we have:
\begin{eqnarray}
\left[ {\Delta}' - (1- \alpha^{-1}) d \delta' \right] A^{(1)}
&=& \frac{\omega^2 + \nu^2 +1}{\alpha } A^{(1)} \:,\nonumber \\
\left[ {\Delta}' - (1- \alpha^{-1})d \delta'  \right] A^{(2)}
&=& (\omega^2 + \nu^2 +1) A^{(2)} \:,\nonumber \\
\left[ {\Delta}'- (1- \alpha^{-1})d \delta'  \right] A^{(J)}
&=& (\nu^2 + \omega^2) A^{(J)}\hspace{1cm}\mbox{if}\:\:\:\:J = 3,4
\nonumber\:.
\end{eqnarray}
Employing the definition in Eq. (\ref{zetafunction}), the above modes
and the definitions given in  Appendix A, we have that (notice that
$\phi^\ast$ and $\phi$ take the same values of ${\bf k}$, $n$,
$\omega$)
\begin{eqnarray}
\zeta(s;x) &=&  (\alpha^s+1) \sum_{n=-\infty}^{\infty} \int d^2{\bf k}
d\omega \frac{\phi^*(x) \phi(x) }{[\omega^2 + \nu^2 +1]^s} \nonumber\\
&&+\sum_{n=-\infty}^{\infty} \int d^2{\bf k} d\omega \frac{2 (1+
\omega^{-2})\phi^*(x)\phi(x) }{[\omega^2 + \nu^2]^s} \nonumber\\
&&\nonumber\\
&=&(\alpha^s+1) \zeta^{\scriptsize \mbox{m.c.s.}}(s;x)
+2\zeta^{\scriptsize \mbox{c.c.s.}}(s;x)+
\zeta^{\scriptsize \mbox{extra}}(s;x)
\label{zetafunction'}\:,
\end{eqnarray}
where we have set
\begin{eqnarray}
\zeta^{\scriptsize \mbox{extra}}(s;x)&=& 2\sum_{n=-\infty}^{\infty}
\int d^2{\bf k} \int \frac{d\omega}{\omega^2} \frac{\phi^*(x)\phi(x)
}{[\omega^2 + \nu^2]^s}\nonumber\\
&=&\frac{\sqrt{\pi}}{\pi^2\beta}\frac{\Gamma(s-\frac{1}{2})}
{\Gamma(s)}\left(\frac{\beta}{2\pi}\right)^{2s-1}\zeta_R(2s-1),
\label{zetanew}
\end{eqnarray}
so that $\zeta^{\scriptsize \mbox{extra}}(s=0;x)=0$ and
$\zeta'^{\scriptsize \mbox{extra}}(s=0;x)=1/3\beta^2$. Notice that the
second and third terms in Eq. (\ref{zetafunction'}) arise from the
transverse modes $A^{(3)}$ and $A^{(4)}$. The first term in
Eq. (\ref{zetafunction'}) is due to the modes with $J=1,2$.

In calculating Eq. (\ref{zetafunction'}), we encountered Kabat's
surface terms similar to those we encountered in \cite{ielmo}.
However, in the present case all these terms vanish automatically and
no further regularization procedure needs. In fact, all these terms
read as
\begin{eqnarray}
D_r \sum_n \int d{\bf k} \int d \omega
r^2 K_{i\omega}(kr)K_{i\omega}(kr) f(\omega,\nu,s)\:,\nonumber
\end{eqnarray}
where $D_r$ is an opportune differential operator in $r$. Passing from
the integration variable ${\bf k}$ to the integration variable $r{\bf
k}$, we see that the term after the operator does not depend on $r$,
and so the differentiation produces a vanishing result.

In order to write the complete local $\zeta$ functions of the
electromagnetic field we have to take account of the ghost
contribution. We have already said that in this approach the $\zeta$
function of the ghosts is just minus two times the $\zeta$ function of
a minimally coupled scalar field, but with a gauge-fixing dependent
scale factor, see Eq. (\ref{fantom1}):
\begin{eqnarray}
\zeta_\alpha^{\scriptsize \mbox{ghosts}}(s;x)= -2\zeta^{\scriptsize
\mbox{m.c.s.}}(s;\mu^{-2}\alpha^{-\frac{1}{2}}L_{\xi=0})(x).
\nonumber
\end{eqnarray}
Using this relation, Eq. (\ref{zetafunction'}) and reintroducing
everywhere the renormalization scale $\mu$, we can write the
complete local $\zeta$ function  of the electromagnetic field as
\begin{eqnarray}
\zeta^{\scriptsize \mbox{em}}(s;x)&=&
(\alpha^s+1) \zeta^{\scriptsize \mbox{m.c.s.}}(s;\mu^{-2}L_{\xi=0})(x)
+2\zeta^{\scriptsize \mbox{c.c.s.}}(s;\mu^{-2}L_{\xi=\frac{1}{6}})(x)
\nonumber\\
&&+\zeta^{\scriptsize \mbox{extra}}(s;\mu^{-2})(x)-2
\zeta^{\scriptsize\mbox{m.c.s.}}
(s;\mu^{-2}\alpha^{-\frac{1}{2}}L_{\xi=0})(x).
\label{zetacompleta}
\end{eqnarray}
It follows that the one-loop effective Lagrangian density is just
\begin{eqnarray}
{\cal{L}}_{\scriptsize \mbox{eff}}(x)&=&\frac{1}{2}
\frac{d}{ds}\left[2\zeta^{\scriptsize \mbox{c.c.s.}}(s;x)+
\zeta^{\scriptsize \mbox{extra}}(s;x)\right]_{s=0}\nonumber\\
&=&\frac{\pi^2}{45\beta^4}+\frac{1}{6\beta^2}.
\label{efflagr}
\end{eqnarray}
We remark the importance of keeping the $\alpha$ dependence of the
action of the ghosts: it gives a contribution proportional to
$\ln\alpha$ which cancels against the ($\ln\alpha$)-dependent term
coming from $(\alpha^s+1)\zeta^{\scriptsize \mbox{m.c.s.}}(s;x)$,
restoring the gauge invariance of the theory. Note also how all the
terms containing $\ln\mu^2$ cancel giving the expected scale invariant
theory.

By integrating this quantity over the manifold and introducing
a cutoff at a distance $\epsilon$ form $r=0$ in order to control
the horizon divergence, we get the one-loop free energy:
\begin{eqnarray}
 F^{(1)}= -\frac{1}{\beta}\int
d^4x\sqrt{g'}{\cal{L}}_{\scriptsize \mbox{eff}}(x)
=-\frac{{\cal A_H}}{1440\pi^2\epsilon^2}
\left[\left(\frac{2\pi}{\beta}\right)^4
+30\left(\frac{2\pi}{\beta}\right)^2\right].
\label{freeen}
\end{eqnarray}
Up to a constant term which is fixed by the subtraction procedure,
the above result coincides with the free energy obtained
integrating the component $T_{00}$  (\ref{tensor2}) of the
energy-momentum tensor computed in the static
manifold by means of the point splitting or the
$\z$ function.

\subsection{Second optical approach}

Let us then consider the second optical approach. We were able to
perform the calculations  in the case $\alpha=1$ only, hence a
complete discussion on the gauge invariance ($\alpha$ invariance) is
not possible. However, the found result contains some interest. As
before, the eigenfunctions of the operator $\Lambda_{\alpha=1}$ are
constructed from the scalar eigenfunctions, Eq. (\ref{phi}):
\begin{eqnarray}
A^{(1)} &=& \frac{1}{|\nu|} (\:\partial_\tau \phi \:|\: + \frac{|\nu
|}{2}\eta \phi\:) =\frac{1}{|\nu|} (\partial_\tau \phi, |\nu|
\frac{\phi}{r}, 0, 0)   \nonumber\\
A^{(2)} &=& \frac{1}{|\nu|}(\:\partial_\tau \phi \:|\: - \frac{|\nu
|}{2}\eta \phi\:) = \frac{1}{|\nu|} (\partial_\tau \phi, -|\nu|
\frac{\phi}{r}, 0, 0)   \nonumber\\
A^{(3)} &=& \frac{1}{k} (\:0\:|\: *d (\xi \phi) \:) =\frac{1}{k}(0,0,
\partial_z \frac{\phi}{r}, -\partial_y \frac{\phi}{r}) \nonumber \\
A^{(4)} &=& \frac{1}{k}(\:0\:|\: * (\xi \wedge *d (\xi \phi))\:)
=\frac{1}{k} (0,0, \partial_y \frac{\phi}{r}, \partial_z
\frac{\phi}{r})
\nonumber\:.
\end{eqnarray}
The following normalization relations hold:
\begin{eqnarray}
\int d^4x \langle  A^{(J,\omega,n,{\bf k})*}, A^{(J',\omega',n',{\bf
k}')}\rangle =\delta^{JJ'} \delta^{nn'}\:\delta^2({\bf k}-{\bf
k}')\:\delta(\omega-\omega').
\end{eqnarray}
As far as the eigenvalues are concerned, we have:
\begin{eqnarray}
\Lambda_{\alpha=1} A^{(J)} &=&
 \{ \omega^2 + [(-1)^J + |\nu|]^2 \} A^{(J)} \:\:\:\:\mbox{if}
\:\:\:\: J =1,2\nonumber\:,\\
\Lambda_{\alpha=1} A^{(J)} &=&
(\nu^2 + \omega^2) A^{(J)} \:\:\:\:\mbox{if}\:\:\:\:J = 3,4
\nonumber\:.
\end{eqnarray}
Employing the definition in Eq. (\ref{zetafunction}) and the found
modes we have (notice that $\phi^\ast$ and $\phi$ take the same
values of ${\bf k}$, $n$, $\omega$)
\begin{eqnarray}
\zeta^{(2)}(s;x) &=&
\sum_{n=1}^{\infty} \int d{\bf k} \int d\omega \frac{2\phi^*(x)
\phi(x) }{[\omega^2 + (\nu+1)^2]^s}  \nonumber\\
&&+\sum_{n=1}^{\infty} \int d{\bf k} \int d\omega \frac{2\phi^*(x)
\phi(x) }{[\omega^2 + (\nu-1)^2]^s}  \nonumber\\
&& + \sum_{n=1}^{\infty} \int d{\bf k} \int d\omega \frac{4\phi^*(x)
\phi(x) }{[\omega^2 + \nu^2]^s}  \label{zetafunction2'}\:.
\end{eqnarray}
For simplicity, we have omitted the terms corresponding to $n=0$,
which contribute only to the temperature-independent part of the free
energy: this part will be changed during the renormalization process
(subtraction of the Minkowski vacuum energy). We also stress that
Kabat's surface terms involved during the calculations disappeared
exactly as in the previous approach. The latter term in Eq.
(\ref{zetafunction'}) is due to the modes with $J=3,4$: this term is
exactly twice the $\zeta$ function of a conformally coupled Euclidean
scalar field propagating in $S^1\times H^3$.

As far as the ghost contribution is concerned, it arises from the
action (\ref{5}).  Since the corresponding small fluctuations operator
involves the curvature of the Euclidean Rindler manifold, which has a
Dirac $\delta$ singularity at $r=0$, mathematically it is not well
defined and is not clear how to deal with it. However, as a try we can
suppose to consider $R=0$ and see the consequences.\footnote{See
footnote number 5.} Under this hypothesis, the ghost contribution is
just minus twice that of a conformally coupled scalar field (see Eq.
(\ref{fantom2})) and so it cancels against the contribution of the
modes $J=3,4$.

After having added the ghost contribution, we can write the complete
$\zeta$ function of the electromagnetic field as
\begin{eqnarray}
\zeta^{\scriptsize\mbox{em}}(s;x)
&=&  \sum_{n=1}^{\infty} \int d{\bf k} \int d\omega \frac{2\phi^*(x)
\phi(x) }{[\omega^2 + (\nu+1)^2]^s} \nonumber\\
&& +  \sum_{n=1}^{\infty} \int d{\bf k} \int d\omega \frac{2\phi^*(x)
\phi(x) }{[\omega^2 + (\nu-1)^2]^s} \label{zetaphotons2}\:.
\end{eqnarray}
The partition function of the photons is obtained employing the
previous function opportunely continued in the variable $s$ in
Eq. (\ref{determinant}). Dealing with it as in
the previous case, we finally find the free energy
\begin{eqnarray}
F^{(2) \scriptsize \mbox{sub}} =
-\frac{{\cal A_H}}{1440 \pi^2 \epsilon^2}
\left[ \left( \frac{2\pi}{\beta}\right)^4 -
30 \left( \frac{2\pi}{\beta}\right)^2
 +29   \right] \:.
\label{freen2}
\end{eqnarray}
In deriving this result we have employed the Riemann zeta function
$\zeta(z, q)$ and its relation with the Bernoulli polynomials
\cite{GR}. This result has the same form as that obtained with the
first approach, Eq. (\ref{freeen}), but the sign in front to the
second term is opposite. The third term is fixed by the
renormalization procedure. The problems arise with  the $\beta^{-2}$
term once again.

In this case it is easy to identify the origin of the discrepancy in our
hypothesis of setting $R=0$ in the ghost action. Nevertheless, there
is some evidence that the origin is not that. In particular, if we
assume that the optical method gives the same results as the
point-splitting one even when $\xi\neq1/6$, then we can suppose
that it is right to substitute the optical result for the ghost
contribution to the above free energy with the point-splitting one
for $\xi=0$. As a result, we get
\begin{eqnarray}
F^{(2) \scriptsize \mbox{sub}} =
-\frac{{\cal A_H}}{1440 \pi^2 \epsilon^2}
\left[ \left( \frac{2\pi}{\beta}\right)^4 -
60 \left( \frac{2\pi}{\beta}\right)^2
+59   \right] \:,
\label{freen3}
\end{eqnarray}
which is different from the previous one but still different from
the first optical  approach result. In
particular, the free energy in Eq. (\ref{freen2}) (or (\ref{freen3}))
would yield a negative entropy at the Unruh-Hawking temperature,
which is very hard to  accept on a physical ground.
Summarizing, it seems to us that this second approach, which is the
natural generalization of the procedure used in the scalar case, does
not yield a correct result.

\section{Summary and Discussion}
\markboth{Optical approach to the cone}{Summary and Discussion}
\label{seccinque}

Let us summarize the main results of this Chapter.
First of all, we have seen that the canonical definition of free
energy is equivalent to the path integral approach in the optical
manifold instead that in the physical static manifold. In general the
two definitions are equivalent, since the difference is a temperature
independent term which does not affect the thermodynamic. However, we
have seen that in the Rindler case, and in general in presence of
conical singularity, the two definitions are not equivalent, leading
to different results. Moreover, we have shown that the partition
function computed in the physical static manifold suffers from some
serious inconsistencies when one tries to apply standard
statistical-mechanical relations, while the optical results are free
from such inconsistencies. The simplest explanation of such problems
would be the existence of anomalous temperature-dependent terms in the
Jacobian of the conformal transformation which relate the static to
the optical manifold, but no definitive computation has been possible to
support this hypothesis, and other explanations are possible. It
is important to notice that these problems do not affect the
discussion of Chapter \ref{BHENTROPY}, and in particular the
criticism to the integrated approach, since also the optical
results show the same kind of horizon divergences as the local
approach employed there, the only difference being in the numerical
coefficient.

The second important result of this Chapter is the proof that the optical method
(the ``first approach'') can be used to compute one-loop quantities in
the Rindler space also in the case of the photon field. The method has
been developed employing a general covariant gauge choice.
It is also important to stress that the partition function arising
from our method is completely free from  ``Kabat's'' surface terms.
This is very important because, as we previously said, the approaches
based on the direct computation in the Euclidean Rindler space using
$\zeta$-function or heat-kernel techniques produces such anomalous
terms \cite{kabat,ielmo} and further regularization procedures seem to be
necessary to get physically acceptable results.

We  have also developed a general optical formalism for the Maxwell
field in the covariant gauges based on Hodge de Rham formalism which,
in principle, can be used in different manifolds than the Rindler
space.

However,  many problems remain to be explained. In particular, both in
the photon and in the scalar case the relation between the optical
approach and the direct approach in the manifold with the conical
singularity remains quite obscure. This is due to difficulties
involved in computing the Jacobian of the conformal transformation
in the presence of conical singularities. Moreover, while
the optical approach can be used in the case of massless
fields without particular difficulties, as soon as the fields have
a mass the optical method becomes much harder to apply.
In this case, the direct computation in the manifold with
conical singularities could show its advantages, provided
one knows how to compute the above Jacobian.

Another general point  which requires further investigation is the
request of self-consistency of the thermodynamics of the gas of
Rindler particle, when the temperature is not the Unruh one. This is a
very important point in calculating the correction to the entropy of a
black hole supposing such corrections due to the fields propagating
around it. We remind one that the Rindler metric approximates the region near
the horizon of a Schwarzschild black hole. The entropy of the fields
is computed using the relation (where $\beta_H$ is the Unruh-Hawking
temperature, $2\pi$ in the Rindler case):
 $S_{\beta_H} = \beta_H^2 \partial_\beta F_\beta |_{\beta_H}$.
In calculating the previous derivative at $\beta=\beta_H$, one has to
consider also the partition function {\em off shell}, namely evaluated
at $\beta \neq 2\pi $ and $\beta$ {\em near} $\beta_H$. It is not so
clear whether it is necessary or not that the thermodynamical laws
hold also for $\beta \neq \beta_H $ and $\beta$ near $\beta_H$ in
order to assure the consistency the procedure followed in calculating
the entropy of the fields at $\beta= \beta_H$. Moreover, it is
well known that the off shell quantum states of a field are affected
by several pathologies on the horizon event.\footnote{Rindler thermal
states with $\beta\neq 2\pi$ violates several axioms of the QFT in
curved backgrounds. For example, see \cite{haaglibro} and ref.s
therein.} Furthermore, they are unstable states in a semiclassical
approach to quantum gravity due to the divergence of the renormalized
stress tensor on the horizon. Thus, it is reasonable to wonder about
the thermodynamical consistency of the  results when one works
off-shell. Although inconsistencies found in the physical
manifold results in  support this view, the optical results do not
show thermodynamical inconsistencies, and so it is likely that
the explanation has different origin.

\medskip
\section{Appendix A}
\markboth{Optical approach to the cone}{Appendix A}

In computing the photon $\zeta$ function on $S^1\times H^3$ one meets
the $\zeta$ function of a scalar field in the same background, both in
conformal and minimal coupling. Therefore, it is useful to report here
these  $\zeta$ functions. The small fluctuations operator for a scalar
field in the optical metric is
\begin{eqnarray}
L_\xi=\Delta-6\xi=-[\partial_\tau^2-r\partial_r
+r^2\partial_r^2+6\xi].
\nonumber
\end{eqnarray}
where $\Delta$ is the Hodge de Rham Laplacian on $S^1\times H^3$. A
complete set of eigenfunctions has been given in the main text,
Eq. (\ref{phi}), with eigenvalue $[\nu_n^2+\omega^2+1-6\xi]$.
Therefore, the local $\zeta$ function is
\begin{eqnarray}
\zeta(s|L_\xi)(x)&=&\sum_{n=-\infty}^\infty
\int_0^\infty d\omega\int d^2{\bf k}\,[\nu_n^2+\omega^2+1-6\xi]^{-s}
\phi^\ast(x)\phi(x)\nonumber\\
&=&\frac{\sqrt{\pi}}{8\pi^2\beta}\frac{\Gamma(s-\frac{3}{2})}
{\Gamma(s)}\sum_{n=-\infty}^\infty
\int_0^\infty d\omega\,\omega^2[\nu_n^2+\omega^2+1-6\xi]^{-s}
\nonumber\\
&=&\frac{\sqrt{\pi}}{8\pi^2\beta}\frac{\Gamma(s-\frac{3}{2})}
{\Gamma(s)}\left(\frac{2\pi}{\beta}\right)^{3-2s}\times \nonumber\\
&&\times\left[2E\left(s-\frac{3}{2};\frac{\beta}{2\pi}
\sqrt{1-6\xi}\right)-
\left(\frac{\beta}{2\pi}\sqrt{1-6\xi}\right)^{3-2s}\right],
\nonumber
\end{eqnarray}
where $E(s;a)=\sum_{n=0}^\infty [n^2+a^2]^{-s}$ is the Epstein
$\zeta$ function. In the conformally coupled case, the Epstein
function becomes a Riemann $\zeta$ function and so
\begin{eqnarray}
\zeta^{\scriptsize \mbox{c.c.s.}}(s;x)\equiv
\zeta(s|L_{\xi=\frac{1}{6}})(x)=\frac{\sqrt{\pi}}{4\pi^2\beta}
\left(\frac{\beta}{2\pi}\right)^{2s-3}\frac{\Gamma(s-\frac{3}{2})}
{\Gamma(s)}\zeta_R(2s-3),\nonumber
\end{eqnarray}
One can easily check that
$\zeta^{\scriptsize \mbox{c.c.s.}}(s;x)|_{s=0}=0$ and
$$
\frac{d}{ds}\zeta^{\scriptsize
\mbox{c.c.s.}}(s;x)|_{s=0}=\frac{\pi^2}{45\beta^4}. $$
Another important case is the minimally coupled one, $\xi=0$, for
which there is not a more explicit form. However, using the identity
$$
E(s;a)=\frac{1}{2a^{2s}}+
\frac{\sqrt{\pi}}{2}\frac{\Gamma(s-\frac{1}{2})}{\Gamma(s)}a^{1-2s}
+\frac{2\sqrt{\pi}}{\Gamma(s)}\sum_{n=1}^\infty (\frac{\pi
n}{a})^{s-\frac{1}{2}} K_{s-\frac{1}{2}}(2\pi n a).
$$
and the  fact that the MacDonald function $K_\nu(x)$ is analytic in
the index $\nu$ and decays exponentially as $|x|\rightarrow\infty$
so that the third term in the previous expansion is analytic in $s$
(and vanishes as $s \rightarrow 0$),
we find that the $\zeta$ function does not vanish in
$s=0$:
$$\zeta^{\scriptsize \mbox{m.c.s.}}(s;x)|_{s=0}=\frac{1}{32\pi^2}.$$
We do not know the value in zero of the derivative, but it is not
required in our computations.

\section{Appendix B: On the Jacobian}
\markboth{Optical approach to the cone}{Appendix B: On the Jacobian}
\label{jacosection}

In this Appendix we want to discuss the dependence on $\b$ of the
Jacobian of the conformal transformations in presence of conical
singularities.   As we have said in the Introduction, on regular
manifolds the logarithm of the Jacobian in Eq. (\ref{foundamental})
is simply proportional to
$\beta$. However, the case of
the Euclidean Rindler space could be more complicated, due to the
presence of a conical singularity at $r=0$, which could yield a
nonlinear dependence on $\beta$. In fact, such a singularity can be
represented as an opportune Dirac $\delta$ function with a coefficient
containing a factor $(2\pi-\beta)$ \cite{SOL95,FUSOL95}, and so
$\beta$ enters not only as integration interval, but also in the
integrand. Of course, only an explicit calculation of the Jacobian can
give an ultimate answer. In two dimensions, the Jacobian
$J[g,g',\beta]$ is the exponential of the well-known Liouville action
\cite{POL81}:
$$
J[g,g',\beta]=\exp -\frac{1}{24}\int d^2x\sqrt{g}\,
g^{\mu\nu}\pa_\mu\sigma \pa_\nu\sigma,
$$
where $\sigma(x)=\ln\Omega=-\frac{1}{2}\ln g_{00}$. In the
Rindler case $g_{00}=r^2$ and the Jacobian is \cite{dAO}
\bea
\ln J[g,g',\beta]=\frac{\b}{24\pi}\ln \frac{R}{\ep},
\label{jacoD2}
\ena
where $\ep$ and $R$ are the usual horizon and volume cutoffs.
Therefore the Jacobian is indeed in the form (\ref{jaco}),
regardless of the conical singularity. It is worth noticing that
the above expression for the Jacobian is exactly the
difference between $F_\b$ and $F'_\b$ at $\b=2\pi$ \cite{dAO}.

Unfortunately, in four dimensions the form of
the Jacobian is far more complicate (see \cite{dAO}  and references
therein) and involves also products of curvature tensors which are ill
defined. A possible way to avoid this problem to compute the Jacobian
in the smoothed conical manifold of section \ref{riemangeom}, but it is
possible to see that  the relevant quantities remain
ill-defined as the regularization is removed.

An alternative approach to the computation of the Jacobian is the
following. The Jacobian can be computed for infinitesimal
transformations (see for example \cite{report} and the references
cited therein): consider the family of conformal transformations
$$
g_{\mu\nu}^q=e^{2q\sigma(x)}g_{\mu\nu},
$$
so that we can pass in a continuous way from the original static
metric $g_{\mu\nu}$ at $q=0$ to the optical metric at $q=1$.
Then, by considering an infinitesimal variation of $q$ one gets
\beas
\ln J[g_q,g_{q+\delta_q}]=\ln\frac{Z_{q+\delta q}}{Z_q}=
\frac{\delta q}{(4\pi)^{D/2}}
\int a_{D/2}(x)\sigma(x)\sqrt{g^q} d^Dx,
\label{infjaco}
\enas
where $a_{D/2}(x)$ is the Seeley-De Witt coefficient (which in the case
of conformal invariant theories is proportional to the trace anomaly).
By integrating over $q$ the above expression one gets the expression
of the Jacobian for a finite transformation:
\bea
\ln J[g,g';\b]=\frac{1}{(4\pi)^{D/2}}\int_0^1dq
\int a_{D/2}(x)\sigma(x)\sqrt{g^q} d^Dx.
\label{jacokernel}
\ena
In subsection \ref{singularsection} we have seen that  in the
case of the conical space  the Seeley-DeWitt coefficients
are distribution at $r=0$:  we can then try to compute the
Jacobian using these singular coefficients $a_{\b,n}(x)$.
In $D=4$ an immediate computation shows that\footnote{ We
inserted an arbitrary constant $\mu$ with the dimension of a
mass to adjust the dimensions.}
\bea
\ln J[g,g';\b]&=&-\frac{\mu^2}{16\pi^2} \int_0^1dq\int
a_{\b,2}(x)\,(\mu r)^{-4q}\,\ln\mu r\,\sqrt{g}\,d^4x\nonumber\\
&=&\frac{{\cal A_H}}{144\pi\mu^2}I_\b(-1)\frac{d^3}{dr^3}
\left[\frac{1}{r^3}-\mu^4r\right]_{r=0},
\label{jacoD4}
\ena
The result is badly divergent in $r=0$, but the interesting fact is
that the overall coefficient has a nontrivial dependence on $\b$: as a
matter of fact, it has the same dependence on $\b$ as the free energy
$F_\b$ computed in the static manifold!

Fortunately, there are few chances that the above result
(\ref{jacoD4}) is correct. First of all, the above Jacobian
vanishes for $\b=2\pi$, while it should have some  finite
value. Furthermore, the same computation in $D=2$ yields
$$
 \ln J[g,g';\b]=-\frac{1}{24\mu^2}
\left(\frac{2\pi}{\b}-\frac{\b}{2\pi}\right)
\frac{d}{dr}\left[\frac{1}{r}-\mu^2 r\right]_{r=0},
$$
which is different from the correct result (\ref{jacoD2}).

Summarizing, it is not yet clear whether the Jacobian has
a trivial dependence on $\b$ even in presence of conical
singularities, although the calculation in two dimensions
shows that  it is the most likely possibility.

\cleardoublepage
\markboth{}{}
\newpage

\chapter*{Conclusions}

To conclude this thesis I would like to point out
which are, in my opinion, the main issues which remain
to be discussed and clarified about quantum field theory
on the cone and the other topics discussed in this work.

There are several interesting problems of technical nature,
the most important being the obscure relation between the local
and the integrated approaches on the cone. In Chapter \ref{BHENTROPY}
I have shown that these two approaches lead to different conclusions
and I have also given an explanation of the origin of the differences
in the context of $\z$-function regularization. Nevertheless, it
remains to be explained from the point of view of other regularizations.
This problem is very important, because the integrated approach
is used by many authors, although we have seen that it yields results
which are physically less reasonable than those of the local approach.

Another interesting problem is that of the inconsistencies
found in Chapter \ref{OPTICCHAP} for the thermodynamical
relations for a thermal gas in the Rindler wedge when the
free energy is computed in the physical metric instead that
in the optical one. Although the reason could be simply
an anomalous dependence on the temperature of the Jacobian
of the conformal transformation which relates the two metrics,
there could be other more interesting explanations.

A less technical problem is that of the Bekenstein-Hawking
entropy: not only the tree-level entropy has no satisfactory
explanation in terms of number of possible states of the system, but
the one-loop corrections to the entropy diverge unless some cutoff is
introduced, and the physical explanation of the origin of this cutoff
is lacking yet. As it was discussed in Chapter \ref{BHENTROPY}, it is
my opinion that these divergences cannot be simply renormalized away
as the usual ultraviolet divergences. It is probable that satisfactory
answers to these problems cannot be found within quantum field
theory on curved spacetimes, but, as the recent progress in string
theory shows,  only within a more fundamental theory  which takes into
account the quantum nature of gravity itself.

\cleardoublepage

\end{document}